\documentclass[12pt]{article}
\usepackage[pdftex]{graphics}
\usepackage{amsmath,amssymb,epsfig,ulem,pdflscape}
\usepackage[square,comma,sort&compress]{natbib}
\allowdisplaybreaks 

\def\bm#1{\mbox{\boldmath$#1$\unboldmath}}
\def\sgn{\mbox{sgn}}
\def\Mkk{M_{\rm KK}}
\def\delslash{\rlap{\hspace{0.02cm}/}{\partial}}
\def\Aslash{\rlap{\hspace{0.08cm}/}{A}}
\newcommand{\beq}{\begin{equation}}
\newcommand{\eeq}{\end{equation}}
\newcommand{\ord}{{\cal O}}
\newcommand{\ie}{{\it i.e.}}
\newcommand{\eg}{{\it e.g.}}
\newcommand{\etc}{{\it etc.}}

\begin{document}

\begin{titlepage}

\begin{flushright}
MZ-TH/08-18\\
October 20, 2008 
\end{flushright}

\vspace{0.5cm}
\begin{center}
\Large\bf
Flavor Physics in the Randall-Sundrum Model\\[0.2cm]
\large\bf
I.~Theoretical Setup and Electroweak Precision Tests
\end{center}

\vspace{0.2cm}
\begin{center}
{\sc S. Casagrande, F. Goertz, U. Haisch, M. Neubert and T. Pfoh}\\
\vspace{0.4cm}
{\sl Institut f\"ur Physik (THEP), Johannes Gutenberg-Universit\"at\\
D-55099 Mainz, Germany}
\end{center}

\vspace{0.2cm}
\begin{abstract}\noindent
A complete discussion of tree-level flavor-changing effects in the Randall-Sundrum (RS) model with brane-localized Higgs sector and bulk gauge and matter fields is presented. The bulk equations of motion for the gauge and fermion fields, supplemented by boundary conditions taking into account the couplings to the Higgs sector, are solved exactly. For gauge fields the Kaluza-Klein (KK) decomposition is performed in a covariant $R_\xi$ gauge. For fermions the mixing between different generations is included in a completely general way. The hierarchies observed in the fermion spectrum and the quark mixing matrix are explained naturally in terms of anarchic five-dimensional Yukawa matrices and wave-function overlap integrals. Detailed studies of the flavor-changing couplings of the Higgs boson and of gauge bosons and their KK excitations are performed, including in particular the couplings of the standard $W^\pm$ and $Z^0$ bosons. A careful analysis of electroweak precision observables including the $S$ and $T$ parameters and the $Z^0 b\bar b$ couplings shows that the simplest RS model containing only Standard Model particles and their KK excitations is consistent with all experimental bounds for a KK scale as low as a few TeV, if one allows for a heavy Higgs boson ($m_h\lesssim 1$\,TeV) and/or for an ultra-violet cutoff below the Planck scale. The study of flavor-changing effects includes analyses of the non-unitarity of the quark mixing matrix, anomalous right-handed couplings of the $W^\pm$ bosons, tree-level flavor-changing neutral current couplings of the $Z^0$ and Higgs bosons, the rare decays $t\to c(u) Z^0$ and $t\to c(u) h$, and the flavor mixing among KK fermions. The results obtained in this work form the basis for general calculations of flavor-changing processes in the RS model and its extensions. 
\end{abstract}
\vfil

\end{titlepage}

\section{Introduction}

The Standard Model (SM) of particle physics has passed every direct experimental test with flying colors. Yet it is not an entirely satisfactory theory, because it raises but leaves unanswered many fundamental questions. In particular, there are many hierarchies built into the SM that have {\it a priori} no explanation. The most famous of these is the huge separation between the electroweak and the Planck scales. Various solutions have been proposed to explain this hierarchy. One particularly appealing possibility is the Randall-Sundrum (RS) scenario \cite{Randall:1999ee}. In this model, four-dimensional (4D) Minkowskian space-time is embedded into a slice of five-dimensional (5D) anti de-Sitter $({\rm AdS}_5)$ space with curvature $k$. The fifth dimension is a $S^1/Z_2$ orbifold of size $r$ labeled by a coordinate $\phi \in [-\pi,\pi]$, such that the points $(x^\mu, \phi)$ and $(x^\mu, -\phi)$ are identified. In its original form, the metric of the RS geometry is given by
\beq
   ds^2 = e^{-2\sigma(\phi)}\,\eta_{\mu\nu}\,dx^\mu dx^\nu 
    - r^2 d\phi^2 \,, \qquad
   \sigma(\phi) = kr|\phi| \,,
\eeq
where $x^\mu$ denote the coordinates on the 4D hyper-surfaces of constant $\phi$ with metric $\eta_{\mu\nu}=\mbox{diag}(1,-1,-1,-1)$, and $e^\sigma$ is called the warp factor. Three-branes are placed at the orbifold fixed points $\phi=0$ as well as $\phi=\pi$ and its reflection at $\phi=-\pi$. The brane at $\phi=0$ is called Planck or ultra-violet (UV) brane, while the brane at $\phi=\pi$ is called TeV or infra-red (IR) brane. The parameters $k$ and $1/r$ are assumed to be of the order of the fundamental Planck scale $M_{\rm Pl}$, with the product $kr\approx 12$ chosen such that the inverse warp factor
\beq
   \epsilon = \frac{\Lambda_{\rm IR}}{\Lambda_{\rm UV}}
   \equiv e^{-kr\pi} \approx 10^{-16}
\eeq
explains the hierarchy between the electroweak and the Planck scales.

In the RS framework the question about the origin of the gauge hierarchy is thus transformed into the question why the logarithm of the warp factor,
\beq\label{Ldef}
   L \equiv -\ln \epsilon = kr\pi\approx 37\gg 1 \,,
\eeq
is moderately large compared to its natural size of the order of a few. Indeed, this small hierarchy can be generated dynamically with reasonable tuning using the Goldberger-Wise stabilization mechanism \cite{Goldberger:1999uk}, and we will take for granted that such a mechanism is at work. The warp factor also sets the mass scale for the low-lying Kaluza-Klein (KK) excitations of the SM fields to be of order of the ``KK scale''
\beq
   \Mkk\equiv k\epsilon = k\,e^{-kr\pi} 
   = \ord(\mbox{few TeV}) \,.
\eeq
For instance, the masses of the first KK gluon and photon are approximately $2.45\Mkk$.

While the original RS model aimed at solving the hierarchy problem up to the Planck scale, there may be good arguments in favor of lowering the UV scale to a value significantly below $M_{\rm Pl}$. Higher-dimensional spaces with warp factors arise naturally in flux compactifications of string theory \cite{Klebanov:2000hb,Giddings:2001yu,Kachru:2003aw,Brummer:2005sh}, and it is thus not unlikely that the RS model will have to be embedded into a more fundamental theory (a UV completion) at some scale $\Lambda_{\rm UV}\ll M_{\rm Pl}$. From a purely phenomenological point of view, it would be possible to lower this cutoff to a scale only few orders of magnitude above the TeV scale, even though in this case the true solution to the hierarchy problem is only postponed to larger energy. Such a scenario has been called the ``little RS'' model \cite{Davoudiasl:2008hx}, in analogy with ``little Higgs'' models, which stabilize the Higgs mass up to scales of order (10--100)\,TeV \cite{ArkaniHamed:2001nc,ArkaniHamed:2002qy}. While less appealing from a conceptual point of view, the possibility that the parameter $L$ in (\ref{Ldef}) could be less than about 37 should not be discarded. As we will see, many amplitudes in the RS model are enhanced by this parameter.

By virtue of the AdS/CFT correspondence \cite{Maldacena:1997re,Gubser:1998bc,Witten:1998qj}, 5D gravitational theories in anti~de-Sitter space have a dual description in terms of strongly coupled 4D conformal field theories. For the case of the RS scenario, in which the conformal symmetry is broken on the IR brane, implications of this correspondence have been explored in \cite{Verlinde:1999fy,ArkaniHamed:2000ds,Rattazzi:2000hs,Contino:2004vy} (for a review, see \cite{Gherghetta:2006ha}). Among other things, the holographic dictionary implies that 5D fields living near the IR brane in the RS setup correspond to composite objects in the conformal field theory, while fields living near the UV brane correspond to elementary particles. In a setup where the Higgs sector is localized on (or near) the IR brane, it is thus natural to think of the Higgs boson as a composite object \cite{Kaplan:1983fs,Kaplan:1983sm}, whose mass is naturally of the order of the KK scale.\footnote{This may be different in gauge-Higgs unification models \cite{Agashe:2004rs}, which will not be considered here.} The fact that a light Higgs boson with $m_H\ll\Mkk$ is (in this sense) unnatural in the RS model will become important for parts of the discussion in this work.

It will often be convenient to introduce a coordinate $t=\epsilon\,e^{\sigma(\phi)}$, which equals $\epsilon$ on the UV brane and 1 on the IR brane \cite{Grossman:1999ra}. Integrals over the orbifold are then obtained using
\beq
   \int_{-\pi}^\pi\!d\phi 
   \to \frac{2\pi}{L} \int_\epsilon^1\!\frac{dt}{t} \,, 
    \qquad
   \int_{-\pi}^\pi\!d\phi\,e^{\sigma(\phi)}
   \to \frac{2\pi}{L\epsilon} \int_\epsilon^1\!dt \,,
    \qquad \etc
\eeq
Another widely used form of the RS background is the conformally flat metric \cite{Csaki:2002gy} 
\beq\label{eq:conformallyflat}
   ds^2 = \left( \frac{R}{z} \right)^2 
   \left( \eta_{\mu\nu}\,dx^\mu dx^\nu - dz^2 \right) ,
\eeq
restricted to the interval $z\in[R,R']$, where $R$ and $R'$ denote the positions of the UV and IR branes, respectively. To facilitate the comparison of our formulae with existing results, we recall that the variables in the two reference frames are related by
\beq
   z = \frac{t}{\Mkk} \,, \qquad 
   R = \frac{1}{k} \,, \qquad 
   R' = \frac{1}{\Mkk} \,, \qquad 
   \ln\frac{R'}{R} =  L \,.
\eeq

In the original formulation of the RS model \cite{Randall:1999ee}, all SM fields were constrained to reside on the IR brane. It was soon realized that, while the Higgs field has to be localized on (or near) the IR brane in order to solve the hierarchy problem, gauge \cite{Davoudiasl:1999tf,Pomarol:1999ad,Chang:1999nh,Gherghetta:2000qt} and matter fields \cite{Grossman:1999ra,Gherghetta:2000qt} can live in the bulk of AdS$_5$. This possibility furnishes both challenges and opportunities for model building. With bulk gauge fields, the compatibility of the model with electroweak precision measurements at first seemed in jeopardy, since delocalized $W^\pm$ and $Z^0$ bosons were found to induce harmful corrections to the Peskin-Takeuchi \cite{Peskin:1990zt,Peskin:1991sw} parameters $S$ and $T$ \cite{Csaki:2002gy,Davoudiasl:1999tf}, which however are tightly constrained by experiments \cite{LEPEWWG:2005ema}. Placing fermions in the bulk allows to significantly relax the constraint from the $S$ parameter \cite{Gherghetta:2000qt,Huber:2000fh,Davoudiasl:2000wi,Huber:2001gw}. The remaining issue of excessive contributions to the $T$ parameter can be cured, \eg, by extending the bulk hypercharge group to $SU(2)_R\times U(1)_X$ and breaking it to $U(1)_Y$ on the UV brane \cite{Agashe:2003zs}. An embedding of the SM fermions into the custodially symmetric $SU(2)_L\times SU(2)_R$ model, under which the left-handed bottom quark is symmetric under the exchange of $SU(2)_L$ and $SU(2)_R$, allows one to protect the left-handed $Z^0 b\bar b$ coupling from vast corrections \cite{Agashe:2006at}.

Delocalized fermions have the further virtue of admitting a natural explanation of the flavor structure of the SM by harnessing the idea of split fermions \cite{ArkaniHamed:1999dc}. In fact, it is perhaps not an overstatement to say that the RS scenario offers the best theory of flavor we have to date. Starting from anarchic 5D Yukawa couplings, the large mass hierarchies of the SM fermions can be generated without flavor symmetries by localizing the SM fermions at different points in the fifth dimension \cite{Grossman:1999ra,Gherghetta:2000qt,Huber:2000ie,Huber:2003tu}. Given the large hierarchy of quark masses in the SM, small mixing angles in the Cabibbo-Kobayashi-Maskawa (CKM) matrix are a natural consequence of this scenario \cite{Huber:2003tu}. This way of generating fermion mass hierarchies also implies a certain amount of suppression of dangerous flavor-changing neutral currents (FCNCs) \cite{Gherghetta:2000qt}, which goes by the name of RS-GIM mechanism \cite{Agashe:2004ay,Agashe:2004cp}. This mechanism successfully suppresses almost all flavor-changing transitions in the quark sector below their experimental limits. 

During the past years various studies of the flavor structure of the RS model have been performed. Properties of the (generalized) CKM matrix, neutral-meson mixing, and CP violation were studied in \cite{Huber:2003tu}. $Z^0$-mediated FCNCs in the kaon system were considered in \cite{Burdman:2002gr}, and effects of KK gauge bosons on CP asymmetries in rare hadronic $B$-meson decays induced by $b\to s$ transitions were explored in \cite{Burdman:2003nt}. A rather detailed survey of $\Delta F=2$ and $\Delta F=1$ processes in the RS framework was presented in \cite{Agashe:2004ay,Agashe:2004cp}. In particular, the second paper explores a variety of possible effects and analyzes several different rare decay processes. The branching ratios for the flavor-changing top-quark decays $t\to c Z^0 (\gamma,g)$ were examined in \cite{Agashe:2006wa}. The first complete study of all operators relevant to $K$--$\bar K$ mixing was presented in \cite{Csaki:2008zd}. Some general, model-independent approaches for studying new physics contributions to $\Delta F=2$ and $\Delta F=1$ operators were developed in \cite{Agashe:2005hk,Davidson:2007si}. It has been recognized that the only observables where some fine-tuning of parameters appears to be unavoidable are CP-violating effects in the neutral kaon system \cite{Bona:2007vi,Davidson:2007si,Csaki:2008zd} and the neutron electric dipole moment \cite{Agashe:2004ay,Agashe:2004cp}, which for generic choices of parameters turn out to be too large unless the masses of the lightest KK gauge bosons lie above (10--20)\,TeV. In view of this problem, several modifications of the quark flavor sector of warped extra-dimension models have been proposed. Most of them try to implement the notion of minimal flavor violation \cite{Buras:2000dm,D'Ambrosio:2002ex} into the RS framework by using a bulk flavor symmetry \cite{Cacciapaglia:2007fw,Fitzpatrick:2007sa,Santiago:2008vq,Csaki:2008eh}. In another approach the idea of textures of the 5D Yukawa matrices is explored \cite{Chang:2008zx}. The problem of too large electric dipole moments has been addressed using the idea of spontaneous CP violation in the context of warped extra dimensions \cite{Cheung:2007bu}.

This paper is the first in a series of articles devoted to a thorough analysis of flavor physics in RS models. It collects a large body of known results in the literature and extends them in various aspects. In a companion paper we apply these results to present a complete discussion of tree-level flavor-violating $\Delta F=2$ and $\Delta F=1$ effects in the quark sector \cite{mytalks,inprep}. As our benchmark scenario we consider the simplest implementation of the RS model, in which all SM gauge and matter fields are allowed to propagate in the bulk. We thus assume that the bulk theory is symmetric under the SM gauge group $SU(3)_C\times SU(2)_L\times U(1)_Y$, which is broken to $SU(3)_C\times U(1)_{\rm EM}$ on the IR brane by coupling the SM fields to a minimal, brane-localized Higgs sector. It would be rather straightforward to extend the discussion of tree-level effects to more complicated setups with an enlarged gauge symmetry or the Higgs sector living in the bulk. At the loop level, the predictions for FCNC processes will become model-dependent and the baroque fermionic content of many extensions of the original RS scenario might lead to more stringent flavor constraints when compared to the minimal model.

This article is organized as follows. In Section~\ref{sec:gauge} we discuss the KK decomposition of the bulk gauge fields in the presence of the brane-localized Higgs sector, working in a covariant $R_\xi$ gauge. We derive exact results for the masses of the SM gauge bosons and their KK excitations, as well as for the profiles of these fields. We also discuss how sums over KK towers of gauge bosons arising in tree-level diagrams at low energy can be evaluated in closed form. The analogous discussion for bulk fermions is presented in Section~\ref{sec:fermions}, where we solve the exact bulk equations of motion in presence of the brane-localized Yukawa sector, which mixes the $N$ fermion generations. In particular, we point out that the exact equations of motion imply that the bulk profiles belonging to different mass eigenstates are not orthogonal on each other. This observation will have important implications for flavor physics. We also comment on dimensional-analysis constraints on the scale of the 5D Yukawa matrices. Section~\ref{sec:hierarchies} is devoted to a review of the implications of the RS setup for the hierarchical structures of fermion masses and mixings. We emphasize the invariance of the results for these masses and mixings under two types of reparametrization transformations of the 5D bulk mass parameters and Yukawa matrices. In Section~\ref{sec:gaugecouplings} we present the main results of our work by analyzing the structure of gauge boson interactions with SM fermions and their KK excitations. In particular, we study in detail the flavor-violating couplings of the $W^\pm$ and $Z^0$ bosons to SM fermions. Some phenomenological implications of these results are discussed in Section~\ref{sec:pheno}. We begin by studying the constraints on the model imposed by various electroweak precision measurements. Even though these constraints have been explored in the past by many authors, generally concluding that they force an extension of the minimal model, we find that even the simplest model provides a consistent framework if one gives up the requirement of a light Higgs boson. As mentioned earlier, this is indeed natural in models with a brane-localized Higgs sector. We also show that this model can be made consistent with the experimental constraints on the $Z^0 b\bar b$ couplings. Allowing again for a heavy Higgs boson, we obtain a significantly better description of the data than in the SM. In our work we concentrate on the leading contributions to the electroweak precision observables, ignoring possible effects of brane-localized kinetic terms \cite{Davoudiasl:2002ua,Carena:2002dz,Carena:2003fx}. Although the UV dynamics is not specified, it is natural to assume that these terms are loop suppressed, so that they can be neglected to first order. We also point out that the KK excitations of the SM fermions have generically large mixings between different generations as well as between $SU(2)_L$ singlets and doublets. The reason is that without the Yukawa couplings the spectra of the KK towers of different fermion fields are nearly degenerate because of the closely spaced 5D bulk mass parameters, so that even small Yukawa couplings can lead to large mixing effects. Finally, we explore predictions for the tree-level FCNC decays $t\to c (u) Z^0$ and $t\to c (u) h$, which might be detectable at the LHC. A comprehensive study of flavor effects in the $B$-, $D$-, and $K$-meson systems will be presented in \cite{inprep}. Section~\ref{sec:concl} contains our conclusions and an outlook. In a series of Appendices we collect details on the textures of the various flavor mixing matrices, our input values for SM parameters, and some numerical results for mixing matrices corresponding to a reference RS parameter point.

\section{Bulk Gauge Fields}
\label{sec:gauge}

In this and the following sections we derive the KK decompositions of bulk gauge and matter fields in the presence of a brane-localized Higgs sector. While most authors treat the couplings of bulk fields to the Higgs sector as a perturbation and expand the theory in powers of $v^2/\Mkk^2$, we instead construct the exact solutions to the bulk equations of motion subject to the boundary conditions imposed by the couplings to the Higgs sector. In that way we obtain exact results for the masses and profiles of the various SM particles and their KK excitations. This approach is more elegant than the perturbative one  and avoids the necessity of diagonalizing infinite-dimensional mass matrices.

We begin our discussion with the gauge sector, considering the simplest case for which the bulk gauge group is that of the SM. We will discuss the KK decomposition for the electroweak sector only. The extension to the strong interaction is straightforward. 

\subsection{Action of the 5D Theory}

We consider bulk gauge fields $W_M^a$ and $B_M$ of $SU(2)_L\times U(1)_Y$, coupled to a scalar sector on the IR brane. We choose the vector components $W_\mu^a$ and $B_\mu$ to be even under the $Z_2$ orbifold symmetry and the scalar components $W_\phi^a$ and $B_\phi$ to be odd. This ensures that the light mass eigenstates (often called ``zero modes'') correspond to the SM gauge bosons. The action can be split up as
\beq
   S_{\rm gauge} = \int d^4x\,r\int_{-\pi}^\pi\!d\phi\, 
   \Big( {\cal L}_{\rm W,B} + {\cal L}_{\rm Higgs}
   + {\cal L}_{\rm GF} + {\cal L}_{\rm FP} \Big) \,,
\eeq
where 
\beq\label{Lgauge}
   {\cal L}_{\rm W,B} = \frac{\sqrt{G}}{r}\,G^{KM} G^{LN}
   \left( - \frac14\,W_{KL}^a W_{MN}^a - \frac14\,B_{KL} B_{MN}
   \right)
\eeq
is the Lagrangian for the 5D gauge theory, while the Higgs-sector Lagrangian
\beq
   {\cal L}_{\rm Higgs} = \frac{\delta(|\phi|-\pi)}{r}
   \left[ (D_\mu\Phi)^\dagger\,(D^\mu\Phi) - V(\Phi) \right] ,
    \qquad
   V(\Phi) = - \mu^2\Phi^\dagger\Phi 
    + \lambda \left( \Phi^\dagger\Phi \right)^2
\eeq
is localized on the IR brane. After electroweak symmetry breaking (EWSB), we decompose the Higgs doublet in terms of real scalar fields $\varphi^i$ as
\beq
   \Phi(x) = \frac{1}{\sqrt2}
   \left( \begin{array}{c}
    -i\sqrt2\,\varphi^+(x) \\
    v + h(x) + i\varphi^3(x)
   \end{array} \right) ,
\eeq
where $v\approx 246$\,GeV is the Higgs vacuum expectation value, and $\varphi^\pm=(\varphi^1\mp i\varphi^2)/\sqrt2$. We also perform the usual field redefinitions of the gauge fields
\begin{eqnarray}
   W_M^\pm &=& \frac{1}{\sqrt2} \left( W_M^1\mp i W_M^2 \right) , 
    \nonumber\\
   Z_M &=& \frac{1}{\sqrt{g_5^2+g_5^{\prime 2}}} 
    \left( g_5 W_M^3 - g_5' B_M \right) , \\
   A_M &=& \frac{1}{\sqrt{g_5^2+g_5^{\prime 2}}} 
    \left( g_5' W_M^3 + g_5 B_M \right) , \nonumber
\end{eqnarray}
where $g_5$ and $g_5'$ are the 5D gauge couplings of $SU(2)_L$ and $U(1)_Y$, respectively. This diagonalizes the 5D mass terms resulting from EWSB, in such a way that the $W^\pm$ and $Z^0$ bosons get ``masses'' (with mass dimension 1/2)
\beq
   M_W = \frac{v g_5}{2} \,, \qquad
   M_Z = \frac{v\sqrt{g_5^2+g_5^{\prime 2}}}{2} \,,
\eeq
while the photon remains massless ($M_A=0$).

The kinetic terms for the Higgs field give rise to mixed terms involving the gauge bosons and the scalar fields $\varphi^\pm$ and $\varphi^3$, which can be read off from
\beq
   D_\mu\Phi = \frac{1}{\sqrt2} \left( \begin{array}{c}
    -i\sqrt2 \left( \partial_\mu\varphi^+ + M_W\,W_\mu^+ \right) \\
    \partial_\mu h 
     + i \left( \partial_\mu\varphi^3 + M_Z\,Z_\mu \right)
   \end{array} \right) 
   + \mbox{terms bi-linear in fields \,.}
\eeq
In addition, the kinetic terms for the gauge fields in (\ref{Lgauge}) contain mixed terms involving the gauge bosons and the scalar components $W_\phi^\pm$, $Z_\phi$, and $A_\phi$. All of these mixed terms can be removed with a suitable choice of the gauge-fixing Lagrangian. We adopt the form 
\beq\label{Sgf}
\begin{split}
   {\cal L}_{\rm GF}
   &= - \frac{1}{2\xi} \left( \partial^\mu A_\mu - \xi \left[ 
    \frac{\partial_\phi\,e^{-2\sigma(\phi)} A_\phi}{r^2} \right] 
    \right)^2 \\
   &\quad\mbox{}- \frac{1}{2\xi} 
    \left( \partial^\mu Z_\mu - \xi \left[ 
    \frac{\delta(|\phi|-\pi)}{r}\,M_Z\,\varphi^3 
    + \frac{\partial_\phi\,e^{-2\sigma(\phi)} Z_\phi}{r^2} \right] 
    \right)^2 \\
   &\quad\mbox{}- \frac{1}{\xi} 
    \left( \partial^\mu W_\mu^+ - \xi \left[ 
    \frac{\delta(|\phi|-\pi)}{r}\,M_W\,\varphi^+
    + \frac{\partial_\phi\,e^{-2\sigma(\phi)} W_\phi^+}{r^2} \right] 
    \right) \\
   &\qquad\times
    \left( \partial^\mu W_\mu^- - \xi \left[ 
    \frac{\delta(|\phi|-\pi)}{r}\,M_W\,\varphi^-
    + \frac{\partial_\phi\,e^{-2\sigma(\phi)} W_\phi^-}{r^2} \right] 
    \right) .
\end{split}
\eeq
More generally, each term could be written with a different gauge-fixing parameter $\xi_i$. Note that, despite appearance, there is no problem in squaring the $\delta$-functions in this expression. We will see below that the derivatives of the scalar components of the gauge fields $W_\phi^\pm$ and $Z_\phi$ also contain a $\delta$-function contribution, which precisely cancels the $\delta$-functions from the Higgs sector. As a result, contrary to the treatment in \cite{Csaki:2005vy}, we do not need to introduce separate gauge-fixing Lagrangians in the bulk and on the IR brane.

Using integration by parts, we now obtain for the quadratic terms in the action
\begin{eqnarray}\label{Sgauge2new}
   &&S_{\rm gauge,2} 
    = \int d^4x\,r\int_{-\pi}^\pi\!d\phi\,\bigg\{
    - \frac14\,F_{\mu\nu} F^{\mu\nu} 
    - \frac{1}{2\xi} \left( \partial^\mu A_\mu \right)^2 
    \nonumber\\
   &&\quad\mbox{}+ \frac{e^{-2\sigma(\phi)}}{2r^2} \left[ 
    \partial_\mu A_\phi\partial^\mu A_\phi
    + \partial_\phi A_\mu\partial_\phi A^\mu \right]
    - \frac{\xi}{2} \left[
    \frac{\partial_\phi\,e^{-2\sigma(\phi)} A_\phi}{r^2} \right]^2 
    \nonumber\\
   &&\quad\mbox{}- \frac14\,Z_{\mu\nu} Z^{\mu\nu} 
    - \frac{1}{2\xi} \left( \partial^\mu Z_\mu \right)^2
    + \frac{e^{-2\sigma(\phi)}}{2r^2} \left[ 
    \partial_\mu Z_\phi\partial^\mu Z_\phi
    + \partial_\phi Z_\mu\partial_\phi Z^\mu \right] 
    \nonumber\\
   &&\quad\mbox{}- \frac12\,W_{\mu\nu}^+ W^{-\mu\nu} 
    - \frac{1}{\xi}\,\partial^\mu W_\mu^+\,\partial^\mu W_\mu^-
    + \frac{e^{-2\sigma(\phi)}}{r^2} \left[ 
    \partial_\mu W_\phi^+\partial^\mu W_\phi^-
    + \partial_\phi W_\mu^+\partial_\phi W^{-\mu} \right] 
    \\
   &&\quad\mbox{}+ \frac{\delta(|\phi|-\pi)}{r} \left[
    \frac12\partial_\mu h\partial^\mu h - \lambda v^2 h^2
    + \partial_\mu\varphi^+\partial^\mu\varphi^- 
    + \frac12\partial_\mu\varphi^3\partial^\mu\varphi^3 
    + \frac{M_Z^2}{2}\,Z_\mu Z^\mu + M_W^2\,W_\mu^+ W^{-\mu}
    \right] 
    \nonumber\\
   &&\quad\mbox{}- \frac{\xi}{2} \left[ 
    \frac{\delta(|\phi|-\pi)}{r}\,M_Z\,\varphi^3 
    + \frac{\partial_\phi\,e^{-2\sigma(\phi)} Z_\phi}{r^2} 
    \right]^2 
    \nonumber\\
   &&\quad\mbox{}- \xi \left[ 
    \frac{\delta(|\phi|-\pi)}{r}\,M_W\,\varphi^+
    + \frac{\partial_\phi\,e^{-2\sigma(\phi)} W_\phi^+}{r^2} \right] 
    \left[ \frac{\delta(|\phi|-\pi)}{r}\,M_W\,\varphi^-
    + \frac{\partial_\phi\,e^{-2\sigma(\phi)} W_\phi^-}{r^2} \right]
    + {\cal L}_{\rm FP} \bigg\} \,. \nonumber
\end{eqnarray}
The form of the Faddeev-Popov ghost Lagrangian ${\cal L}_{\rm FP}$ will be discussed after the KK decomposition.

Before proceeding, a comment is in order concerning the precise meaning of the above expressions. For the consistency of the theory it is important that one can integrate by parts in the action without encountering boundary terms. Otherwise the Lagrangian is not hermitian. Yet, the presence of $\delta$-function terms on the IR brane gives rise to discontinuities of some of the fields at $|\phi|=\pi$, which appears to jeopardize this crucial feature. In order to define the model properly, we will always understand the $\delta$-functions via the limiting procedure
\beq
   \delta(|\phi|-\pi)\equiv \lim_{\theta\to 0^+}\,
   \frac12\,\Big[ \delta(\phi-\pi+\theta) + \delta(\phi+\pi-\theta)
   \Big] \,.
\eeq
In this way the discontinuities are moved into the bulk, and the fields can be assigned proper boundary conditions on the branes, which are consistent with integration by parts. We perform all calculations at small but finite $\theta$ and find that at the end the limit $\theta\to 0^+$ is smooth, giving rise to well-defined jump conditions for the fields and their derivatives on the IR brane. When necessary, we will use the notation $f(\pi^-)\equiv\lim_{\theta\to 0^+}\,f(\pi-\theta)$ to indicate the value of a function $f$ that is discontinuous at $|\phi|=\pi$.

\subsection{Kaluza-Klein Decomposition}

We write the KK decompositions of the various 5D fields in the form
\beq
\begin{aligned}
   A_\mu(x,\phi) 
   &= \frac{1}{\sqrt r} \sum_n A_\mu^{(n)}(x)\,\chi_n^A(\phi) \,,
    &\qquad 
   A_\phi(x,\phi) 
   &= \frac{1}{\sqrt r} \sum_n a_n^A\,\varphi_A^{(n)}(x)\,
    \partial_\phi\,\chi_n^A(\phi) \,, \\
   Z_\mu(x,\phi) 
   &= \frac{1}{\sqrt r} \sum_n Z_\mu^{(n)}(x)\,\chi_n^Z(\phi) \,,
    &\qquad
   Z_\phi(x,\phi) 
   &= \frac{1}{\sqrt r} \sum_n a_n^Z\,\varphi_Z^{(n)}(x)\,
    \partial_\phi\,\chi_n^Z(\phi) \,, \\
   W_\mu^\pm(x,\phi) 
   &= \frac{1}{\sqrt r} \sum_n W_\mu^{\pm(n)}(x)\,\chi_n^W(\phi)
    \,, &\qquad
   W_\phi^\pm(x,\phi) 
   &= \frac{1}{\sqrt r} \sum_n a_n^W\,\varphi_W^{\pm(n)}(x)\,
    \partial_\phi\,\chi_n^W(\phi) \,,
\end{aligned}
\eeq
where $A_\mu^{(n)}$ \etc\ are the 4D mass eigenstates, and the various $\chi_n^a$ profiles form complete sets of even functions on the orbifold, which can be taken to obey the orthonormality condition
\beq\label{chinorm}
   \int_{-\pi}^\pi\!d\phi\,\chi_m^a(\phi)\,\chi_n^a(\phi)
   = \delta_{mn} \,.
\eeq
The 4D scalar fields can also be expanded in the basis of mass eigenstates, and we write these expansions in the form
\beq
   \varphi^\pm(x) = \sum_n b_n^W\,\varphi_W^{\pm(n)}(x) \,, \qquad 
   \varphi^3(x) = \sum_n b_n^Z\,\varphi_Z^{(n)}(x) \,.
\eeq
We will denote the masses of the 4D vector fields by $m_n^a\ge 0$ (with $a=A,Z,W$). The masses of the scalar fields $\varphi_a^{(n)}$ will be related to these by gauge invariance.

Inserting these decompositions into the action, one finds that the profiles $\chi_n^a$ obey the equation of motion \cite{Davoudiasl:1999tf,Pomarol:1999ad}
\beq\label{gaugeeom}
   - \frac{1}{r^2}\,\partial_\phi\,e^{-2\sigma(\phi)}\,
   \partial_\phi\,\chi_n^a(\phi) = (m_n^a)^2\,\chi_n^a(\phi) 
   - \frac{\delta(|\phi|-\pi)}{r}\,M_a^2\,\chi_n^a(\phi) \,.
\eeq
The boundary conditions are
\beq\label{bcs}
\begin{aligned}
   \partial_\phi\,\chi_n^a(0) &= 0 \,, 
   &&\mbox{(UV brane)} \\
   \partial_\phi\,\chi_n^a(\pi^-) 
   &= - \frac{r M_a^2}{2\epsilon^2}\,\chi_n^a(\pi) \,.
   &\quad &\mbox{(IR brane)}
\end{aligned}
\eeq
From these conditions one derives the eigenvalues $m_n^a$.

We find that the action takes the desired form
\beq\label{gaugefinal}
\begin{split}
   S_{\rm gauge,2} 
   &= \sum_n \int d^4x\,\bigg\{
    - \frac14\,F_{\mu\nu}^{(n)} F^{\mu\nu(n)}
    - \frac{1}{2\xi} \left( \partial^\mu A_\mu^{(n)} \right)^2 
    + \frac{(m_n^A)^2}{2}\,A_\mu^{(n)} A^{\mu(n)} \\
   &\quad\mbox{}- \frac14\,Z_{\mu\nu}^{(n)} Z^{\mu\nu(n)}
    - \frac{1}{2\xi} \left( \partial^\mu Z_\mu^{(n)} \right)^2
    + \frac{(m_n^Z)^2}{2}\,Z_\mu^{(n)} Z^{\mu(n)} \\
   &\quad\mbox{}- \frac12\,W_{\mu\nu}^{+(n)} W^{-\mu\nu(n)}
    - \frac{1}{\xi}\,\partial^\mu W_\mu^{+(n)}\,
    \partial^\mu W_\mu^{-(n)} 
    + (m_n^W)^2\,W_\mu^{+(n)} W^{-\mu(n)} \\
   &\quad\mbox{}
    + \frac12\partial_\mu\varphi_A^{(n)}
    \partial^\mu\varphi_A^{(n)} 
    - \frac{\xi (m_n^A)^2}{2}\,\varphi_A^{(n)}\varphi_A^{(n)} 
    + \frac12\partial_\mu\varphi_Z^{(n)}
    \partial^\mu\varphi_Z^{(n)} 
    - \frac{\xi (m_n^Z)^2}{2}\,\varphi_Z^{(n)}\varphi_Z^{(n)} \\
   &\quad\mbox{}+ \partial_\mu\varphi_W^{+(n)}
    \partial^\mu\varphi_W^{-(n)} 
    - \xi (m_n^W)^2\,\varphi_W^{+(n)}\varphi_W^{-(n)} 
    \bigg\} \\
   &\quad\mbox{}+ \int d^4x \left( 
    \frac12\partial_\mu h\partial^\mu h - \lambda v^2 h^2 \right) 
    + \sum_n \int d^4x\,{\cal L}_{\rm FP}^{(n)} \,,
\end{split}
\eeq
if and only if
\beq\label{coefs}
   a_n^a = - \frac{1}{m_n^a} \,, \qquad
   b_n^a = \frac{M_a}{\sqrt r}\,\frac{\chi_n^a(\pi^-)}{m_n^a} \,.
\eeq
The resulting theory contains a tower of massive gauge bosons with masses $m_n^a$, accompanied by a tower of massive scalars with masses $\sqrt{\xi}\,m_n^a$, as well as the Higgs field $h$ with mass $\sqrt{2\lambda} v$. Note that with (\ref{coefs}) the 4D gauge-fixing Lagrangian derived from (\ref{Sgf}) takes the simple form
\beq
   r\int_{-\pi}^\pi\!d\phi\,{\cal L}_{\rm GF}
   = \sum_n\,{\cal L}_{\rm GF}^{(n)} \,,
\eeq
with
\beq
\begin{split}
   {\cal L}_{\rm GF}^{(n)}
   &= - \frac{1}{2\xi} \left( \partial^\mu A_\mu^{(n)}
    - \xi m_n^A\varphi_A^{(n)} \right)^2 
    - \frac{1}{2\xi} \left( \partial^\mu Z_\mu^{(n)}
    - \xi m_n^Z\varphi_Z^{(n)} \right)^2 \\
   &\quad\mbox{}- \frac{1}{\xi} \left( \partial^\mu W_\mu^{+(n)}
    - \xi m_n^W\varphi_W^{+(n)} \right)
    \left( \partial^\mu W_\mu^{-(n)}
    - \xi m_n^W\varphi_W^{-(n)} \right) .
\end{split}
\eeq
For each KK mode these expressions are identical to those of the SM. 
It follows that the form of the Faddeev-Popov ghost Lagrangians ${\cal L}_{\rm FP}^{(n)}$ in (\ref{gaugefinal}) is analogous to that of the SM, with the only generalization that a ghost field is required for every KK mode. 

\subsection{Summing over Kaluza-Klein Modes}

The tree-level exchange of a SM gauge boson accompanied by its KK excitations in a generic Feynman diagram leads to a combination of propagator and vertex functions, which in the low-energy limit, \ie, for small momentum transfer $q^2$, can be expanded as\footnote{From now on we omit the superscript $a$ on the gauge-boson profiles unless it is required for clarity.}
\beq\label{expandedprop}
   \sum_n\,\frac{\chi_n(\phi)\,\chi_n(\phi')}{m_n^2-q^2}
   = \sum_{N=1}^\infty \left( q^2 \right)^{N-1}
   \sum_n\,\frac{\chi_n(\phi)\,\chi_n(\phi')}%
                {\left( m_n^2 \right)^N} \,.
\eeq
The gauge-boson profiles $\chi_n$ are integrated with other profiles at each vertex. The sums over profiles weighted by inverse powers of $m_n^2$ can be evaluated in closed form by generalizing a method developed in \cite{Hirn:2007bb}.

The simplest sum is obtained using that the bulk profiles $\chi_n$ form a complete set of orthonormal, even functions on the orbifold, subject to the boundary conditions (\ref{bcs}). Relation (\ref{chinorm}) then implies that
\beq\label{complete}
   \sum_n\,\chi_n(\phi)\,\chi_n(\phi')
   = \frac12 \left[ \delta(\phi-\phi') + \delta(\phi+\phi') \right] , 
\eeq
where $n$ runs from 0 to $\infty$. Here and below, the index $n=0$ refers to the light SM particle, which may or may not be massless.

To proceed, we integrate the equation of motion (\ref{gaugeeom}) twice, accounting for the boundary condition $\partial_\phi\chi_n(0)=0$ on the UV brane. This yields 
\beq\label{key}
   \chi_n(\phi) - \chi_n(0)
   = - r^2 m_n^2 \int_0^\phi\!d\phi'\,e^{2\sigma(\phi')}
   \int_0^{\phi'}\!d\phi''\,\chi_n(\phi'') \,.
\eeq
Note that this formula holds even in the presence of the boundary term on the right-hand side of the equation of motion, which results from EWSB. This term affects the eigenvalues $x_n$ determining the gauge-boson masses, but it does not affect the functional form of the profiles beyond that change. The appearance of the profile $\chi_n(0)$ on the UV brane in (\ref{key}) is a new element, which has not been considered in \cite{Hirn:2007bb}.

Using the result (\ref{key}) along with the completeness relation (\ref{complete}), the sums over profiles in (\ref{expandedprop}) can be evaluated using an iterative procedure, which can be generalized to arbitrary order. For the first and most important sum we obtain
\beq\label{nice}
   \sum_n\,\frac{\chi_n(\phi)\,\chi_n(\phi')}{m_n^2}
   = \sum_n\,\frac{\chi_n^2(0)}{m_n^2}
   - \frac{L\epsilon^2}{4\pi\Mkk^2} 
   \left[ e^{2\sigma(\phi_>)} - 1 \right] , 
\eeq
where $\phi_>\equiv\mbox{max}(|\phi|,|\phi'|)$. As it stands, this relation holds only for the case where the gauge symmetry is spontaneously broken, so that the lowest mass eigenvalue satisfies $m_0>0$. 

The remaining sum over profiles $\chi_n^2(0)$ on the UV brane can be performed by integrating $\chi_0(\phi)$ times the expression on the left-hand side of (\ref{nice}) over the entire orbifold and using the orthonormality condition (\ref{chinorm}). We find
\begin{eqnarray}\label{wonder1b}
\begin{split}
   \sum_n\,\frac{\chi_n^2(0)}{m_n^2}
   &= \frac{1}{2\pi m_0^2}
    + \frac{1}{4\pi\Mkk^2} \left[ \left( 1 - \frac{1}{2L} \right) 
    - \epsilon^2 \left( L - \frac{1}{2L} \right) \right] \\
   &\hspace{-2cm} \quad\mbox{}- \frac{m_0^2}{32\pi\Mkk^4} \left[
    \left( L - \frac52 + \frac{21}{8L} - \frac{1}{L^2} \right)
    -2\epsilon^2 \left( \frac{1}{L} - \frac{1}{L^2} \right)
    - \epsilon^4 \left( \frac{5}{8L} + \frac{1}{L^2} \right)
    \right] + \ord \left ( \frac{m_0^2}{\Mkk^6} \right ) \,.
\end{split}
\end{eqnarray}

For the case where a massless zero mode exists ($m_0=0$), one must subtract the contribution of the ground state from the sum over states. This can either be done by subtracting this contribution from both sides of the completeness relation (\ref{complete}), or by inserting the explicit form of the ground-state profile (see Section~\ref{sec:gaugeprofiles} below) and then taking the limit $m_0\to 0$. We find in this way 
\beq\label{nice1}
\begin{split}
   {\sum_n}^\prime\,\frac{\chi_n(\phi)\,\chi_n(\phi')}{m_n^2}
   &= \frac{1}{4\pi\Mkk^2} \left[ \frac{1}{2L} 
   - \epsilon^2 \left( L + 1 + \frac{1}{2L} \right) \right] 
    - \frac{L\epsilon^2}{4\pi\Mkk^2} 
    \left[ e^{2\sigma(\phi_>)} - 1 \right] \\
   &\quad\mbox{}+ \frac{\epsilon^2}{4\pi\Mkk^2} \left[
    e^{2\sigma(\phi)} \left( \frac{L|\phi|}{\pi}- \frac12 \right)
    + e^{2\sigma(\phi')} \left( \frac{L|\phi'|}{\pi} - \frac12 \right) 
    + 1 \right] ,
\end{split}
\eeq
where the prime on the sum indicates that $n$ runs from 1 to $\infty$.

Note that $\ord(\epsilon^n)$ terms on the right-hand sides of (\ref{nice}), (\ref{wonder1b}), and (\ref{nice1}) not accompanied by $n$ powers of a warp factor are Planck-scale suppressed and can be dropped for all practical purposes, in which case the relations simplify considerably. Introducing the variables $t=\epsilon\,e^{\sigma(\phi)}$ and $t'=\epsilon\,e^{\sigma(\phi')}$, we obtain for $m_0>0$ and $m_0=0$ the important results
\beq\label{important1}
   \sum_n\,\frac{\chi_n(\phi)\,\chi_n(\phi')}{m_n^2}
   = \frac{1}{2\pi m_0^2} + \frac{1}{4\pi\Mkk^2}
   \left[ L\,t_<^2 - L \left( t^2 + t'^2 \right) + 1 - \frac{1}{2L} 
   + \ord\left( \frac{m_0^2}{\Mkk^2} \right) \right] ,
\eeq
and
\beq\label{important2}
   {\sum_n}^\prime\,\frac{\chi_n(\phi)\,\chi_n(\phi')}{m_n^2}
   = \frac{1}{4\pi\Mkk^2} \left[ L\,t_<^2 
   - t^2 \left( \frac12 - \ln t \right) 
   - t'^2 \left( \frac12 - \ln t' \right) + \frac{1}{2L} \right] ,
\eeq
respectively, where $t_<\equiv\mbox{min}(t,t')$. The neglected terms in the first relation only affect the constant, \ie, $t$-independent contribution. Notice that the terms proportional to $t^2$ and $t'^2$ in the first sum are enhanced by a factor $L$, whereas this is not the case for the second sum. This fact has important consequences for the phenomenology of flavor-violating processes \cite{mytalks,inprep}. It implies that in the RS model new physics contributions to $\Delta F=2$ processes such as $B$--$\bar B$ or $K$--$\bar K$ mixing are dominated by tree-level KK gluon exchange, whereas those to $\Delta F=1$ processes such as rare $B$-meson decays arise predominantly from the FCNC couplings of the $Z^0$ boson to fermions.

It is possible to extend the above procedure in an iterative way to sums with $N>1$ in (\ref{expandedprop}). As an example, we give the relations for $N=2$, neglecting $\ord(\epsilon)$ terms. They read
\beq
\begin{split}
   \sum_n\,\frac{\chi_n(\phi)\,\chi_n(\phi')}{m_n^4}
   &= \frac{1}{2\pi m_0^4}
    + \frac{1}{4\pi m_0^2\Mkk^2} \left( 1 - \frac{1}{L} \right) 
    - \frac{1}{32\pi\Mkk^4} \left( L - 5 + \frac{29}{4L} 
    - \frac{3}{L^2} \right) \\
   &\hspace{-3.2cm} \quad\mbox{}- \left[ \frac{1}{4\pi m_0^2\Mkk^2}
    + \frac{1}{8\pi\Mkk^4} \left( 1 - \frac{1}{2L} \right) \right] 
    \left[ t^2 \left( L - \frac12 + \ln t \right)
    + t'^2 \left( L - \frac12 + \ln t' \right) \right] \\
   &\hspace{-3.2cm} \quad\mbox{}+ \frac{L}{32\pi\Mkk^4} \left[
    t_>^4 + 4 t^2 t'^2 \left( L - \frac12 + \ln t_< \right) 
    \right] + \ord\left( \frac{m_0^2}{\Mkk^6} \right) ,
\end{split}
\eeq
and
\begin{eqnarray}\label{m4rela2}
\begin{split}
   {\sum_n}^\prime\,\frac{\chi_n(\phi)\,\chi_n(\phi')}{m_n^4}
   &= \frac{1}{32\pi\Mkk^4} \left( \frac{5}{8L} 
    - \frac{1}{L^2} \right) \\
   &\quad\mbox{}- \frac{1}{32\pi\Mkk^4} \left[
    t^4 \left( L - \frac54 + \ln t \right)
    + \frac{2t^2}{L} \left( L - \frac12 + \ln t \right) \right] \\
   &\quad\mbox{}- \frac{1}{32\pi\Mkk^4} \left[
    t'^4 \left( L - \frac54 + \ln t' \right)
    + \frac{2t'^2}{L} \left( L - \frac12 + \ln t' \right) \right] \\
   &\quad\mbox{}+ \frac{1}{32\pi\Mkk^4} \left[ L\,t_>^4 
    + 4 t^2 t'^2 \left[ L\ln t_< 
    - \left( L - \frac12 \right) 
    \left( \ln t t' - \frac12 \right) 
    - \ln t\ln t' \right] \right] .
\end{split}
\end{eqnarray}

For completeness, we note that for the case without EWSB the gauge-boson propagator in (\ref{expandedprop}) coincides with the 5D mixed position/momentum-space propagator derived in \cite{Randall:2001gb,Randall:2001gc}. Our relations (\ref{important2}) and (\ref{m4rela2}) can also be derived by expanding this quantity about $q^2=0$.

\subsection{Bulk Profiles}
\label{sec:gaugeprofiles}

The explicit form of the profiles $\chi_n$ was first obtained in \cite{Davoudiasl:1999tf,Pomarol:1999ad} for the case of an unbroken gauge symmetry. The same solution remains valid in the case of a spontaneously broken symmetry. However, the boundary conditions on the IR brane must be modified in this case so as to obey (\ref{bcs}). In terms of the variable $t$, which takes values between $t=\epsilon$ (UV brane) and $t=1$ (IR brane), we write the solution in the form
\beq\label{chin}
   \chi_n(\phi) = N_n\sqrt{\frac{L}{\pi}}\,t\,c_n^+(t) \,,
\eeq
where
\beq\label{cplcmi}
\begin{split}
   c_n^+(t) &= Y_0(x_n\epsilon)\,J_1(x_n t)
    - J_0(x_n\epsilon)\,Y_1(x_n t) \,, \\
   c_n^-(t) &= \frac{1}{x_n t}\,
    \frac{d}{dt} \left[ t\,c_n^+(t) \right]
    = Y_0(x_n\epsilon)\,J_0(x_n t)
    - J_0(x_n\epsilon)\,Y_0(x_n t) \,.
\end{split}
\eeq
Here
\beq
   x_n\equiv \frac{m_n}{\Mkk}
\eeq
are dimensionless parameters related to the masses of the gauge bosons and their KK excitations in the 4D theory. The normalization condition (\ref{chinorm}) fixes the constant $N_n$ to obey
\beq\label{Nngauge}
   N_n^{-2} = \left[ c_n^+(1) \right]^2 + \left[ c_n^-(1) \right]^2
   - \frac{2}{x_n}\,c_n^+(1)\,c_n^-(1)
   - \epsilon^2 \left[ c_n^+(\epsilon) \right]^2 .
\eeq

Obviously, equation~(\ref{chin}) satisfies the boundary condition $\partial_\phi\chi_n(0)=0$ on the UV brane, since $c_n^-(\epsilon)=0$. The boundary condition (\ref{bcs}) on the IR brane imposes the relation\footnote{This relation holds for the profiles of $W^\pm$ bosons and their KK excitations. For the case of the $Z^0$ boson, one must replace $g^2$ by $(g^2+g'^2)$.}
\beq\label{condi}
   x_n\,c_n^-(1) 
   = - \frac{g^2 v^2}{4\Mkk^2}\,L\,c_n^+(1) \,,
\eeq
from which the eigenvalues $x_n$ can be derived. Here we have introduced the 4D gauge coupling $g$, which is related to the 5D coupling $g_5$ via
\beq\label{g4def}
   g = \frac{g_5}{\sqrt{2\pi r}} \,.
\eeq
Analogous relations hold for all gauge couplings \cite{Davoudiasl:1999tf}. 

Without EWSB, \ie, for $v=0$, there is a zero mode ($m_0=0$) with flat profile
\beq\label{chi0flat}
   \chi_{\gamma,g}(\phi) = \frac{1}{\sqrt{2\pi}} \,.
\eeq
In the case where $m_0\ne 0$, the results for the SM gauge bosons can be simplified, since $x_0\ll 1$. Expanding (\ref{condi}) in powers of $x_0$ leads to
\beq\label{eq:m02}
   m_W^2 = \frac{g^2 v^2}{4} \left[
   1 - \frac{g^2 v^2}{8\Mkk^2} 
   \left( L - 1 + \frac{1-\epsilon^2}{2L} \right) 
   + \ord\left( \frac{v^4}{\Mkk^4} \right) \right] ,
\eeq
and an analogous relation with $g^2$ replaced by $(g^2+g'^2)$ holds for the $Z^0$-boson mass. This result may be compared with the SM relations $m_W^2=g^2 v^2/4$ and $m_Z^2=(g^2+g'^2)v^2/4$. It will also be useful to have an approximate expression for the ground-state profile $\chi_0$ valid for $x_0\ll 1$. We find
\beq\label{chi0WZ}
   \chi_{W,Z}(\phi) = \frac{1}{\sqrt{2\pi}} \left[
   1 + \frac{m_{W,Z}^2}{4\Mkk^2} \left( 1 - \frac{1-\epsilon^2}{L} 
   + t^2 \left( 1 - 2L - 2\ln t\right) \right) 
   + \ord\left( \frac{m_{W,Z}^4}{\Mkk^4} \right) \right] .
\eeq
This agrees with a result given in \cite{Csaki:2002gy} apart form the normalization, which in that reference is chosen such that $\chi_{W,Z}(\pi)=1$.

We finish this section by collecting useful expressions for the bulk profiles of KK gauge bosons. The fact that $\epsilon\approx 10^{-16}$ is extremely small allows us to replace 
\beq 
   J_0(x_n\epsilon)\approx 1 \,, \qquad
   Y_0(x_n\epsilon)\approx - \frac{2}{\pi}
    \left( L - \ln\frac{x_n}{2} - \gamma_E \right)
\eeq
in (\ref{cplcmi}) and (\ref{Nngauge}), where $\gamma_E\approx 0.5772$ is the Euler-Mascheroni constant. This can be used to simplify the form of the bulk profiles in (\ref{chin}) to
\beq
   \chi_n(\phi) = - \bar N_n\sqrt{\frac{L}{\pi}}\,t \left[
   \left( L - \ln\frac{x_n}{2} - \gamma_E \right) J_1(x_n t)
   + \frac{\pi}{2}\,Y_1(x_n t) \right] ,
\eeq
where 
\beq 
   \bar N_n^{-2} = \left[ 
   \left( L - \ln\frac{x_n}{2} - \gamma_E \right) J_1(x_n)
   + \frac{\pi}{2}\,Y_1(x_n) \right]^2
   \left[ 1 - \frac{g^2 v^2}{2\Mkk^2}\,\frac{L}{x_n^2}
   + \left( \frac{g^2 v^2}{4\Mkk^2}\,\frac{L}{x_n} \right)^2
   \right] - \frac{1}{x_n^2} \,.
\eeq 
As before this relation holds for the KK excitations of the $W^\pm$ bosons. For the $Z^0$-boson case one replaces $g^2$ by $(g^2+g'^2)$. The KK masses follow from the solutions to (\ref{condi}). To good approximation they are given by the zeros of the Bessel function $J_0(x_n)$.

\section{Bulk Fermions and Flavor Mixing}
\label{sec:fermions}

We consider $N$ generations of 5D fermions in the bulk. They are grouped into $SU(2)_L$ doublets $Q$ and singlets $u^c$ and $d^c$, each of which is an $N$-component vector in flavor space. After EWSB on the IR brane, these fields are coupled by Yukawa matrices $\bm{Y}_u$ and $\bm{Y}_d$, like in the SM. 

\subsection{Action of the 5D Theory}

The quadratic terms in the 5D action can be written in the form \cite{Grossman:1999ra,Gherghetta:2000qt} 
\beq\label{Sferm2}
\begin{split}
   S_{\rm ferm,2} 
   &= \int d^4x\,r\int_{-\pi}^\pi\!d\phi\,\bigg\{ e^{-3\sigma(\phi)}
    \bigg( \bar Q\,i\delslash\,Q 
    + \sum_{q=u,d} \bar q^c\,i\delslash\,q^c \bigg) \\
   &\quad\mbox{}- e^{-4\sigma(\phi)}\,\sgn(\phi) \bigg(
    \bar Q\,\bm{M}_Q\,Q + \sum_{q=u,d} \bar q^c\,\bm{M}_q\,q^c 
    \bigg) \\
   &\quad\mbox{}- \frac{e^{-2\sigma(\phi)}}{r} \Bigg[ 
    \bar Q_L\,\partial_\phi\,e^{-2\sigma(\phi)}\,Q_R 
    - \bar Q_R\,\partial_\phi\,e^{-2\sigma(\phi)}\,Q_L \\
   &\hspace{2.4cm}\mbox{}+ \sum_{q=u,d} \bigg( 
    \bar q_L^c\,\partial_\phi\,e^{-2\sigma(\phi)}\,q_R^c 
    - \bar q_R^c\,\partial_\phi\,e^{-2\sigma(\phi)}\,q_L^c 
    \bigg) \Bigg] \\
   &\quad\mbox{}- \delta(|\phi|-\pi)\,e^{-3\sigma(\phi)} 
    \frac{v}{\sqrt2 r} \left[ \bar u_L\,\bm{Y}_u^{\rm (5D)}\,u_R^c
    + \bar d_L\,\bm{Y}_d^{\rm (5D)}\,d_R^c + \mbox{h.c.} \right] 
    \bigg\} \,,
\end{split}
\eeq
where $\bm{M}_{Q,q}$ are diagonal matrices containing the (real) bulk masses, and $\bm{Y}_q^{\rm (5D)}$ are the 5D Yukawa matrices. We define the dimensionless 4D Yukawa matrices via 
\beq\label{Y4Ddef}
   \bm{Y}_q^{\rm (5D)}\equiv \frac{2\bm{Y}_q}{k} \,,
   \qquad q=u,d \,. 
\eeq
Note that the bulk masses can be positive or negative. Unlike in 4D, the sign of the Dirac mass term cannot be reversed by a field redefinition in the 5D theory. In fact, phenomenology requires that the bulk masses are clustered around the values $M_{Q_i}\approx -k/2$ and $M_{q_i}\approx +k/2$.

The chiral components of each spinor field are as usual defined as $Q=Q_L+Q_R$ \etc\ The left-handed (right-handed) components of the $SU(2)_L$ doublet $Q$ are even (odd) under the $Z_2$ orbifold symmetry. Likewise, the right-handed (left-handed) components of the singlets $u^c$ and $d^c$ are even (odd). This assignment of $Z_2$ parities is such that the zero modes of the even fields correspond to the SM particles. Without the Yukawa interactions, each 5D fermion would give rise to a massless Weyl fermion in the 4D effective theory, accompanied by a tower of massive KK excitations \cite{Grossman:1999ra}. After EWSB, the Yukawa couplings remove the massless modes and replace them by light (compared with the KK scale) SM fermions, each accompanied by two towers of heavy KK states.

The choice of diagonal bulk masses in (\ref{Sferm2}) can be justified as follows. In a more general setup, the action contains positive, hermitian kinetic matrices $\bm{Z}_A$ and hermitian bulk mass matrices $\bm{M}_A$, with $A=Q,u,d$. Unitary transformations can be used to bring the kinetic terms into diagonal form, \ie, $\bm{U}_A^\dagger\,\bm{Z}_A\,\bm{U}_A=\mbox{diag}(Z_{A_1},\dots,Z_{A_N})\equiv\bm{D}_A$ with $Z_{A_i}>0$. The fields are then rescaled to obtain the canonical normalization. In the process, the bulk mass matrices get transformed into $\bm{M}_A'=\bm{D}_A^{-1/2}\,\bm{U}_A^{-1}\,\bm{M}_A\,\bm{U}_A\,\bm{D}_A^{-1/2}$. These new mass matrices can be diagonalized by another set of unitary transformations, \ie, $\bm{U}_A'^\dagger\,\bm{M}_A'\,\bm{U}_A'=\mbox{diag}(M_{A_1},\dots,M_{A_N})$. It follows that, without loss of generality, it is always possible to switch to a basis in which the bulk mass terms are diagonal in flavor space. We will refer to this as the bulk mass basis. If not stated otherwise, from now on $\bm{Y}_u$ and $\bm{Y}_d$ will always denote the Yukawa matrices in this specific basis.

\subsection{Kaluza-Klein Decomposition}
\label{sec:KKferm}

We will discuss the solution of the eigenvalue problem to find the KK
decomposition for the case of the up-type quark sector. An analogous discussion holds for the down-type quark sector. Let $m_n>0$ be the masses of the Dirac fermions in the effective 4D theory and $u^{(n)}=u_L^{(n)}+u_R^{(n)}$ the corresponding spinor fields. We write the KK decomposition of the 5D fields in the form
\begin{eqnarray}\label{KKdecomp}
\begin{split}
   u_L(x,\phi) 
   &= \frac{e^{2\sigma(\phi)}}{\sqrt r} \sum_n
    \bm{C}_n^{(Q)}(\phi)\,a_n^{(U)}\,u_L^{(n)}(x) \,, \qquad
   u_R(x,\phi) = \frac{e^{2\sigma(\phi)}}{\sqrt r} \sum_n
    \bm{S}_n^{(Q)}(\phi)\,b_n^{(U)}\,u_R^{(n)}(x) \,, 
    \hspace{4mm} \\
   u^c_L(x,\phi) &= \frac{e^{2\sigma(\phi)}}{\sqrt r} \sum_n
    \bm{S}_n^{(u)}(\phi)\,b_n^{(u)}\,u_L^{(n)}(x) \,, \hspace{1.05cm}
   u^c_R(x,\phi) = \frac{e^{2\sigma(\phi)}}{\sqrt r} \sum_n
    \bm{C}_n^{(u)}(\phi)\,a_n^{(u)}\,u_R^{(n)}(x) \,, 
\end{split}
\end{eqnarray}
where $\bm{C}_n^{(Q,u)}$ are even profiles under $Z_2$, while $\bm{S}_n^{(Q,u)}$ are odd. The index $n$ labels the mass eigenstates with fermion masses $m_n$ (in this case $m_u, m_c, m_t$ as well as KK excitations in the up-type quark sector) and spinor fields $u^{(n)}$. The spinor fields on the left-hand side of the equations as well as the objects $a_n^{(U,u)}$ and $b_n^{(U,u)}$ are $N$-component vectors in flavor space, and the profiles $\bm{C}_n^{(Q,u)}$, $\bm{S}_n^{(Q,u)}$ are diagonal $N\times N$ matrices, where each entry refers to a different bulk mass parameter (in the bulk mass basis). Note that while the bulk profiles $\bm{C}_n^{(Q)}$ and $\bm{S}_n^{(Q)}$ are the same for the up- and down-type quarks belonging to the same $SU(2)_L$ doublet, the vectors $a_n^{(U)}$ and $a_n^{(D)}$ associated with these profiles in the KK decomposition of the up- and down-type 5D quark fields will be different objects, as indicated by our notation. 

Inserting these relations into the action, one derives the equations of motion 
\beq 
\begin{split} 
   \left( \frac{1}{r}\,\partial_\phi - \bm{M}_Q\,\sgn(\phi) \right)
   \bm{C}_n^{(Q)}(\phi)\,a_n^{(U)}
   &= - m_n\,e^{\sigma(\phi)}\,\bm{S}_n^{(Q)}(\phi)\,a_n^{(U)} \,,
    \\
   \left( - \frac{1}{r}\,\partial_\phi - \bm{M}_Q\,\sgn(\phi) \right)
    \bm{S}_n^{(Q)}(\phi)\, 
   b_n^{(U)} 
   &= - m_n\,e^{\sigma(\phi)}\,\bm{C}_n^{(Q)}(\phi)\,b_n^{(U)} \\
   &\quad\mbox{}+ \delta(|\phi|-\pi)\,e^{\sigma(\phi)}\,
    \frac{\sqrt2\,v}{kr}\,\bm{Y}_u\,\bm{C}_n^{(u)}(\phi)\,
    a_n^{(u)} \,,
\end{split}
\eeq
and similarly 
\beq 
\begin{split}
   \left( - \frac{1}{r}\,\partial_\phi - \bm{M}_u\,\sgn(\phi) \right)
   \bm{C}_n^{(u)}(\phi)\,a_n^{(u)} 
   &= - m_n\,e^{\sigma(\phi)}\,\bm{S}_n^{(u)}(\phi)\,a_n^{(u)} \,,
    \\
   \left( \frac{1}{r}\,\partial_\phi - \bm{M}_u\,\sgn(\phi) \right)
   \bm{S}_n^{(u)}(\phi)\,b_n^{(u)} 
   &= - m_n\,e^{\sigma(\phi)}\,\bm{C}_n^{(u)}(\phi)\,b_n^{(u)} \\
   &\quad\mbox{}+ \delta(|\phi|-\pi)\,e^{\sigma(\phi)}\,
    \frac{\sqrt2\,v}{kr}\,\bm{Y}_u^\dagger\,
    \bm{C}_n^{(Q)}(\phi)\,a_n^{(U)} \,.
\end{split}
\eeq
In the bulk, \ie, for $|\phi|\ne\pi$, and for the special case of a single generation, these relations reduce to the equations first obtained in \cite{Grossman:1999ra}, where it has been shown that the general solutions can be written as linear combinations of Bessel functions (see Section~\ref{sec:fermionprofiles} below). The presence of the brane-localized terms only affects the boundary conditions for the solutions. At $\phi = 0$ one has 
\beq\label{bcUV}
   \bm{S}_n^{(Q,u)}(0) = 0 \,, \qquad \mbox{(UV brane)}
\eeq
and integrating the equations of motion over an infinitesimal interval around $|\phi|=\pi$ we find
\beq\label{bcIR}
\begin{split}
   \bm{S}_n^{(Q)}(\pi^-)\,b_n^{(U)}
   &= \frac{v}{\sqrt2\Mkk}\,\bm{Y}_u\,\bm{C}_n^{(u)}(\pi)\,
    a_n^{(u)} \,, \\
   - \bm{S}_n^{(u)}(\pi^-)\,b_n^{(u)} 
   &= \frac{v}{\sqrt2\Mkk}\,\bm{Y}_u^\dagger\,\bm{C}_n^{(Q)}(\pi)\,
    a_n^{(U)} \,.
\end{split}  
   \qquad \mbox{(IR brane)}
\eeq

Without the brane-localized Yukawa terms, the profiles $\bm{C}_n^{(Q,u)}$ and $\bm{S}_n^{(Q,u)}$ form complete sets of even and odd functions on the orbifold, which can be chosen to obey orthonormality conditions with respect to the measure $d\phi\,e^\sigma$ \cite{Grossman:1999ra}. However, it is not difficult to show that the $\delta$-function terms in the equations of motion are inconsistent with these orthonormality relations. We thus impose the generalized orthonormality conditions 
\beq\label{orthonorm}
\begin{split}
   \int_{-\pi}^\pi\!d\phi\,e^{\sigma(\phi)}\,
   \bm{C}_m^{(Q,u)}(\phi)\,\bm{C}_n^{(Q,u)}(\phi)
   &= \delta_{mn}\,\bm{1} + \bm{\Delta C}_{mn}^{(Q,u)} \,, \\
   \int_{-\pi}^\pi\!d\phi\,e^{\sigma(\phi)}\,
   \bm{S}_m^{(Q,u)}(\phi)\,\bm{S}_n^{(Q,u)}(\phi) 
   &= \delta_{mn}\,\bm{1} + \bm{\Delta S}_{mn}^{(Q,u)} \,.
\end{split}
\eeq
We then find that the 4D action reduces to the desired form
\beq
   S_{\rm ferm,2} = \sum_n \int d^4x \left[ 
   \bar u^{(n)}(x)\,i\delslash\,u^{(n)}(x)
   - m_n\,\bar u^{(n)}(x)\,u^{(n)}(x) \right] ,
\eeq
if and only if, in addition to the boundary conditions (\ref{bcUV})
and (\ref{bcIR}), the relations
\beq\label{abrel}
   a_n^{(U,u)} = b_n^{(U,u)} \,, \qquad 
   a_n^{(U)\dagger}\,a_n^{(U)} + a_n^{(u)\dagger}\,a_n^{(u)} = 1 \,,
\eeq
and
\beq\label{magicCS}
   a_m^{(U,u)\dagger}\,\bm{\Delta C}_{mn}^{(Q,u)}\,a_n^{(U,u)} 
   + a_m^{(u,U)\dagger}\,\bm{\Delta S}_{mn}^{(u,Q)}\,a_n^{(u,U)}
   = 0
\eeq
hold. With the help of the relations (\ref{abrel}) it is straightforward to show that the equations of motion imply 
\beq\label{eq56}
   m_m\,\bm{\Delta C}_{mn}^{(Q,u)} - m_n\,\bm{\Delta S}_{mn}^{(Q,u)} 
   = \pm\frac{2}{r}\,\bm{C}_n^{(Q,u)}(\pi)\,
   \bm{S}_m^{(Q,u)}(\pi^-) \,.
\eeq
Using the symmetry of the relations (\ref{orthonorm}) in $m$ and $n$,
we obtain for $m \ne n$ 
\beq 
\begin{split}   
   \bm{\Delta C}_{mn}^{(Q,u)} 
   &= \pm\frac{2}{r}\,
    \frac{m_m\,\bm{C}_n^{(Q,u)}(\pi)\,\bm{S}_m^{(Q,u)}(\pi^-) 
          - m_n\,\bm{C}_m^{(Q,u)}(\pi)\,\bm{S}_n^{(Q,u)}(\pi^-)}%
         {m_m^2-m_n^2} \,, \\
   \bm{\Delta S}_{mn}^{(Q,u)} &= \mp\frac{2}{r}\,
    \frac{m_m\,\bm{C}_m^{(Q,u)}(\pi)\,\bm{S}_n^{(Q,u)}(\pi^-)
          - m_n\,\bm{C}_n^{(Q,u)}(\pi)\,\bm{S}_m^{(Q,u)}(\pi^-)}%
         {m_m^2-m_n^2} \,.
\end{split}
\eeq
Finally, using the explicit results for the bulk profiles derived in Section~\ref{sec:fermionprofiles}, one finds that the correct expressions for $m=n$ are 
\beq
   \bm{\Delta C}_{nn}^{(Q,u)} = - \bm{\Delta S}_{nn}^{(Q,u)} 
   = \pm\frac{1}{r m_n}\,\bm{C}_n^{(Q,u)}(\pi)\,
   \bm{S}_n^{(Q,u)}(\pi^-) \,.
\eeq
One would naively expect that the extra terms in the generalized
orthonormality conditions (\ref{orthonorm}) are small corrections of
order $v/\Mkk$. However, as we will see below, these terms are in fact $\ord(1)$ for the profiles of the light SM fields.

The boundary conditions (\ref{bcIR}) on the IR brane can now be simplified as 
\beq\label{bcIRn1}
\begin{split}
   \bm{S}_n^{(Q)}(\pi^-)\,a_n^{(U)} 
   &= \frac{v}{\sqrt2 \Mkk}\,\bm{Y}_u\,\bm{C}_n^{(u)}(\pi)\,
    a_n^{(u)} \,, \\
   - \bm{S}_n^{(u)}(\pi^-)\,a_n^{(u)} 
   &= \frac{v}{\sqrt2 \Mkk}\,\bm{Y}_u^\dagger\,
    \bm{C}_n^{(Q)}(\pi)\,a_n^{(U)} \,.
\end{split}
\eeq
These relations can be written in the form of system of $2N$ linear equations for the components of the vectors $a_n^{(U,u)}$, and the eigenvalues are thus determined by the zeros of the determinant of a $(2N)\times(2N)$ matrix. The problem can be simplified by noting that the matrices $\bm{C}_n^{(Q,u)}$, $\bm{S}_n^{(Q,u)}$ are non-singular, so that the inverse matrices exist. We can thus replace (\ref{bcIRn1}) by the decoupled equations 
\beq\label{bcIRn2}
\begin{split}
   \bm{S}_n^{(Q)}(\pi^-)\,a_n^{(U)} 
   &= - \frac{v^2}{2\Mkk^2}\,\bm{Y}_u\,\bm{C}_n^{(u)}(\pi)
    \left[ \bm{S}_n^{(u)}(\pi^-) \right]^{-1}
    \bm{Y}_u^\dagger\,\bm{C}_n^{(Q)}(\pi)\,a_n^{(U)} \,, \\
   \bm{S}_n^{(u)}(\pi^-)\,a_n^{(u)} 
   &= - \frac{v^2}{2\Mkk^2}\,\bm{Y}_u^\dagger\,
    \bm{C}_n^{(Q)}(\pi) \left[ \bm{S}_n^{(Q)}(\pi^-)\right]^{-1}
    \bm{Y}_u\,\bm{C}_n^{(u)}(\pi)\,a_n^{(u)} \,.
\end{split}
\eeq
The mass eigenvalues follow from the solutions to the equation
\beq\label{fermeigenvals}
   \det\left( \bm{1} - \frac{v^2}{2\Mkk^2}
   \left[ \bm{S}_n^{(Q)}(\pi^-) \right]^{-1} \bm{Y}_u\,
   \bm{C}_n^{(u)}(\pi) \left[ - \bm{S}_n^{(u)}(\pi^-) \right]^{-1} 
   \bm{Y}_u^\dagger\,\bm{C}_n^{(Q)}(\pi) \right) = 0 \,.
\eeq
Once they are known, the eigenvectors $a_n^{(U,u)}$ can be
determined from (\ref{bcIRn2}). Note that, while it is always
possible to work with real profiles $\bm{C}_n^{(Q,u)}$ and $\bm{S}_n^{(Q,u)}$, these eigenvectors are, in general, complex-valued objects.

\subsection{Bulk Profiles}
\label{sec:fermionprofiles}

The explicit form of the profiles $(C_n^{(Q,q)})_i$ and $(S_n^{(Q,q)})_i$ associated with bulk mass parameters $M_{Q_i,q_i}$ (with $q=u,d$) was obtained in \cite{Grossman:1999ra,Gherghetta:2000qt}. We will drop the index $i$ for the purposes of this discussion. In terms of the variable $t=\epsilon\,e^\sigma$, one finds
\beq\label{fermprofiles}
\begin{split}
   C_n^{(Q,q)}(\phi) 
   &= {\cal N}_n(c_{Q,q})\,\sqrt{\frac{L\epsilon t}{\pi}}\,
    f_n^+(t,c_{Q,q}) \,, \\
   S_n^{(Q,q)}(\phi) 
   &= \pm {\cal N}_n(c_{Q,q})\,\sgn(\phi)\,
    \sqrt{\frac{L\epsilon t}{\pi}}\,f_n^-(t,c_{Q,q}) \,,
\end{split}
\eeq
where $c_{Q,q}\equiv\pm M_{Q,q}/k$ are dimensionless parameters derived from the bulk mass terms, and
\beq\label{fplmi}
   f_n^\pm(t,c) 
   = J_{-\frac12-c}(x_n\epsilon)\,J_{\mp\frac12+c}(x_n t) 
   \pm J_{\frac12+c}(x_n\epsilon)\,J_{\pm\frac12-c}(x_n t) \,,
\eeq
where as before $x_n=m_n/\Mkk$. The orthonormality relations (\ref{orthonorm}) imply the normalization conditions
\beq
   2\int_\epsilon^1\!dt\,t \left[ f_n^\pm(t,c) \right]^2
   = \frac{1}{{\cal N}_n^2(c)} 
   \pm \frac{f_n^+(1,c)\,f_n^-(1,c)}{x_n} \,, 
\eeq
from which we derive
\beq\label{fermnorm}
   {\cal N}_n^{-2}(c) 
   = \left[ f_n^+(1,c) \right]^2 + \left[ f_n^-(1,c) \right]^2 
    - \frac{2c}{x_n}\,f_n^+(1,c)\,f_n^-(1,c) 
    - \epsilon^2 \left[ f_n^+(\epsilon,c) \right]^2 \,.
\eeq
For the special cases where $c+1/2$ is an integer, the profiles must be obtained from the above expressions by a limiting procedure. 

For the SM fermions it is a very good approximation to expand the above results in the limit $x_n\ll 1$, since even the top-quark mass is much lighter than the KK scale. We find 
\beq\label{SMfermions}
\begin{split}
   C_n^{(Q,q)}(\phi) 
   &\approx \sqrt{\frac{L\epsilon}{\pi}}\,
    \frac{F(c_{Q,q})}{\sqrt{1+\delta_n(c_{Q,q})}}\,
    \left[ t^{c_{Q,q}} - \delta_n(c_{Q,q})\,t^{1-c_{Q,q}} \right] , \\
   S_n^{(Q,q)}(\phi) 
   &\approx \pm\sgn(\phi)\,\sqrt{\frac{L\epsilon}{\pi}}\,
    \frac{x_n F(c_{Q,q})}{\sqrt{1+\delta_n(c_{Q,q})}}\,
    \frac{t^{1+c_{Q,q}} - \epsilon^{1+2c_{Q,q}}\,t^{-c_{Q,q}}}%
         {1+2c_{Q,q}} \,,
\end{split}
\eeq
where we have introduced the ``zero-mode profile'' \cite{Grossman:1999ra,Gherghetta:2000qt}
\beq\label{Fdef}
   F(c) \equiv \sgn[\cos(\pi c)]\,
   \sqrt{\frac{1+2c}{1-\epsilon^{1+2c}}} \,,
\eeq
and the parameters (valid for $c^2\ne 1/4$)
\beq\label{deltan}
   \delta_n(c) \equiv \frac{x_n^2}{4c^2-1}\,\epsilon^{1+2c} \,.
\eeq
The sign factor in (\ref{Fdef}) is chosen such that the signs in (\ref{SMfermions}) agree with those derived from the exact profiles (\ref{fermprofiles}). 

The quantities $F(c)$ and $\delta_n(c)$ strongly depend on the
values of $c$ and $x_n$. For $-1/2<c<1/2$, one has to an excellent approximation 
\beq
   F(c)\approx\sqrt{1+2c} \,, \qquad 
   \delta_n(c)\approx 0 \,.
\eeq
For $-3/2<c<-1/2$, on the other hand,  
\beq
   F(c)\approx -\sqrt{-1-2c}\,\,
   \epsilon^{-c-\frac12}
\eeq
is exponentially small, while $\delta_n$ is strictly positive and can be large in magnitude provided that $x_n^2\gg\ord(\epsilon^{-1-2c})$. For the case of one fermion generation this can never happen, since the eigenvalue equation (\ref{fermeigenvals}) ensures that $x_n^2\,\epsilon^{1+2c}$ can be at most of $\ord(v^2/\Mkk^2)$ for the lightest mass eigenstate, which we would identify with the SM fermion. For more than one generation, however, some of the $\delta_n$ parameters become large, of order $\epsilon^a\,v^2/\Mkk^2$ with a negative coefficient $a$. It is therefore not justified to drop the corresponding terms in (\ref{SMfermions}). However, we find that after the fermion profiles are combined with the mixing parameters $a_n^{(U,u)}$ in (\ref{KKdecomp}), the $\delta_n$ terms always have a very small effect on the fermion masses and, in particular, on the flavor-changing couplings of the RS model.

For the KK excitations of the SM fermions the general relations for the bulk profiles given above can be simplified using that $\epsilon\approx 10^{-16}$ is extremely small, so that we can take the limit $\epsilon\to 0$ in (\ref{fermprofiles}), (\ref{fplmi}), and (\ref{fermnorm}). We then obtain for $c>-1/2$
\beq
\begin{split}
   {\cal N}_n(c)\,f_n^\pm(t,c)
   &= \bar{\cal N}_n(c)\,J_{\mp\frac12+c}(x_n t) \,, \\
   \left[ \bar{\cal N}_n(c) \right]^{-2}
   &= J_{\frac12+c}^2(x_n) + J_{-\frac12+c}^2(x_n)
    - \frac{2c}{x_n}\,J_{\frac12+c}(x_n)\,J_{-\frac12+c}(x_n) \,,
\end{split}
\eeq
whereas for $c<-1/2$
\beq
\begin{split}
   {\cal N}_n(c)\,f_n^\pm(t,c)
   &= \pm\bar{\cal N}_n(c)\,J_{\pm\frac12-c}(x_n t) \,, \\
   \left[ \bar{\cal N}_n(c) \right]^{-2}
   &= J_{\frac12-c}^2(x_n) + J_{-\frac12-c}^2(x_n)
    + \frac{2c}{x_n}\,J_{\frac12-c}(x_n)\,J_{-\frac12-c}(x_n) \,.
\end{split}
\eeq

\subsection{Normalization of the 4D Yukawa Couplings}
\label{sec:before}

Our definition of the 4D Yukawa couplings in (\ref{Y4Ddef}) was based on making the 5D Yukawa couplings dimensionless with the help of the AdS$_5$ curvature $k$. In that regard we followed the convention most frequently adopted in the literature and used in the original papers \cite{Grossman:1999ra,Gherghetta:2000qt}. It is important to emphasize that this convention is neither unique nor particularly natural. While the relation between the 5D and 4D gauge couplings in (\ref{g4def}) is dictated by the fact that the couplings of the zero-mode gauge bosons to fermions take the familiar form (see relation (\ref{eq115}) in Section~\ref{subsec:gluons} below), no such argument can be given for the Yukawa couplings. In the SM the couplings of the Higgs boson to fermions are proportional to the masses of the fermions. However, in the RS framework the fermion masses arise from products of Yukawa couplings with strongly hierarchical fermion profiles. 

Given the form of the Yukawa interactions in (\ref{Sferm2}) and the fact that the KK decomposition (\ref{KKdecomp}) of the 5D fermion fields introduces an additional factor $1/r$, it would appear logical to define dimensionless 4D Yukawa couplings by using the scale $r$ instead of $k$ in (\ref{Y4Ddef}), \eg,
\beq\label{Y4Dalt}
   \bar{\bm{Y}}_q \equiv \frac{\bm{Y}_q^{\rm (5D)}}{2\pi r}
   = \frac{1}{L}\,\bm{Y}_q \,, 
   \qquad q=u,d \,,
\eeq
which resembles (\ref{g4def}). The so-defined 4D Yukawa couplings are more than an order of magnitude smaller than the ones defined in (\ref{Y4Ddef}) and used in most of the literature. The question arises: should we expect that $\bm{Y}_q=\ord(1)$, or $\bar{\bm{Y}}_q=\ord(1)$, or something else?

Notice that with the conventional definition (\ref{Y4Ddef}) the 4D Yukawa matrices absorb a factor of $L$ arising from the fermion profiles evaluated on the IR brane, which appear in the Higgs-boson couplings to fermions. However, the same factor of $L$ appears in the Higgs-boson couplings to KK gauge bosons, after the 5D gauge coupling is rescaled as in (\ref{g4def}). More generally, all interaction terms in the effective 4D Lagrangian that result from interactions on (or close) to the IR brane are enhanced by powers of $\sqrt{L}$. For instance, it is well known that the couplings of KK gauge bosons to heavy fermions contain such a factor \cite{Davoudiasl:1999tf,Pomarol:1999ad}. It seems {\it ad hoc\/} to remove the factor $L$ in the case of the Yukawa couplings, when similar factors are present and cannot be removed for all other interactions near the IR brane.

Consider, for example, the Feynman rule for the Higgs coupling to a pair of neutral weak gauge bosons of KK levels $n_1$ and $n_2$, which can be written as
\beq\label{hZZcoupl}
\raisebox{-0.85cm}{
\makebox[3.5cm]{
\includegraphics[height=2cm]{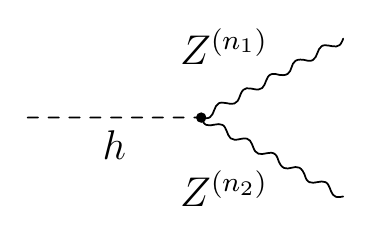}}
\raisebox{0.85cm}{
$\displaystyle = \quad 
 \frac{2im_Z^2}{v}\,z_{n_1} z_{n_2}\,g^{\mu\nu}\,,$}}
\eeq
where to very good approximation
\beq
   z_0 = 1 \,, \qquad
   z_{n\ge 1} = (-1)^n\,\sqrt{2L} \,.
\eeq
For comparison, the Feynman rule for the Higgs coupling to a pair of  quarks of KK levels $n_1$ and $n_2$ reads\footnote{We neglect flavor mixing for simplicity here, \ie, we assume a single fermion generation.}
\beq\label{hff}
\raisebox{-0.85cm}{
\makebox[3.5cm]{
\includegraphics[height=2cm]{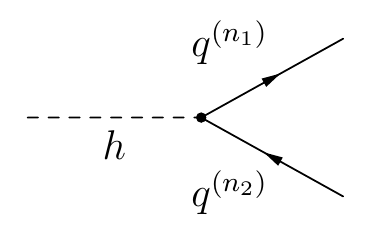}}
\raisebox{0.85cm}{
$\displaystyle = \quad - \frac{i}{\sqrt2} \left[ 
 \left( f_{n_1}^Q \right)^* Y_q\,f_{n_2}^q\,\frac{1+\gamma_5}{2} 
 + \left( f_{n_1}^q \right)^* Y_q^* f_{n_2}^Q\,\frac{1-\gamma_5}{2} 
 \right] ,$}}
\eeq
where to good approximation (for $A=Q,q$)
\beq
   f_0^A = F(c_A)\,\sqrt{2} a_0^{(A)} \,, \qquad
   f_{n\ge 1}^A = (-1)^n\,\sgn[\cos(\pi c_A)]\,\sqrt{2} a_n^{(A)} \,.
\eeq
The complex coefficients $a_n^{(A)}$ are $\ord(1)$ parameters determined by the conditions (\ref{abrel}) and (\ref{bcIRn1}), with $\sqrt{2}\,|a_0^{(A)}|\approx 1$ for the SM quarks. For the case of the Higgs-boson couplings to two SM particles, the Feynman rules in (\ref{hZZcoupl}) and (\ref{hff}) coincide with those of the SM once we identify $(f_0^Q)^*\,Y_q\,f_0^q=\sqrt2\,m_q/v$ with the effective Yukawa coupling of the SM fermion, which is close to 1 for the top quark. For the Higgs-boson couplings to two (or one) KK particles, on the other hand, the vertex (\ref{hZZcoupl}) is enhanced by a factor of $L$ (or $\sqrt{L}$), while no such factor appears in (\ref{hff}). However, if we were to replace the Yukawa couplings $Y_q$ by $L\,\bar Y_q$ according to (\ref{Y4Dalt}), then this would make (\ref{hff}) look more similar to (\ref{hZZcoupl}). 

\begin{figure}[!t]
\begin{center}
\raisebox{0.2cm}{\makebox[5cm]{
\includegraphics[height=2.6cm]{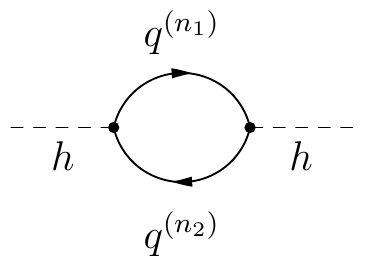}}}
\makebox[4cm]{
\includegraphics[height=3cm]{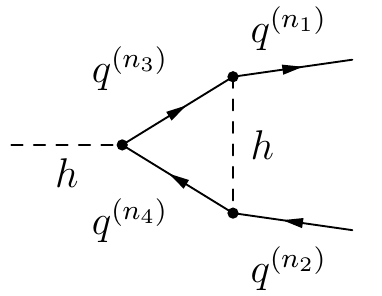}}
\makebox[5cm]{
\includegraphics[height=3cm]{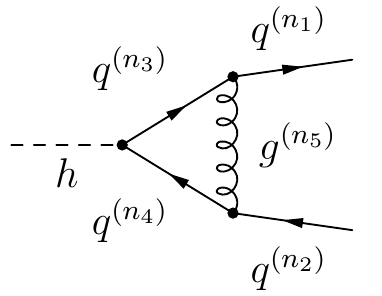}}
\vspace{-2mm}
\parbox{15.5cm}{\caption{\label{fig:higgsyukawa}
Examples of one-loop contributions to the renormalization of the
Higgs-boson mass (left) and of the Yukawa couplings, involving a Higgs-boson (middle) and a gluon as well as its KK excitations (right).}}
\end{center}
\end{figure}

One may try to derive an upper bound on the scale of the Yukawa couplings by invoking perturbativity of the effective 4D theory up to a cutoff scale of a few TeV. For instance, naive dimensional analysis shows that at one-loop order the Yukawa interactions receive a multiplicative correction of order \cite{Csaki:2008zd}
\beq\label{NDA1}
   \frac{|Y_q|^2}{16\pi^2}\,\frac{\Lambda_{\rm UV}^2}{\Mkk^2} \,,
\eeq
where $\Lambda_{\rm UV}$ is the cutoff scale of the theory on the IR brane. The graph in the middle in Figure~\ref{fig:higgsyukawa} shows an example of a diagram giving rise to such a correction. Requiring that this correction be at most of $\ord(1)$ for a cutoff a factor of 4 above the KK scale gives the upper bound $|Y_q|<\pi$, which has been used in \cite{Csaki:2008zd,Santiago:2008vq}. 

A potential weakness of this argument is that it is not clear {\it a priori\/} if the theory should be weakly coupled in the Yukawa sector (or in any other sector), and up to what cutoff scale weak coupling should hold. We note in this context that an explicit calculation of the graph on the right in Figure~\ref{fig:higgsyukawa} suggests that there is a QCD correction to the Yukawa interaction of order
\beq\label{NDA2}
   \frac{L\alpha_s}{4\pi}\,\frac{\Lambda_{\rm UV}^2}{\Mkk^2} \,,
\eeq
where the factor $L$ reflects the enhanced strength of the coupling of KK gluons to KK fermions. There is no point in requiring that the correction (\ref{NDA1}) be smaller than (\ref{NDA2}), and therefore $|Y_q|<\sqrt{4\pi L\alpha_s}\approx\sqrt{L}$ appears to be the strongest reasonable bound one should impose. 

Even this bound may be too restrictive, however. Using once more naive dimensional analysis, we expect a correction to the Higgs mass from Yukawa interactions that scales like
\beq
   \delta m_h^2\sim n_f\,\Lambda_{\rm UV}^2\,
   \frac{|Y_q|^2}{16\pi^2}\,\frac{\Lambda_{\rm UV}^2}{\Mkk^2} \,,
\eeq
which stems from the graph on the left in Figure~\ref{fig:higgsyukawa}. Here $n_f=6$ is the number of quark flavors. Additional corrections will arise from the lepton sector. We emphasize that this correction grows like the fourth power of the cutoff scale, not like the second power as in 4D. The ``little hierarchy problem'' is thus more severe than in 4D extensions of the SM. Choosing the cutoff far above the KK scale would inevitably induce a very large correction to the Higgs-boson mass, which would reintroduce a fine-tuning problem to the RS model. In order to avoid this problem, we must require that the cutoff be of order the KK scale, as is the natural expectation for the Higgs-boson mass. For $\Lambda_{\rm UV}\sim\Mkk$, however, relation (\ref{NDA1}) allows Yukawa couplings as large as $|Y_q|<4\pi$.

\section{Hierarchies of Fermion Masses and Mixings}
\label{sec:hierarchies}

In order to analyze the implications of the RS setup for the masses and mixings of the SM fermions, it is instructive to use approximate formulae for the bulk profiles. The approach most commonly adopted in the literature is to employ a ``zero-mode approximation'' (ZMA), in which one first solves for the fermion bulk profiles without the Yukawa couplings and then treats these couplings as a perturbation \cite{Grossman:1999ra,Gherghetta:2000qt,Huber:2000ie}. At the technical level, this corresponds to an expansion of the exact profiles in powers of $v^2/\Mkk^2$, in which also the quantities $\delta_n$ in (\ref{deltan}) are treated as $\ord{(v^2/\Mkk^2)}$ parameters. In this approximation the values of the SM fermion profiles on the IR brane simplify considerably. From (\ref{SMfermions}) we obtain
\beq
   C_n^{(Q,q)}(\pi)\to \sqrt{\frac{L\epsilon}{\pi}}\,F(c_{Q,q}) \,,
    \qquad
   S_n^{(Q,q)}(\pi^-)\to \pm\sqrt{\frac{L\epsilon}{\pi}}\,
    \frac{x_n}{F(c_{Q,q})} \,.
\eeq
Using these expressions, the system of equations (\ref{bcIRn1}) can be recast into the form
\beq\label{eq60}
   \frac{\sqrt2\,m_n}{v}\,\hat a_n^{(U)} 
   = \bm{Y}_u^{\rm eff}\,\hat a_n^{(u)} \,, \qquad 
   \frac{\sqrt2\,m_n}{v}\,\hat a_n^{(u)} 
   = (\bm{Y}_u^{\rm eff})^\dagger\,\hat a_n^{(U)} \,,
\eeq
where
\beq\label{eq:Yueff}
   \left( Y_u^{\rm eff} \right)_{ij} 
   \equiv F(c_{Q_i}) \left( Y_u \right)_{ij} F(c_{u_j})
\eeq
are effective Yukawa matrices, and the rescaled vectors $\hat a_n^{(A)}\equiv\sqrt{2}\,a_n^{(A)}$ obey the normalization conditions
\beq\label{ahatnorm}
   \hat a_n^{(U)\dagger}\,\hat a_n^{(U)} 
   = \hat a_n^{(u)\dagger}\,\hat a_n^{(u)} = 1 \,.
\eeq
Furthermore, we obtain from (\ref{eq60}) the equalities
\beq\label{eq61}
   \left( m_n^2\,\bm{1} - \frac{v^2}{2}\,
    \bm{Y}_u^{\rm eff}\,(\bm{Y}_u^{\rm eff})^\dagger \right) 
    \hat a_n^{(U)} = 0 \,, \qquad 
   \left( m_n^2\,\bm{1} - \frac{v^2}{2}\,
    (\bm{Y}_u^{\rm eff})^\dagger\,\bm{Y}_u^{\rm eff} \right) 
    \hat a_n^{(u)} = 0 \,,
\eeq
and the mass eigenvalues are the solutions to the equation 
\beq
   \det\left( m_n^2\,\bm{1} - \frac{v^2}{2}\, 
   \bm{Y}_u^{\rm eff}\,(\bm{Y}_u^{\rm eff})^\dagger \right) = 0 \,.
\eeq
Notice that in the ZMA, but not in general, the vectors $a_n^{(A)}$ and $\hat a_n^{(A)}$ belonging to different $n$ are orthogonal on each other.

The eigenvectors $\hat{a}_n^{(Q)}$ and $\hat{a}_n^{(q)}$ of the matrices ${\bm Y}_q^{\rm eff} \left( {\bm Y}_q^{\rm eff} \right)^\dagger$ and $\left( {\bm Y}_q^{\rm eff} \right)^\dagger {\bm Y}_q^{\rm eff}$ (with $n=1,2,3$ and $Q=U,D$, $q=u,d$) form the columns of the unitary matrices ${\bm U}_q$ and ${\bm W}_q$ appearing in the singular-value decomposition
\beq\label{eq:singular}
   \bm{Y}_q^{\rm eff} 
   = \bm{U}_q\,\bm{\lambda}_q\,\bm{W}_q^\dagger \,,
\eeq
where
\beq\label{eq:lambdaud}
   \bm{\lambda}_u = \frac{\sqrt{2}}{v}\,
    \mbox{diag}(m_u,m_c,m_t) \,, \qquad 
   \bm{\lambda}_d = \frac{\sqrt{2}}{v}\,
    \mbox{diag}(m_d,m_s,m_b) \,.
\eeq
It follows that in this approximation the relations between the original 5D fields and the SM mass eigenstates involve the matrices $\bm{U}_q$ and $\bm{W}_q$. In particular, the CKM mixing matrix is given by
\beq\label{eq:VCKM}
   \bm{V}_{\rm CKM} = \bm{U}_u^\dagger\,\bm{U}_d \,.
\eeq

\subsection{Warped-Space Froggatt-Nielsen Mechanism}			
\label{sec:quark}

Concentrating on the well-motivated case of anarchic 5D Yukawa couplings, \ie, complex-valued matrices $\bm{Y}_q$ with random elements, it turns out that the up- and down-type quark mass hierarchies can be reproduced by assuming a hierarchical structure of the elements of the zero-mode profiles of the form\footnote{The quarks are assumed to be ordered in such a way that these relations hold. In the case of degenerate values $|F(c_{A_i})|=|F(c_{A_{i+1}})|$, the following discussion is only order-of-magnitude wise correct.}
\beq\label{eq:hierarchy}
   |F(c_{A_1})| < |F(c_{A_2})| < |F(c_{A_3})| \,.
\eeq
In the RS framework such a hierarchy is natural, since it results from small differences in the bulk mass parameters $c_{A_i}$.

Given effective Yukawa matrices of the form (\ref{eq:Yueff}), the hierarchies of fermion masses and mixings result without further assumption from the Froggatt-Nielsen mechanism \cite{Froggatt:1978nt}. To verify this statement, note that the products of up- and down-type quark masses are given by
\beq\label{eq:detdu}
\begin{split}
   m_u\,m_c\,m_t 
   &= \frac{v^3}{2\sqrt2} \left| \det\left( \bm{Y}_u \right ) \right|
    \prod_{i=1,2,3} \left| F(c_{Q_i})\,F(c_{u_i}) \right| , \\
   m_d\,m_s\,m_b 
   &= \frac{v^3}{2\sqrt2} \left| \det\left( \bm{Y}_d \right ) \right|
    \prod_{i=1,2,3} \left| F(c_{Q_i})\,F(c_{d_i}) \right| .
\end{split}
\eeq
Since $|F(c_{A_i})|<|F(c_{A_{i+1}})|$, one can consistently evaluate
all the eigenvalues to leading order in hierarchies. We obtain
\beq\label{eq:quarkmasses} 
\begin{aligned}
   m_u &= \frac{v}{\sqrt2}\,
    \frac{|\det(\bm{Y}_u)|}{|(M_u)_{11}|}\,
    |F(c_{Q_1}) F(c_{u_1})| \,, & \qquad
   m_d &= \frac{v}{\sqrt2}\,
    \frac{|\det(\bm{Y}_d)|}{|(M_d)_{11}|}\,
    |F(c_{Q_1}) F(c_{d_1})| \,, \\
   m_c & = \frac{v}{\sqrt2}\,
    \frac{|(M_u)_{11}|}{|(Y_u)_{33}|}\,
    |F(c_{Q_2}) F(c_{u_2})| \,, & \qquad
   m_s & = \frac{v}{\sqrt2}\,
    \frac{|(M_d)_{11}|}{|(Y_d)_{33}|}\,
    |F(c_{Q_2}) F(c_{d_2})| \,, \\
   m_t & = \frac{v}{\sqrt2}\,|(Y_u)_{33}|\,
    |F(c_{Q_3}) F(c_{u_3})| \,, & \qquad 
   m_b & = \frac{v}{\sqrt2}\,|(Y_d)_{33}|\,
    |F(c_{Q_3}) F(c_{d_3})| \,.
\end{aligned}
\eeq
Here $(M_q)_{ij}$ denotes the minor of $\bm{Y}_q$, \ie, the determinant of the square matrix formed by removing the $i^{\rm th}$
row and the $j^{\rm th}$ column from $\bm{Y}_q$.

The elements of the matrices $\bm{U}_q$ and $\bm{W}_q$ are given to
leading order in hierarchies by
\beq\label{eq:UWq}
   (U_q)_{ij} = (u_q)_{ij}
   \begin{cases} 
    \frac{\displaystyle F(c_{Q_i})}{\displaystyle F(c_{Q_j})} \,,
    & i\le j \,, \\[5mm]
    \frac{\displaystyle F(c_{Q_j})}{\displaystyle F(c_{Q_i})} \,, 
    & i>j \,, 
   \end{cases} \qquad 
   (W_q)_{ij} = (w_q)_{ij}\;e^{i\phi_j}
   \begin{cases} 
    \frac{\displaystyle F(c_{q_i})}{\displaystyle F(c_{q_j})} \,, 
    & i\le j \,, \\[5mm]
    \frac{\displaystyle F(c_{q_j})}{\displaystyle F(c_{q_i})} \,, 
    & i>j \,. 
   \end{cases}
\eeq
Expressed through the elements $(Y_q)_{ij}$ of the original Yukawa matrices and their minors $(M_q)_{ij}$, the coefficient matrices $\bm{u}_q$ and $\bm{w}_q$ read
\beq\label{eq:uwq} 
   \bm{u}_q = \begin{pmatrix}
    1 & \frac{\displaystyle \left( M_q \right)_{21}}%
             {\displaystyle \left( M_q \right)_{11}} &
    \frac{\displaystyle \left( Y_q \right)_{13}}%
         {\displaystyle \left( Y_q \right)_{33}} \\ \\[-2mm]
    - \frac{\displaystyle \left( M_q \right)_{21}^\ast}
          {\displaystyle \left( M_q \right)_{11}^\ast} & 1 &
    \frac{\displaystyle \left( Y_q \right)_{23}}%
         {\displaystyle \left( Y_q \right)_{33}} \\ \\[-2mm]
    \frac{\displaystyle \left( M_q \right)_{31}^\ast}%
         {\displaystyle \left( M_q \right)_{11}^\ast} & 
    -\frac{\displaystyle \left( Y_q \right)_{23}^\ast}%
          {\displaystyle \left( Y_q \right)_{33}^\ast} & 1
    \end{pmatrix} , \qquad
   \bm{w}_q = \begin{pmatrix}
    1 & \frac{\displaystyle \left( M_q \right)_{12}^\ast}%
             {\displaystyle \left( M_q \right)_{11}^\ast} &
    \frac{\displaystyle \left( Y_q \right)_{31}^*}%
         {\displaystyle \left( Y_q \right)_{33}^*} \\ \\[-2mm]
    - \frac{\displaystyle \left( M_q \right)_{12}}%
           {\displaystyle \left( M_q \right)_{11}} & 1 & 
    \frac{\displaystyle \left( Y_q \right)_{32}^\ast}%
         {\displaystyle \left( Y_q \right)_{33}^\ast} \\ \\[-2mm]
    \frac{\displaystyle \left( M_q \right)_{13}}%
         {\displaystyle \left( M_q \right)_{11}} & 
    - \frac{\displaystyle \left( Y_q \right)_{32}}%
          {\displaystyle \left( Y_q \right)_{33}} & 1
    \end{pmatrix} .
\eeq
The phase factors $e^{i\phi_j}$ entering $\bm{W}_q$ are given by
\beq\label{eq:expphij}
   e^{i\phi_j} = \sgn\big[ F(c_{Q_j})\,F(c_{q_j})\big]\, 
   e^{-i(\rho_j-\rho_{j+1})} \,,
\eeq
with 
\beq
   \rho_1 = \arg\left( \det(\bm{Y}_q) \right) , \qquad 
   \rho_2 = \arg\left( (M_q)_{11} \right) , \qquad 
   \rho_3 = \arg\left( (Y_q)_{33} \right) , 
\eeq
and $\rho_4=0$. Note that to leading order the matrices $\bm{U}_q$ and therefore also the CKM mixing matrix do not depend on the right-handed profiles $F(c_{q_i})$. This feature has first been pointed out in \cite{Huber:2003tu}. Exploiting the invariance of the singular-value decomposition (\ref{eq:singular}) under field redefinitions allows to make either the diagonal elements $(U_q)_{ii}$ or $(W_q)_{ii}$ real. In (\ref{eq:UWq}) we have chosen $(U_q)_{ii}$ to be real, so that all phase factors $e^{i\phi_j}$ appear in the elements $(W_q)_{ij}$.

Recalling the definitions of the Wolfenstein parameters of the CKM matrix,
\beq\label{eq:lAre}
   \lambda 
   = \frac{\left|V_{us}\right|}%
          {\sqrt{\left|V_{ud}\right|^2 + \left|V_{us}\right|^2}} ,
    \qquad
   A = \frac{1}{\lambda} \left| \frac{V_{cb}}{V_{us}} \right| ,
    \qquad
   \bar\rho - i\bar\eta 
   = - \frac{V_{ud}^* V_{ub}}{V_{cd}^* V_{cb}} \,,
\eeq
it is straightforward to derive from (\ref{eq:VCKM}), (\ref{eq:UWq}), and (\ref{eq:uwq}) the leading-order expressions for $\lambda$, $A$,
$\bar\rho$, and $\bar\eta$. We find
\beq\label{eq:wolfenstein}
\begin{split}
   \lambda = \frac{|F(c_{Q_1})|}{|F(c_{Q_2})|} 
    \left| \frac{\left( M_d \right)_{21}}{\left( M_d \right)_{11}}
    - \frac{\left( M_u \right)_{21}}{\left( M_u \right)_{11}} \right|
    \,, \qquad 
   A = \frac{|F(c_{Q_2})|^3}{|F(c_{Q_1})|^2\,|F(c_{Q_3})|}
    \left| \frac{%
     \frac{\displaystyle \left( Y_d \right)_{23}}%
          {\displaystyle \left( Y_d \right)_{33}}
     - \frac{\displaystyle \left( Y_u \right)_{23}}%
            {\displaystyle \left(Y_u \right)_{33}}}%
     {\left[ \frac{\displaystyle \left( M_d \right)_{21}}%
                  {\displaystyle \left( M_d \right)_{11}} 
     - \frac{\displaystyle \left( M_u \right)_{21}}%
            {\displaystyle \left( M_u \right)_{11}} \right]^2}
     \right| , \\
   \bar\rho - i\bar\eta
   = \frac{\left( Y_d \right)_{33} \left( M_u \right)_{31}
           - \left( Y_d \right)_{23} \left( M_u \right)_{21} 
           + \left( Y_d \right)_{13} \left( M_u \right)_{11}}%
          {\left( Y_d \right)_{33} \left( M_u \right)_{11} 
           \left[ \frac{\displaystyle \left( Y_d \right)_{23}}%
                       {\displaystyle \left( Y_d \right)_{33}} 
           - \frac{\displaystyle \left( Y_u \right)_{23}}%
                  {\displaystyle \left( Y_u \right)_{33}} \right]
           \left[ \frac{\displaystyle \left( M_d \right)_{21}}%
                       {\displaystyle \left( M_d \right)_{11}} 
           - \frac{\displaystyle \left( M_u \right)_{21}}%
                  {\displaystyle \left( M_u \right)_{11}} \right]}
   \,. \hspace{2.0cm}
\end{split}
\eeq
Notice that $\bar{\rho}$ and $\bar{\eta}$ are to first order independent of the zero-mode profiles $F(c_{A_i})$. Like in the case of the Froggatt-Nielsen mechanism \cite{Froggatt:1978nt}, the RS setup predicts that these parameters are thus not suppressed by any small parameters, while their precise $\ord(1)$ values remain unexplained. 

The relations given in (\ref{eq:quarkmasses}) and (\ref{eq:lAre}) do not allow one to determine the zero-mode profiles solely in terms of the quark masses and Wolfenstein parameters. Expressing them through $F(c_{Q_2})$, one finds for the left-handed quark profiles 
\beq\label{eq:Qwave}
   |F(c_{Q_1})| 
   = \frac{\lambda}{\left| 
    \frac{\displaystyle \left( M_d \right)_{21}}%
         {\displaystyle \left( M_d \right)_{21}} 
    - \frac{\displaystyle \left( M_u \right)_{21}}%
           {\displaystyle \left( M_u \right)_{21}}\right|}\,
    |F(c_{Q_2})| \,, \qquad 
   |F(c_{Q_3})| 
   = \frac{\left| 
    \frac{\displaystyle \left( Y_d \right)_{23}}%
         {\displaystyle \left( Y_d \right)_{33}} 
    - \frac{\displaystyle \left( Y_u \right)_{23}}%
           {\displaystyle \left( Y_u \right)_{33}}\right|}%
          {A\lambda^2}\,|F(c_{Q_2})| \,.
\eeq
In the case of the right-handed up- and down-type quark profiles, one
obtains 
\beq\label{eq:uwave}
\begin{split}
   |F(c_{u_1})| &= \frac{\sqrt{2} m_u}{v}\,
    \frac{\left| \left( M_u \right)_{11} \right| 
          \left| \frac{\displaystyle \left( M_d \right)_{21}}%
                      {\displaystyle \left( M_d \right)_{11}} 
          - \frac{\displaystyle \left( M_u \right)_{21}}%
                 {\displaystyle \left( M_u \right)_{11}} \right|}%
         {\lambda\left| \det(\bm{Y}_u) \right|}\,
    \frac{1}{|F(c_{Q_2})|} \,, \\
   |F(c_{u_2})| &= \frac{\sqrt{2} m_c}{v}\,
    \frac{\left| \left( Y_u \right)_{33} \right|}%
         {\left| \left( M_u \right)_{11} \right|}\,
    \frac{1}{|F(c_{Q_2})|} \,, \\
   |F(c_{u_3})| &= \frac{\sqrt{2} m_t}{v}\,
    \frac{A\lambda^2}%
         {\left| \left( Y_u \right)_{33} \right|
          \left| \frac{\displaystyle \left( Y_d \right)_{23}}%
                      {\displaystyle \left( Y_d \right)_{33}} 
          - \frac{\displaystyle \left( Y_u \right)_{23}}%
                 {\displaystyle \left( Y_u \right)_{33}}\right|}\,
    \frac{1}{|F(c_{Q_2})|} \,,
\end{split}
\eeq
and 
\beq
\begin{split}
   |F(c_{d_1})| &= \frac{\sqrt{2} m_d}{v}\,
    \frac{\left| \left( M_d \right)_{11} \right| 
          \left| \frac{\displaystyle \left( M_u \right)_{21}}%
                      {\displaystyle \left( M_u \right)_{11}} 
          - \frac{\displaystyle \left( M_d \right)_{21}}%
                 {\displaystyle \left( M_d \right)_{11}} \right|}%
         {\lambda\left| \det(\bm{Y}_d) \right|}\,
    \frac{1}{|F(c_{Q_2})|} \,, \\
   |F(c_{d_2})| &= \frac{\sqrt{2} m_s}{v}\,
    \frac{\left| \left( Y_d \right)_{33} \right|}%
         {\left| \left( M_d \right)_{11} \right|}\,
    \frac{1}{|F(c_{Q_2})|} \,, \\
   |F(c_{d_3})| &= \frac{\sqrt{2} m_b}{v}\,
    \frac{A\lambda^2}%
         {\left| \left( Y_d \right)_{33} \right|
          \left| \frac{\displaystyle \left( Y_u \right)_{23}}%
                      {\displaystyle \left( Y_u \right)_{33}} 
          - \frac{\displaystyle \left( Y_d \right)_{23}}%
                 {\displaystyle \left( Y_d \right)_{33}}\right|}\,
    \frac{1}{|F(c_{Q_2})|} \,.
\end{split}
\eeq
These relations imply a hierarchical structure among the various quark profiles \cite{Huber:2003tu}. For the left-handed profiles one finds
\beq\label{eq:WFLhierrarchy}
   \frac{|F(c_{Q_1})|}{|F(c_{Q_2})|} \sim \lambda \,, \qquad
   \frac{|F(c_{Q_2})|}{|F(c_{Q_3})|} \sim \lambda^2 \,, \qquad 
   \frac{|F(c_{Q_1})|}{|F(c_{Q_3})|} \sim \lambda^3 \,.
\eeq
The values of the right-handed up- and down-type quark profiles are
then fixed by the observed quark-mass hierarchies. One obtains 
\beq\label{eq:Fvalues}
\begin{split}
   \frac{|F(c_{u_1})|}{|F(c_{u_3})|} 
    \sim \frac{m_u}{m_t}\,\frac{1}{\lambda^3} \,, \qquad 
   \frac{|F(c_{u_2})|}{|F(c_{u_3})|}
    \sim \frac{m_c}{m_t}\,\frac{1}{\lambda^2} \,, \hspace{1.7cm} \\ 
   \frac{|F(c_{d_1})|}{|F(c_{u_3})|}
    \sim \frac{m_d}{m_t}\,\frac{1}{\lambda^3} \,, \qquad 
   \frac{|F(c_{d_2})|}{|F(c_{u_3})|}
    \sim \frac{m_s}{m_t}\,\frac{1}{\lambda^2} \,, \qquad 
   \frac{|F(c_{d_3})|}{|F(c_{u_3})|}
    \sim \frac{m_b}{m_t} \,.
\end{split}
\eeq

It is now straightforward to deduce the hierarchical structures of the flavor mixing matrices $\bm{U}_q$ and $\bm{W}_q$ in (\ref{eq:UWq}). The matrices in the left-handed quark sector have the same structure as the CKM matrix, \ie,
\beq\label{eq:Uud} 
   \bm{U}_{u,d} \sim \bm{V}_{\rm CKM} \sim 
   \begin{pmatrix} 1 & ~\lambda~ & \lambda^3 \\ 
                        \lambda & 1 & \lambda^2 \\ 
                        \lambda^3 & \lambda^2 & 1 
   \end{pmatrix} \sim
   \begin{pmatrix} 1 & ~0.23~ & 0.01 \\ 
                   0.23 & 1 & 0.05 \\ 
                  0.01 & 0.05 & 1 
   \end{pmatrix} , 
\eeq
while in the right-handed quark sector one has
\beq\label{eq:WudVR}
\begin{split}
   \bm{W}_u &\sim 
    \begin{pmatrix} 1 & 
     \frac{\displaystyle m_u}{\displaystyle m_c}\,
      \frac{\displaystyle 1}{\displaystyle\lambda} & 
     \frac{\displaystyle m_u}{\displaystyle m_t}\,
      \frac{\displaystyle 1}{\displaystyle \lambda^3} \\ \\[-2mm]
     \frac{\displaystyle m_u}{\displaystyle m_c}\,
      \frac{\displaystyle 1}{\displaystyle \lambda} & 1 &
     \frac{\displaystyle m_c}{\displaystyle m_t}\,
      \frac{\displaystyle 1}{\displaystyle \lambda^2} \\ \\[-2mm]
     \frac{\displaystyle m_u}{\displaystyle m_t}\,
      \frac{\displaystyle 1}{\displaystyle \lambda^3} & 
     \frac{\displaystyle m_c}{\displaystyle m_t}\,
      \frac{\displaystyle 1}{\displaystyle \lambda^2} & 1
    \end{pmatrix} \sim 
    \begin{pmatrix} 1 & ~0.012~ & 0.001 \\ 
                    0.012 & 1 & 0.077 \\ 
                    0.001 & ~0.077~ & 1 
    \end{pmatrix} , \\[2mm]
   \bm{W}_d & \sim 
    \begin{pmatrix} 1 & 
     \frac{\displaystyle m_d}{\displaystyle m_s}\,
      \frac{\displaystyle 1}{\displaystyle \lambda} & 
     \frac{\displaystyle m_d}{\displaystyle m_b}\,
      \frac{\displaystyle 1}{\displaystyle \lambda^3} \\ \\[-2mm]
     \frac{\displaystyle m_d}{\displaystyle m_s}\,
      \frac{\displaystyle 1}{\displaystyle \lambda} & 1 & 
     \frac{\displaystyle m_s}{\displaystyle m_b}\,
      \frac{\displaystyle 1}{\displaystyle \lambda^2} \\ \\[-2mm]
     \frac{\displaystyle m_d}{\displaystyle m_b}\,
      \frac{\displaystyle 1}{\displaystyle \lambda^3} & 
     \frac{\displaystyle m_s}{\displaystyle m_b}\,
      \frac{\displaystyle 1}{\displaystyle \lambda^2} & 1
    \end{pmatrix} \sim 
    \begin{pmatrix} 1 & 0.26 & 0.12 \\ 
                    0.26 & 1 & 0.44 \\ 
                    0.12 & ~0.44~ & 1 
    \end{pmatrix} .
\end{split}
\eeq
The default values for the quark masses and CKM parameters used to obtain these estimates are collected in Appendix~\ref{app:masses}.

It is worth emphasizing at this point that while the fermion profiles are strongly hierarchical in the left-handed quark sector as well as in the right-handed up-quark sector, the hierarchy in the right-handed down-quark sector is much weaker, namely
\beq
   \frac{|F(c_{d_1})|}{|F(c_{d_2})|}
    \sim \frac{m_d}{m_s}\,\frac{1}{\lambda} \sim 0.3 \,, 
    \qquad 
   \frac{|F(c_{d_2})|}{|F(c_{d_3})|}
    \sim \frac{m_s}{m_b}\,\frac{1}{\lambda^2} \sim 0.4 \,.
\eeq
In this particular sector (and only there), it is thus a viable possibility to assume equal profiles $F(c_{d_i})$ and explain the required modest splittings in terms of $\ord(1)$ variations of the Yukawa couplings. While such a choice might seem {\it ad hoc}, it has the effect of strongly suppressing some particularly dangerous FCNC couplings in $K$--$\bar K$ mixing \cite{Santiago:2008vq}.

\subsection{Reparametrization Invariance}

The results for quark masses and mixing matrices obtained in the ZMA and discussed in the previous section are invariant under a set of  reparametrization transformations, which change the values of the 5D bulk mass parameters and Yukawa couplings. The first type of reparametrization invariance (RPI-1) refers to a simultaneous rescaling of the fermion profiles for $SU(2)_L$ doublet and singlet fields by opposite factors, while leaving the 5D Yukawa couplings invariant. Specifically, 
\beq\label{RPI1}
   F(c_{Q_i}) \to e^{-\xi}\,F(c_{Q_i}) \,, \qquad
   F(c_{q_i}) \to e^{+\xi}\,F(c_{q_i}) \,, \qquad
   \bm{Y}_q \to  \bm{Y}_q \,.
   \qquad \mbox{(RPI-1)}
\eeq
Provided the $c_i$ are all below $-1/2$, these transformations approximately correspond to the following shifts of the bulk mass parameters: $c_{Q_i}\to c_{Q_i}-\xi/L$ and $c_{q_i}\to c_{q_i}+\xi/L$. The second type of reparametrization invariance (RPI-2) refers to a simultaneous rescaling of all fermion profiles by a common factor, while the 5D Yukawa couplings are rescaled with an opposite factor. Specifically, 
\beq\label{RPI2}
   F(c_{Q_i}) \to \eta\,F(c_{Q_i}) \,, \qquad
   F(c_{q_i}) \to \eta\,F(c_{q_i}) \,, \qquad
   \bm{Y}_q\to \frac{1}{\eta^2}\,\bm{Y}_q \,.
   \qquad \mbox{(RPI-2)}
\eeq
The corresponding shifts of the bulk mass parameters are approximately universal and given by $c_{Q_i}\to c_{Q_i}+L^{-1}\ln\eta$ and $c_{q_i}\to c_{q_i}+L^{-1}\ln\eta$, provided again the $c_i$ are all below $-1/2$. A stronger form of this relation is
\beq\label{RPI2a}
   F(c_{Q_i}) \to \eta\,F(c_{Q_i}) \,, \qquad
   F(c_{q_i}) \to \eta_q\,F(c_{q_i}) \,, \qquad
   \bm{Y}_q\to \frac{1}{\eta\eta_q}\,\bm{Y}_q
   \qquad \mbox{(RPI-$2^{\,\prime}$)}
\eeq
with different parameters $\eta_u\ne\eta_d$. Of course, it is possible to combine (\ref{RPI1}) and (\ref{RPI2a}) in arbitrary ways.

While the masses and mixing angles are not affected by these transformations (in the ZMA), the shapes of the fermion bulk profiles change, and so do the results obtained for flavor-changing interactions (apart from the CKM matrix) derived in later sections.

\subsection{Basis Transformations}

In a recent paper \cite{Fitzpatrick:2007sa}, the authors expand the
5D bulk mass matrices under the assumption of minimal flavor violation. Their ansatz is\footnote{The universal term and $\ord(1)$ coefficients were omitted in \cite{Fitzpatrick:2007sa} for simplicity.}
\beq\label{eq:massexpansion}
\begin{split}
   \bm{M}_Q &= m_Q\,\bm{1}
    + m_Q' \left( r\,\bm{Y}_u \bm{Y}_u^\dagger 
    + \bm{Y}_d \bm{Y}_d^\dagger \right) + \dots \,, \\
   \bm{M}_q &= m_q\,\bm{1} + m_q'\,\bm{Y}_q^\dagger \bm{Y}_q
    + \dots \,, \qquad q=u,d \,,
\end{split}
\eeq
where the ellipses stand for subdominant higher-order terms. The quantity $r$ is a small parameter, which is adjusted to reproduce the observed mixings in the quark sector. 

It is instructive to work out to what ansatz for the Yukawa matrices the model (\ref{eq:massexpansion}) corresponds after we transform it to the bulk mass basis, \ie, the basis in which the bulk mass terms are diagonal in flavor space. Using unitary transformations on the 5D fields, it is easy to show that the 5D Yukawa matrix $\bm{Y}_u$ remains anarchic in this case, while $\bm{Y}_d$ becomes diagonal (with hierarchical diagonal elements) up to terms of $\ord(r)$. In the limit $r\to 0$, flavor violation in the down-quark sector is thus eliminated by hand, and flavor-changing effects including those described by the CKM matrix only arise from the up-quark sector. Obviously, to explain the hierarchical structure of $\bm{Y}_d$ in the bulk basis either requires an additional flavor symmetry or mere fine tuning. Unless a dynamical mechanism giving rise to relations of the form (\ref{eq:massexpansion}) can be found, one thus gives up the natural explanation of the quark mass and mixing hierarchies offered by the RS framework.

\section{Gauge and Higgs-Boson Interactions with Fermions}
\label{sec:gaugecouplings}

The results of the previous sections allow us to derive exact expressions for all Feynman rules in the RS model. Of particular importance are, of course, the couplings of gauge bosons to fermions. They will be discussed in Sections~\ref{subsec:gluons} to \ref{sec:chargedcurrents}. The couplings of the Higgs boson to fermions are the subject of Section~\ref{sec:Hcouplings}.

\subsection{Fermion Couplings to Gluons, Photons, and KK Excitations}
\label{subsec:gluons}

Consider, for example, the coupling of gluons to up-type quarks. Analogous relations hold for down-type quarks. The 4D Lagrangian contains the terms
\beq\label{gluoncouplings}
\begin{split}
   {\cal L}_{\rm 4D} &\ni \sum_{n_1, n_2, n_3} \bigg\{ \left[ 
     a_{n_2}^{(U)\dagger}\,\bm{I}_{n_1 n_2 n_3}^{C(Q)}\,a_{n_3}^{(U)}
     + a_{n_2}^{(u)\dagger}\,\bm{I}_{n_1 n_2 n_3}^{S(u)}\,
     a_{n_3}^{(u)} \right]
    \bar u_L^{(n_2)} g_s\Aslash^{(n_1)a}\,t^a\,u_L^{(n_3)} \\
   &\hspace{1.6cm}\mbox{}+ \left[ 
    a_{n_2}^{(u)\dagger}\,\bm{I}_{n_1 n_2 n_3}^{C(u)}\,a_{n_3}^{(u)} 
    + a_{n_2}^{(U)\dagger}\,\bm{I}_{n_1 n_2 n_3}^{S(Q)}\,
    a_{n_3}^{(U)} \right]
    \bar u_R^{(n_2)} g_s\Aslash^{(n_1)a}\,t^a\,u_R^{(n_3)} \bigg\} \,,
\end{split}
\eeq
where $g_s=g_{s,5}/\sqrt{2\pi r}$ is the 4D gauge coupling of QCD, and we have defined the overlap integrals
\beq\label{eq:overlap}
   \bm{I}_{n_1 n_2 n_3}^{C(A)} 
   = \int_{-\pi}^\pi\!d\phi\,\sqrt{2\pi}\,\chi_{n_1}(\phi)\,
   e^{\sigma(\phi)}\,\bm{C}_{n_2}^{(A)}(\phi)\,
   \bm{C}_{n_3}^{(A)}(\phi) \,,
   \qquad A = Q, u, d \,,
\eeq
and similarly for $\bm{I}_{n_1 n_2 n_3}^{S(A)}$ in terms of integrals over $\bm{S}_n^{(A)}$ profiles. These objects are $N\times N$ diagonal matrices in generation space. For the gluon zero mode, we obtain using (\ref{chi0flat}) and (\ref{orthonorm})
\beq
   \bm{I}_{0 n_2 n_3}^{C(A)} 
   = \delta_{n_2 n_3}\,\bm{1} + \Delta \bm{C}_{n_2 n_3}^{(A)} \,, 
    \qquad
   \bm{I}_{0 n_2 n_3}^{S(A)} 
   = \delta_{n_2 n_3}\,\bm{1} + \Delta \bm{S}_{n_2 n_3}^{(A)} \,.
\eeq
The relations (\ref{abrel}) and (\ref{magicCS}) then imply that these couplings are flavor diagonal and take the same form as in the SM, \ie,
\beq\label{eq115}
   {\cal L}_{\rm 4D}\ni \sum_n 
   \bar u^{(n)} g_s\Aslash^{(0)a}\,t^a\,u^{(n)} \,,
   \vspace{-1mm}
\eeq
with $u^{(n)}=u_L^{(n)}+u_R^{(n)}$. Note that the couplings of KK gluons are not flavor diagonal and must be worked out from the general relation (\ref{gluoncouplings}). It follows from the structure of the overlap integrals (\ref{eq:overlap}) that the effective coupling strength of KK gluons to heavy fermions, which live near the IR brane, is not $g_s$ but $\sqrt{L}\,g_s$ \cite{Davoudiasl:1999tf,Pomarol:1999ad}.

The couplings of photons and their KK excitations to fermions are obtained from the above relations by replacing the strong coupling constant and color matrices by the appropriate electromagnetic couplings, \ie, $g_s t^a\to e Q_f$, where $f$ can be any charged fermion species.

\subsection{Fermion Couplings to the $\bm{Z^0}$ Boson}

For the weak interactions, overlap integrals similar to those in (\ref{gluoncouplings}) arise, but in this case the gauge bosons couple differently to the $SU(2)_L$ doublet and singlet fermions, and so different combinations of overlap integrals contribute. Interestingly, in this case even the couplings of the light $W^\pm$ and $Z^0$ bosons  to fermions get corrected compared with their standard form. These non-universal corrections are not diagonal in flavor space, giving rise to tree-level FCNC couplings of the $Z^0$ boson. Indeed, these couplings are parametrically enhanced by a factor $L$ compared to the FCNC couplings of KK gauge bosons \cite{mytalks,inprep}. 

FCNC couplings of the $Z^0$ boson arise from two effects: first, the bulk profiles (\ref{chi0WZ}) of the lowest-lying massive gauge bosons are not flat, giving rise to non-trivial overlap integrals with the fermion profiles; secondly, as seen from (\ref{orthonorm}), the overlap integrals of profiles of only the $SU(2)_L$ doublet fermions do not obey exact orthonormality conditions. The latter effect has so far not been studied in great detail in the literature. In the perturbative approach, in which the Yukawa couplings are treated as a small correction, it would be interpreted as an $SU(2)_L$ singlet admixture in the wave functions of the $SU(2)_L$ doublet SM fermions due to mixing with their KK excitations \cite{Huber:2003tu, delAguila:2000kb, Hewett:2002fe} (see also \cite{delAguila:2000rc}). This effect is often neglected, since it is proportional to the masses of the light SM fermions. However, we will see that it is parametrically as well as numerically as important as the first one. For the first time we will present exact expressions for the corresponding corrections and compact analytical results valid at first non-trivial order in the ZMA.

Including corrections up to $\ord(m_Z^2/\Mkk^2)$, the $Z^0$-boson couplings to fermions and their KK excitations can be written in the form
\beq\label{Zff}
\begin{split}
   {\cal L}_{\rm 4D}
   &\ni \frac{g}{\cos\theta_W}
    \left[ 1 + \frac{m_Z^2}{4\Mkk^2} \left( 1 - \frac{1}{L} \right) 
    \right] Z_\mu^0 \\
   &\quad\times \sum_f \sum_{n_1,n_2}
    \left[ \big( g_L^f \big)_{n_1 n_2}\,
    \bar f_{L,n1}\gamma^\mu f_{L,n_2}
    + \big( g_R^f \big)_{n_1 n_2}\,\bar f_{R,n1}\gamma^\mu f_{R,n_2} 
    \right] ,
\end{split}
\eeq
where $g$ is the 4D weak $SU(2)_L$ gauge coupling defined in (\ref{g4def}). As in the SM, the weak mixing angle is given by
\beq\label{sWcWrela}
   \cos\theta_W = \frac{g}{\sqrt{g^2+g'^2}} \,, \qquad
   \sin\theta_W = \frac{g'}{\sqrt{g^2+g'^2}} \,.
\eeq
The factor in brackets in the first line of (\ref{Zff}) accounts for a universal correction due to the constant terms in the bulk profile (\ref{chi0WZ}). The left- and right-handed couplings $\bm{g}_{L,R}^f$ are infinite-dimensional matrices in the space of fermion modes, which can be parametrized as
\begin{eqnarray}\label{gLR}
\begin{split}
   \bm{g}_L^f 
   &= \left( T_3^f - \sin^2\theta_W\,Q_f \right)
    \left[ \bm{1} - \frac{m_Z^2}{2\Mkk^2}
    \left( L\,\bm{\Delta}_F - \bm{\Delta}'_F \right) \right] 
    - T_3^f \left[ \bm{\delta}_F - \frac{m_Z^2}{2\Mkk^2} 
    \left( L\,\bm{\varepsilon}_F - \bm{\varepsilon}'_F \right)
    \right] , \hspace{6mm} \\
   \bm{g}_R^f 
   &= - \sin^2\theta_W\,Q_f
    \left[ \bm{1} - \frac{m_Z^2}{2\Mkk^2}
    \left( L\,\bm{\Delta}_f - \bm{\Delta}'_f \right) \right] 
    + T_3^f \left[ \bm{\delta}_f - \frac{m_Z^2}{2\Mkk^2} 
    \left( L\,\bm{\varepsilon}_f - \bm{\varepsilon}'_f \right)
    \right] .
\end{split}
\end{eqnarray}
Here $T_3^f$ and $Q_f$ denote the weak isospin and electric charge (in units of $e$) of the fermion $f$.\footnote{The discussion of this section refers to quarks as well as leptons. In the remainder of the paper we will, however, focus on the quark sector of the theory.} 
On the matrices $\bm{\Delta}$, $\bm{\Delta}'$, $\bm{\delta}$, $\bm{\varepsilon}$, and $\bm{\varepsilon}'$, a subscript $F$ refers to a fermion from an $SU(2)_L$ doublet ($F=U,D$ in the quark sector, and $F=\nu,E$ in the lepton sector), while $f$ refers to a singlet ($f=u,d$ or $f=\nu_R,e$, respectively). The matrices $\bm{\Delta}^{(\prime)}$ and $\bm{\varepsilon}^{(\prime)}$ arise due to the effects of the $t$-dependent terms in the gauge-boson profile (\ref{chi0WZ}). The elements of the former are given by
\begin{eqnarray}\label{overlapints1}
\begin{split}
   \left( \Delta_F \right)_{mn}
   &= \frac{2\pi}{L\epsilon} \int_\epsilon^1\!dt\,t^2
    \left[ a_m^{(F)\dagger}\,\bm{C}_m^{(F)}(\phi)\, 
    \bm{C}_n^{(F)}(\phi)\,a_n^{(F)} 
    + a_m^{(f)\dagger}\,\bm{S}_m^{(f)}(\phi)\, 
    \bm{S}_n^{(f)}(\phi)\,a_n^{(f)} \right] , \\
   \left( \Delta_f \right)_{mn}
   &= \frac{2\pi}{L\epsilon} \int_\epsilon^1\!dt\,t^2
    \left[ a_m^{(f)\dagger}\,\bm{C}_m^{(f)}(\phi)\, 
    \bm{C}_n^{(f)}(\phi)\,a_n^{(f)} 
    + a_m^{(F)\dagger}\,\bm{S}_m^{(F)}(\phi)\, 
    \bm{S}_n^{(F)}(\phi)\,a_n^{(F)} \right] , \\
   \left( \Delta'_F \right)_{mn}
   &= \frac{2\pi}{L\epsilon} \int_\epsilon^1\!dt\,t^2
    \left( \frac12 - \ln t \right)
    \left[ a_m^{(F)\dagger}\,\bm{C}_m^{(F)}(\phi)\, 
    \bm{C}_n^{(F)}(\phi)\,a_n^{(F)} 
    + a_m^{(f)\dagger}\,\bm{S}_m^{(f)}(\phi)\, 
    \bm{S}_n^{(f)}(\phi)\,a_n^{(f)} \right] , \\
   \left( \Delta'_f \right)_{mn}
   &= \frac{2\pi}{L\epsilon} \int_\epsilon^1\!dt\,t^2
    \left( \frac12 - \ln t \right)
    \left[ a_m^{(f)\dagger}\,\bm{C}_m^{(f)}(\phi)\, 
    \bm{C}_n^{(f)}(\phi)\,a_n^{(f)} 
    + a_m^{(F)\dagger}\,\bm{S}_m^{(F)}(\phi)\, 
    \bm{S}_n^{(F)}(\phi)\,a_n^{(F)} \right] . \hspace{4mm}
\end{split}
\end{eqnarray}
Recall that on the profiles of $SU(2)_L$ doublet fermions no distinction is made between $Q=U,D$ and $L=\nu,E$. The matrices $\bm{\varepsilon}^{(\prime)}$ are given in terms of analogous expressions, in which the contributions from the even profiles $\bm{C}_n^{(A)}$ are omitted. Finally, the matrices $\bm{\delta}$ arise from the fact that the fermion profiles are not orthonormal on each other. They are defined as 
\beq\label{defdelta}
   \left( \delta_F \right)_{mn} 
   = a_m^{(f)\dagger} \left( \delta_{mn}
    + \Delta\bm{S}_{mn}^{(f)} \right) a_n^{(f)} \,, \qquad
   \left( \delta_f \right)_{mn}
   = a_m^{(F)\dagger} \left( \delta_{mn} 
    + \Delta\bm{S}_{mn}^{(F)} \right) a_n^{(F)} . 
\eeq

Since our main focus in this work is on tree-level processes, we are particularly interested in the couplings of the gauge bosons to the SM fermions. These are described by the upper-left $3\times 3$ blocks of the matrices $\bm{g}_{L,R}^f$ for $f=u$ (up-type quarks), $f=d$ (down-type quarks), $f=\nu$ (neutrinos), and $f=e$ (charged leptons). With the exception of the heavy top quark and (to a lesser extent) the left-handed component of the bottom quark, the profiles of the SM fermions are all localized near the UV brane, where $t=\ord(\epsilon)$. On the other hand, the presence of the weight factor $t^2$ in the overlap integrals in (\ref{overlapints1}) emphasizes the region near the IR brane, where $t=\ord(1)$. As a result, these overlap integrals are strongly suppressed for the light SM fermions, giving rise to a corresponding suppression of FCNC processes in the RS model \cite{Gherghetta:2000qt}. This mechanism is referred to as the RS-GIM suppression \cite{Agashe:2004ay,Agashe:2004cp,Agashe:2005hk}, even though its dynamical origin is rather different from the GIM mechanism of the SM. 

It is very instructive to have approximate formulae for the overlap integrals at hand, from which interesting information about the structure and magnitude of flavor-changing effects can be read off. To this end, we employ the ZMA for the fermion profiles given in (\ref{ZMA}), which corresponds to working to leading order in $v^2/\Mkk^2$. The even fermion profiles in (\ref{SMfermions}) then reduce to the familiar zero-mode profiles obtained in a theory without Yukawa couplings \cite{Grossman:1999ra}, while the odd profiles are suppressed by an extra factor of $x_n=m_n/\Mkk$ and thus can be neglected to first approximation. This yields
\beq\label{ZMA}
   \bm{C}_n^{(Q,u)}(\phi)\,a_n^{(U,u)}
   \to \sqrt{\frac{L\epsilon}{2\pi}}\,
   \mbox{diag}( F(c_{Q_i,u_i})\,
    t^{c_{Q_i,u_i}} )\,\hat a_n^{(U,u)} \,, \qquad
   \bm{S}_n^{(Q,u)}(\phi)\,a_n^{(U,u)} \to 0 \,,
\eeq
and similarly for the down-type quarks. It is then straightforward to find the expressions 
\beq\label{ZMA1}
\begin{split}
   \bm{\Delta}_F
   &\to \bm{U}_f^\dagger\,\,\mbox{diag} \left[ 
    \frac{F^2(c_{F_i})}{3+2c_{F_i}} \right] \bm{U}_f \,, \\
   \bm{\Delta}_f
   &\to \bm{W}_f^\dagger\,\,\mbox{diag} \left[ 
    \frac{F^2(c_{f_i})}{3+2c_{f_i}} \right] \bm{W}_f \,, \\
   \bm{\Delta}'_F
   &\to \bm{U}_f^\dagger\,\,\mbox{diag} \left[ 
    \frac{5+2c_{F_i}}{2(3+2c_{F_i})^2}\,F^2(c_{F_i}) \right] 
    \bm{U}_f \,, \\
   \bm{\Delta}'_f
   &\to \bm{W}_f^\dagger\,\,\mbox{diag} \left[ 
    \frac{5+2c_{f_i}}{2(3+2c_{f_i})^2}\,F^2(c_{f_i}) \right] 
    \bm{W}_f \,, \\
\end{split}
\eeq
where all quantities are $3\times 3$ matrices in generation space, and the diagonal matrices contain the elements shown in brackets. Note further that, to a good approximation, we have $\bm{\Delta}_A\approx\bm{\Delta}'_A$ for $A=F,f$, since all $c_i$ parameters are near $-1/2$. The approximate relations (\ref{ZMA1}) are not new. They have been presented first in \cite{Burdman:2002gr,Agashe:2003zs}. Corresponding results valid for a general warped metric can be found in \cite{Delgado:2007ne}.

The matrices $\bm{\varepsilon}_A^{(\prime)}$ vanish at leading order in the ZMA, meaning that they are suppressed by an extra factor of $v^2/\Mkk^2$. The same is true for the matrices $\bm{\delta}_A$, which are also given in terms of overlap integrals containing $\bm{S}_n^{(A)}$ profiles. However, since the contributions of these matrices in (\ref{gLR}) are not suppressed by $v^2/\Mkk^2$, it is necessary to go beyond the leading order in the ZMA. From (\ref{SMfermions}) we then obtain
\beq\label{ZMAforS}
   \bm{S}_n^{(Q,u)}(\phi)\,a_n^{(U,u)}
   \to \pm\sgn(\phi)\,\sqrt{\frac{L\epsilon}{2\pi}}\,x_n\,
   \mbox{diag}\left( F(c_{Q_i,u_i})\,
   \frac{t^{1+c_{Q_i,q_i}} - \epsilon^{1+2c_{Q_i,q_i}}\,
         t^{-c_{Q_i,q_i}}}%
        {1+2c_{Q_i,q_i}} \right) \hat a_n^{(U,u)} \,,
\eeq
and using these results gives
\beq\label{ZMA2}
\begin{split}
   \bm{\delta}_F
   &\to \bm{x}_f\,\bm{W}_f^\dagger\,\,
    \mbox{diag}\left[ \frac{1}{1-2c_{f_i}} 
    \left( \frac{1}{F^2(c_{f_i})} 
    - 1 + \frac{F^2(c_{f_i})}{3+2c_{f_i}} \right) \right] 
    \bm{W}_f\,\bm{x}_f \,, \\
   \bm{\delta}_f
   &\to \bm{x}_f\,\bm{U}_f^\dagger\,\,
    \mbox{diag}\left[ \frac{1}{1-2c_{F_i}} 
    \left( \frac{1}{F^2(c_{F_i})} 
    - 1 + \frac{F^2(c_{F_i})}{3+2c_{F_i}} \right) \right] 
    \bm{U}_f\,\bm{x}_f \,, 
\end{split}
\eeq
where $\bm{x}_f=\mbox{diag}(m_{f_1},m_{f_2},m_{f_3})/\Mkk$ is a diagonal matrix containing the masses of the SM fermions. To the best of our knowledge these important expressions have not been presented before. Approximate results for the singlet admixture in the wave functions of the $SU(2)_L$ doublet SM fermions due to mixing with their KK excitations can be found in \cite{Csaki:2008zd,Chang:2008zx,delAguila:2000kb,Hewett:2002fe}, but this effect has never been discussed systematically in connection with flavor-changing effects. Using the scaling relations derived in Section~\ref{sec:quark}, it is straightforward to find that to leading power in hierarchies
\beq
\begin{split}
   \left( \Delta_F \right)_{ij}
   &\sim \left( \Delta'_F \right)_{ij} 
    \sim F(c_{F_i})\,F(c_{F_j}) \,, \\
   \left( \Delta_f \right)_{ij}
   &\sim \left( \Delta'_f \right)_{ij}
    \sim F(c_{f_i})\,F(c_{f_j}) \,, \\
   \left( \delta_F \right)_{ij}
   &\sim \frac{m_{f_i} m_{f_j}}{\Mkk^2}\,
    \frac{1}{F(c_{f_i})\,F(c_{f_j})}
    \sim \frac{v^2\,Y_f^2}{\Mkk^2}\,F(c_{F_i})\,F(c_{F_j}) \,, \\
   \left( \delta_f \right)_{ij}
   &\sim \frac{m_{f_i} m_{f_j}}{\Mkk^2}\,
    \frac{1}{F(c_{F_i})\,F(c_{F_j})}
    \sim \frac{v^2\,Y_f^2}{\Mkk^2}\,F(c_{f_i})\,F(c_{f_j}) \,,
\end{split}
\eeq
where $Y_f$ represents an element (or a combination of elements) of the Yukawa matrix $\bm{Y}_f$. These relations make the RS-GIM suppression factors explicit. Note that the contributions of the $\bm{\delta}_A$ matrices in (\ref{gLR}), which have often been neglected in the literature, are of the same order as the effects proportional to the $\bm{\Delta}^{(\prime)}_A$ matrices. More accurate expressions for the mixing matrices, in which the relevant combinations of Yukawa matrices are included, can be found in Appendix~\ref{app:textures}.

It is interesting to study how the various flavor-changing couplings transform under the two types of reparametrizations, which leave the fermion masses and CKM mixing angles unchanged in the ZMA. From the relations (\ref{RPI1}), we obtain for a RPI-1 transformation
\beq\label{RPI1Delta}
\begin{aligned}
   \bm{\Delta}_F &\to e^{-2\xi}\,\bm{\Delta}_F \,,
   &\qquad
   \bm{\Delta}_f &\to e^{+2\xi}\,\bm{\Delta}_f \,, \\
   \bm{\delta}_F &\to e^{-2\xi}\,\bm{\delta}_F \,,
   &\qquad
   \bm{\delta}_f &\to e^{+2\xi}\,\bm{\delta}_f \,.
\end{aligned}
\eeq
This redistributes effects between the left- and right-handed sectors. For a RPI-$2^{\, \prime}$ transformation, on the other hand, we find from (\ref{RPI2a})
\beq\label{RPI2Delta}
\begin{aligned}
   \bm{\Delta}_F &\to \eta^2\,\bm{\Delta}_F \,,
   &\qquad
   \bm{\Delta}_f &\to \eta_q^2\,\bm{\Delta}_f \,, \\
   \bm{\delta}_F &\to \frac{1}{\eta_q^2}\,\bm{\delta}_F \,,
   &\qquad
   \bm{\delta}_f &\to \frac{1}{\eta^2}\,\bm{\delta}_f \,.
\end{aligned}
\eeq
This type of reparametrization acts in a similar way on the left- and right-handed couplings, while reshuffling effects between the two sources of flavor violations: those arising from the gauge-boson profiles ($\bm{\Delta}_A$) and those arising from the fermion profiles ($\bm{\delta}_A$). When one type of effect is enhanced, the other is reduced. 

\subsection{Fermion Couplings to $\bm{W^\pm}$ Bosons}
\label{sec:chargedcurrents}

The couplings of the charged weak bosons $W^\pm$ to SM fermions and their KK excitations can be derived in an analogous way. In this case, of course, flavor-changing effects are unsuppressed already in the SM. Focusing only on the quark sector,\footnote{An analogous discussion holds for leptons.} and including corrections up to $\ord(m_W^2/\Mkk^2)$, we find
\beq\label{Wff}
   {\cal L}_{\rm 4D}\ni \frac{g}{\sqrt 2}\,W_\mu^+ 
   \sum_{n_1,n_2} \left[ \big( V_L \big)_{n_1 n_2}\,
   \bar u_{L,n1}\gamma^\mu d_{L,n_2}
   + \big( V_R \big)_{n_1 n_2}\,\bar u_{R,n1}\gamma^\mu d_{R,n_2} 
   \right] + \mbox{h.c.,}
\eeq
where in terms of the overlap integrals defined in (\ref{eq:overlap}) one has
\beq\label{VLVRdef}
   \left( V_L \right)_{n_1 n_2} 
   = a_{n_1}^{(U)\dagger}\,
   \bm{I}_{0 n_1 n_2}^{C(Q)}\,a_{n_2}^{(D)} \,,
    \qquad
   \left( V_R \right)_{n_1 n_2} 
   = a_{n_1}^{(U)\dagger}\,
   \bm{I}_{0 n_1 n_2}^{S(Q)}\,a_{n_2}^{(D)} \,.
\eeq
The couplings to the SM fermions are encoded in the upper-left $3\times 3$ blocks of these matrices. In the ZMA, one finds
\beq
   \bm{V}_L\to \bm{U}_u^\dagger\,\bm{U}_d \equiv \bm{V}_{\rm CKM} \,,
    \qquad
   \bm{V}_R\to 0 \,.
\eeq

At $\ord(v^2/\Mkk^2)$ corrections to the two matrices arise, which give rise to a non-unitarity of the CKM matrix and to right-handed charged currents. The matrix $\bm{V}_R$ can be estimated using the ZMA as described in the previous section, \ie, by including the leading contribution to the $S_n^{(A)}$ profiles shown in (\ref{SMfermions}) but ignoring the $\delta_n$ terms. In this way we obtain
\beq\label{VRres}
   \bm{V}_R\to \bm{x}_u\,\bm{U}_u^\dagger\,\,
   \mbox{diag}\left[ \frac{1}{1-2c_{Q_i}} 
   \left( \frac{1}{F^2(c_{Q_i})} - 1 
   + \frac{F^2(c_{Q_i})}{3+2c_{Q_i}} \right) \right] \bm{U}_d\,
   \bm{x}_d \,.
\eeq
The scaling relations from Section~\ref{sec:quark} imply that
\beq
   \left( V_R \right)_{ij}
   \sim \frac{v^2}{\Mkk^2}\,F(c_{u_i})\,F(c_{d_j}) 
   \sim \frac{m_{u_i} m_{d_j}}{\Mkk^2}\,
   \frac{1}{F(c_{Q_i})\,F(c_{Q_j})} \,.
\eeq
The deviations of the CKM matrix from a unitary matrix are sensitive to the $t$-dependent terms in the gauge-boson profile (\ref{chi0WZ}), and to the deviations of the even fermion profiles and the eigenvectors $a_n^{(A)}$ from the expressions valid in the ZMA. These effects are best studied using the exact results for the fermion profiles and eigenvectors derived in Section~\ref{sec:fermionprofiles}. Will we present numerical estimates in Section~\ref{sec:63}. The non-unitarity of the CKM matrix has been analyzed previously in \cite{Huber:2003tu,Cheung:2007bu}. However, a thorough discussion of all relevant effects is missing in these articles.

\subsection{Fermion Couplings to the Higgs Boson}
\label{sec:Hcouplings}

Within the SM, the tree-level interactions of fermions with the Higgs-boson are flavor diagonal in the mass eigenstate basis of the fermion fields. Due to the mixing of fermion zero-modes and their KK excitations, this is not the case in the RS scenario \cite{Agashe:2006wa}. The flavor-changing $hf\bar f$ couplings can be expressed in terms of the Yukawa matrices $\bm{Y}_q$, the even profiles $\bm{C}_n^{(A)}$ evaluated at the IR brane, and certain combinations of the eigenvectors $a_n^{(A)}$. Working in unitary gauge, the relevant terms in the 4D Lagrangian describing the coupling of the Higgs boson to up-type quarks read
\beq\label{eq:hff}
   {\cal L}_{\rm 4D} \ni - \sum_{n_1,n_2}\,(g_h^u)_{n_1 n_2}\,
   h\,\bar u_L^{(n_1)}\,u_R^{(n_2)} + {\rm h.c.} \,,
\eeq
where we have defined the couplings 
\beq\label{eq:ghn1n2}
   (g_h^u)_{mn} 
   = \frac{\sqrt2\,\pi}{\epsilon L}\,
    a_{m}^{(U)\dagger}\,\bm{C}_{m}^{(Q)}(\pi)\,\bm{Y}_u\,
    \bm{C}_{n}^{(u)}(\pi)\,a_{n}^{(u)} 
   \equiv \delta_{mn}\,\frac{m_{u_m}}{v} 
   - (\Delta g_h^u)_{mn} \,.
\eeq
The first term in the last expression gives the SM couplings, while $(\Delta g_h^u)_{mn}$ contains the RS contributions. Analogous expressions describe the interactions of the Higgs boson and
down-type quarks.

The couplings $(\Delta g_h^q)_{mn}$ are not independent from the flavor matrices derived earlier in this work. Using relations (\ref{abrel}), (\ref{magicCS}), (\ref{eq56}), (\ref{bcIRn1}), and (\ref{defdelta}), it is not difficult to show that (for $q=u,d$ and $Q=U,D$)
\beq
   (\Delta g_h^q)_{mn} = \frac{m_{q_m}}{v}\,(\delta_q)_{mn}
   + (\delta_Q)_{mn}\,\frac{m_{q_n}}{v} \,.
\eeq
The flavor-changing Higgs-boson couplings are thus parametrized in terms of the matrices $\bm{\delta}_A$ entering also the $Z^0$-boson couplings to SM fermions. The expressions for these matrices valid in the ZMA have been given in (\ref{ZMA2}). Note that the extra suppression by factors of $m_q/v$ implies that FCNC processes mediated by Higgs-boson exchange are suppressed compared with those mediated by the exchange of a $Z^0$ boson.

\subsection{Parameter Counting in the Flavor Sector}

Assuming $N$ quark generations, one starts out with $N_Y=2\,(N^2,N^2)$ real moduli and CP-odd phases for the 5D Yukawa matrices $\bm{Y}_{u,d}^{({\rm 5D})}$, plus $N_c=3\,(N(N+1)/2,N(N-1)/2)$ parameters for the hermitian bulk mass matrices $\bm{c}_{Q,u,d}$. The Yukawa matrices break the global bulk flavor symmetry $G=U(N)_Q\times U(N)_u\times U(N)_d$ with $N_G=3\,(N(N-1)/2,N(N+1)/2)$ parameters to the subgroup $H=U(1)_B$ with $N_H=(0,1)$ parameters. One thus ends up with $N_{\rm phys}=N_Y+N_c-N_G+N_H=(N(2N+3),(N-1)(2N-1))$ physical parameters. 

For $N=3$ quark generations this leads to 27 moduli and ten phases \cite{Agashe:2004cp}. In the ZMA the real parameters consist of six quark masses, twelve mixing angles appearing in the matrices ${\bm U}_{u,d}$ and ${\bm W}_{u,d}$, and the nine eigenvalues of the zero-mode profiles ${\bm F}_{Q,u,d}$. Out of the ten phases, one is the phase of the CKM matrix, and nine new phases enter (in different combinations) the various matrices in (\ref{overlapints1}) and (\ref{defdelta}), which parametrize the flavor-changing interactions in the RS model. 

For $N=2$ quark generations one has 14 moduli and three phases, all of which can be chosen to reside in the new mixing matrices, since the two-generation CKM matrix can be made real by phase redefinitions. It is instructive to illustrate this point using the explicit (approximate) forms for the mixing matrices collected in Appendix~\ref{app:textures}. For $N=2$ these matrices depend on the four complex quantities $(Y_q)_{12}/(Y_q)_{22}$ and $(Y_q)_{21}/(Y_q)_{22}$ for $q=u,d$. Adopting a phase convention in which the two-generation Cabibbo matrix is real imposes the constraint $(Y_u)_{12}/(Y_u)_{22}=(Y_d)_{12}/(Y_d)_{22}$. 

We emphasize that the RS model allows for CP-violating effects which do not involve all three fermion generations, and which can be restricted to either the up- or down-quark sectors. In that sense CP violation is much less suppressed that in the SM. It would be interesting to work out the implications of this observation for baryogenesis.

As a final comment, let us mention that the new CP-odd phases encountered in the minimal RS model induce electric dipole moments of the electron and neutron at the one-loop level. To avoid the stringent experimental limits on the neutron electric dipole moment, the KK gauge-boson masses typically have to be larger than 10\,TeV \cite{Agashe:2004cp}. Several solutions \cite{Fitzpatrick:2007sa,Santiago:2008vq,Cheung:2007bu} to this ``CP problem'' have been proposed, all of which postpone the appearance of a non-vanishing neutron electric dipole moment to the two-loop level by reducing the number of CP-violating phases in the two-generation case to one.

\section{Phenomenological Applications}
\label{sec:pheno}

We now discuss some simple applications of the results derived so far, beginning with the definition and determination of important SM parameters. We then study the constraints imposed by the electroweak precision measurements encoded by the $S$, $T$, and $U$ parameters and the $Z^0b \bar b$ couplings. In this context, we point out that the minimal RS scenario provides a consistent framework in the case of a heavy Higgs boson for fairly low KK gauge-boson masses. Next we deal with the mixing matrices in the charged- and neutral-current sectors, paying special attention to the non-unitarity of the CKM matrix and the potential size of the anomalous right-handed $Wtb$ coupling. Then follows a discussion of the mass spectrum of KK fermions. We find that due to the near degeneracy of 5D fermionic bulk mass parameters, even small Yukawa couplings generically lead to large mixing effects among the fermion excitations of the same KK level. We close this section by analyzing FCNC top-quark decays. After determining the preferred chirality of the $Z^0tc(u)$ interactions we perform a detailed study of the $t\to c(u) Z^0$ and $t\to c(u) h$ branching ratios. Comments on correlations between flavor-diagonal and non-diagonal $Z^0$ vertices as well as anomalous $ht \bar t$ couplings round off our phenomenological survey.

\subsection{Modifications of SM Parameters}
\label{sec:mod}

Since the couplings of the photon and gluon zero modes to fermions are universal and have the same form as in the SM, the low-energy extractions of the fine-structure constant $\alpha$ (defined in the Thomson limit) and of the strong coupling $\alpha_s$ are, to very good approximation, unaffected from higher-dimensional effects in the RS model. The weak mixing angle $\theta_W$ is related to the 4D gauge couplings as usual by (\ref{sWcWrela}). It follows that
\beq\label{SU2gdef}
   g^2 = \frac{4\pi\alpha}{\sin^2\theta_W} \,.
\eeq 
The mixing angle defined in this way can be extracted from measurements of the left-right polarization asymmetries of light SM fermions on the $Z^0$ pole. In this case the RS-GIM mechanism ensures that the modifications of the $Z^0 f\bar f$ couplings are given to excellent approximation by the universal prefactor in (\ref{Zff}), which cancels in the standard formula for the asymmetries,
\beq\begin{split}
   A_f 
   &= \frac{\Gamma(Z^0\to f_L\bar f_R) - \Gamma(Z^0\to f_R\bar f_L)}%
           {\Gamma(Z^0\to f_L\bar f_R) + \Gamma(Z^0\to f_R\bar f_L)}
   = \frac{(g_L^f)^2 - (g_R^f)^2}{(g_L^f)^2 + (g_R^f)^2} \\
   &\approx 
    \frac{(1/2-|Q_f|\sin^2\theta_W)^2 - (Q_f\sin^2\theta_W)^2}%
         {(1/2-|Q_f|\sin^2\theta_W)^2 + (Q_f\sin^2\theta_W)^2}  \,.
\end{split}
\eeq

\begin{figure}[!t]
\begin{center}
\mbox{\includegraphics[height=3cm]{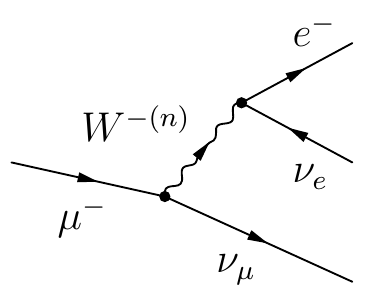}}
\vspace{-2mm}
\parbox{15.5cm}{\caption{\label{fig:muondecay}
Tree-level contributions to $\mu^-\to e^-\nu_\mu\bar\nu_e$ arising from the exchange of a $W^-$ boson and its KK excitations.}}
\end{center}
\end{figure}

We next turn to the determination of the Fermi constant $G_F$ from muon decay. As shown in Figure~\ref{fig:muondecay}, at tree level in the RS model this process is mediated by the exchange of the entire tower of the $W^-$ boson and its KK excitations. We have calculated the relevant sum over these intermediate states in (\ref{important1}). The terms proportional to $t^2$ or $t'^2$ in this relation give rise to non-universal effects suppressed by the fermion profiles near the IR brane, which to excellent approximation can be neglected for the light leptons involved in muon decay. This leaves a universal correction given by the constant terms in (\ref{important1}). We obtain
\beq\label{GFcor}
   \frac{G_F}{\sqrt2} = \frac{g^2}{8m_W^2} 
   \left[ 1 + \frac{m_W^2}{2\Mkk^2} \left( 1 - \frac{1}{2L} \right) 
   + \ord\left( \frac{m_W^4}{\Mkk^4} \right) \right] .
\eeq
The correction term receives a contribution $(1-1/L)$ from the ground-state $W^-$ boson and $1/(2L)$ from the tower of KK excitations.

\begin{figure}[!t]
\begin{center}
\mbox{\includegraphics[height=2.85in]{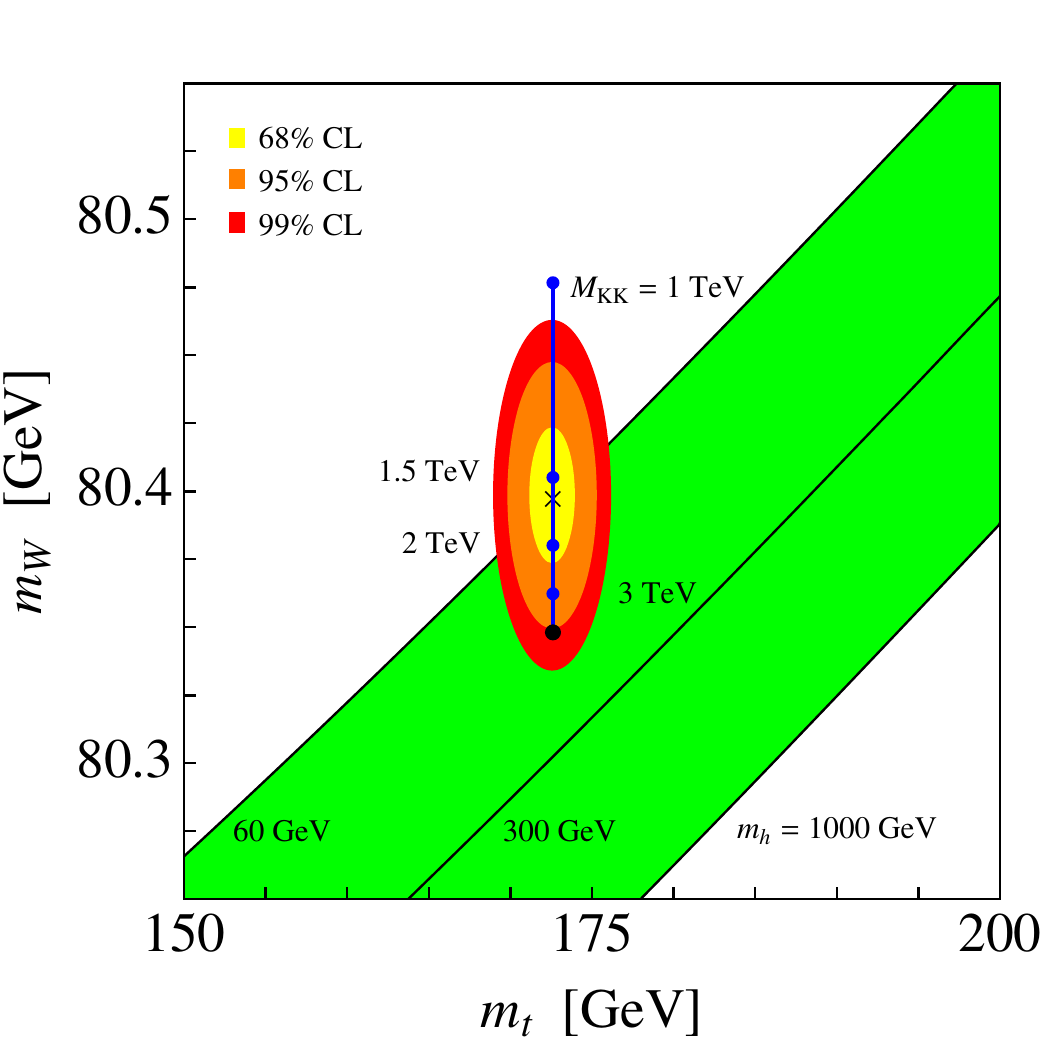}}
\vspace{-2mm}
\parbox{15.5cm}{\caption{\label{fig:MWplot}
Regions of 68\%, 95\%, and 99\% probability in the $m_t$--$m_W$ plane following from the direct measurements of $m_W$ and $m_t$ at LEP2 and the Tevatron. The black dot corresponds to the SM prediction based on the value of $G_F$ for our reference input values, while the green (medium gray) shaded band shows the SM expectation for values of the Higgs-boson mass $m_h\in [60,1000]$\,GeV. The blue (dark gray) line and points represent the RS prediction for $\Mkk\in [1,10]$\,TeV. See text for details.}}
\end{center}
\end{figure}

The SM relation between the weak mixing angle and the masses of the weak gauge bosons receives important corrections in the RS model. From (\ref{eq:m02}) it follows that
\beq\label{rho1}
   \frac{m_W^2}{m_Z^2} = \cos^2\theta_W \left[
   1 + \sin^2\theta_W\,\frac{m_Z^2}{2\Mkk^2} 
   \left( L - 1 + \frac{1}{2L} \right) 
   + \ord\left( \frac{m_Z^4}{\Mkk^4} \right) \right] .
\eeq
Notice that in this case the correction term is enhanced by a factor of $L$. It is customary to encode the corrections to the SM (tree-level) relation between $m_W/m_Z$ and $\cos\theta_W$ in a parameter $\varrho\equiv m_W^2/(m_Z^2\cos^2\theta_W)$. The prediction for this parameter obtained in the RS model depends on how the $W^\pm$-boson mass is determined. If $m_W$ is obtained from a direct measurement, then the $\varrho$ parameter is given by the expression in brackets on the right-hand side of the relation (\ref{rho1}). Alternatively, $m_W$ can be derived from precise measurements of $\alpha$, $G_F$, and $\sin^2\theta_W$ using the SM relation 
\beq\label{eq:mwindirect}
   (m_W^2)_{\rm indirect}\equiv 
   \frac{\pi\alpha}{\sqrt{2}G_F\sin^2\theta_W} \,.
\eeq
In the RS model we have
\beq\label{eq:RSMWrelation}
   (m_W)_{\rm indirect} 
   = m_W \left[ 1 - \frac{m_W^2}{4\Mkk^2} 
   \left( 1 - \frac{1}{2L} \right) 
   + \ord\left( \frac{m_W^4}{\Mkk^4} \right) \right] .
\eeq
When this is done, one obtains instead
\beq\label{varrho}
   \varrho = 1 + \frac{m_Z^2}{2\Mkk^2}
   \left( L\sin^2\theta_W - 1 + \frac{1}{2L} \right)  
   + \ord\left( \frac{m_Z^4}{\Mkk^4} \right) .
\eeq

The comparison between the indirect constraint and the direct measurements of $m_W$ and $m_t$ is shown in Figure~\ref{fig:MWplot}. The regions of 68\%, 95\%, and 99\% probability following from the direct measurements performed at LEP2 \cite{LEPEWWG:2005ema} and the Tevatron \cite{Yao:2006px} are shown by the ellipses. The shaded band shows the SM prediction based on (\ref{eq:mwindirect}) for Higgs-boson masses in the range $m_h\in [60,1000]$\,GeV. The central values and 68\% confidence level (CL) ranges of the direct and indirect $W^\pm$-boson mass determinations are given by
\beq\label{eq:mwmasses}
   m_W = (80.398\pm 0.025) \, \mbox{GeV} \,, \qquad 
   (m_W)_{\rm indirect} = (80.348\pm 0.015) \, \mbox{GeV} \,.
\eeq
The value of $(m_W)_{\rm indirect}$ has been derived using {\tt{ZFITTER}} \cite{Bardin:1999yd, Arbuzov:2005ma}. The quoted
central value and error correspond to the SM reference values for $\Delta\alpha^{(5)}_{\rm had}(m_Z)$, $\alpha_s(m_Z)$, $m_Z$, and $m_t$ collected in Appendix~\ref{app:masses}, as well as the Higgs-boson mass $m_h=150$\,GeV. The vertical line and points correspond to the RS prediction of $m_W$ following from (\ref{eq:RSMWrelation}) for $L=\ln(10^{16})$ and $\Mkk\in [1,10]$\,TeV. It is evident from the plot that the RS corrections allow to explain the shift of around 50\,MeV between the direct measurement and the indirect determination for KK scales slightly above $\Mkk=1.5$\,TeV. Motivated by this observation, we will take $\Mkk=1.5$\,TeV as one of the KK reference scales in our paper. Our second choice will be $\Mkk=3$\,TeV. Notice also that even for a heavy Higgs boson, agreement between $m_W$ and $(m_W)_{\rm indirect}$ at the 99\% CL can always be reached for KK scales above 1\,TeV. Taking for example $m_h=400$\,GeV ($m_h=1000$\,GeV) would allow for KK scales as low as 1.5\,TeV (1.0\,TeV).

\subsection{\boldmath$S$, $T$, and $U$ Parameters\unboldmath}

The $S$, $T$, and $U$ parameters measure deviations from the electroweak radiative corrections expected in the SM from new physics effects in universal electroweak corrections, \ie, those entering through vacuum polarization diagrams \cite{Peskin:1990zt,Peskin:1991sw,Golden:1990ig,Holdom:1990tc,Altarelli:1990zd,Altarelli:1993sz,Altarelli:1993bh}. These parameters are defined as shifts relative to a fixed set of SM values, so that $S$, $T$, and $U$ are identical to zero at that point. In the entire discussion of the electroweak precision measurements we focus on the leading contributions to the $S$, $T$, and $U$ parameters and thus restrict ourselves to tree-level effects in the framework of the minimal RS model. We further assume that SM loop effects are not significantly modified by the presence of KK excitations. A calculation of the complete one-loop contributions to $S$, $T$, and $U$, extending the work of \cite{Carena:2006bn, Carena:2007ua}, would be worthwhile, but is beyond the scope of this paper.

Using the relations (\ref{eq:m02}) and (\ref{chi0WZ}), it is a simple exercise to derive the $S$, $T$, and $U$ parameters in the minimal RS scenario at tree-level. In agreement with \cite{Carena:2003fx,Delgado:2007ne}, we find the positive corrections
\beq\label{eq:STURS}
   S = \frac{2\pi v^2}{\Mkk^2} \left( 1 - \frac{1}{L} \right) ,
    \qquad 
   T = \frac{\pi v^2}{2\cos^2\theta_W\,\Mkk^2}
    \left( L - \frac{1}{2L} \right) ,
\eeq
while $U$ vanishes. Placing the fermion fields in the bulk greatly softens the strong constraint from $S$ present in the RS models with bulk gauge fields and brane-localized fermions, for which $S, T\sim -L\pi v^2/\Mkk^2$ are both large and negative \cite{Csaki:2002gy}.

\begin{figure}[!t]
\begin{center} 
\hspace{-2mm}
\mbox{\includegraphics[height=2.85in]{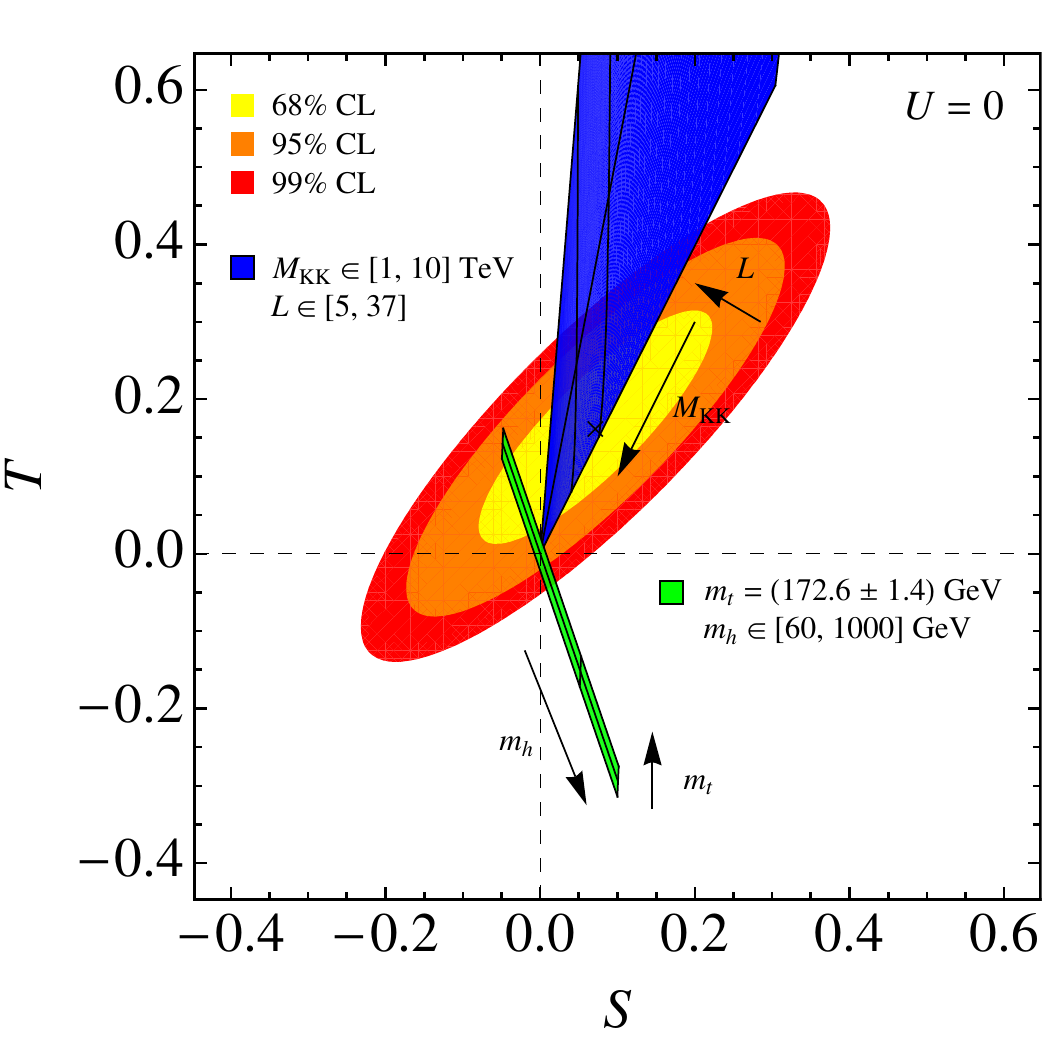}} 
\qquad 
\mbox{\includegraphics[height=2.85in]{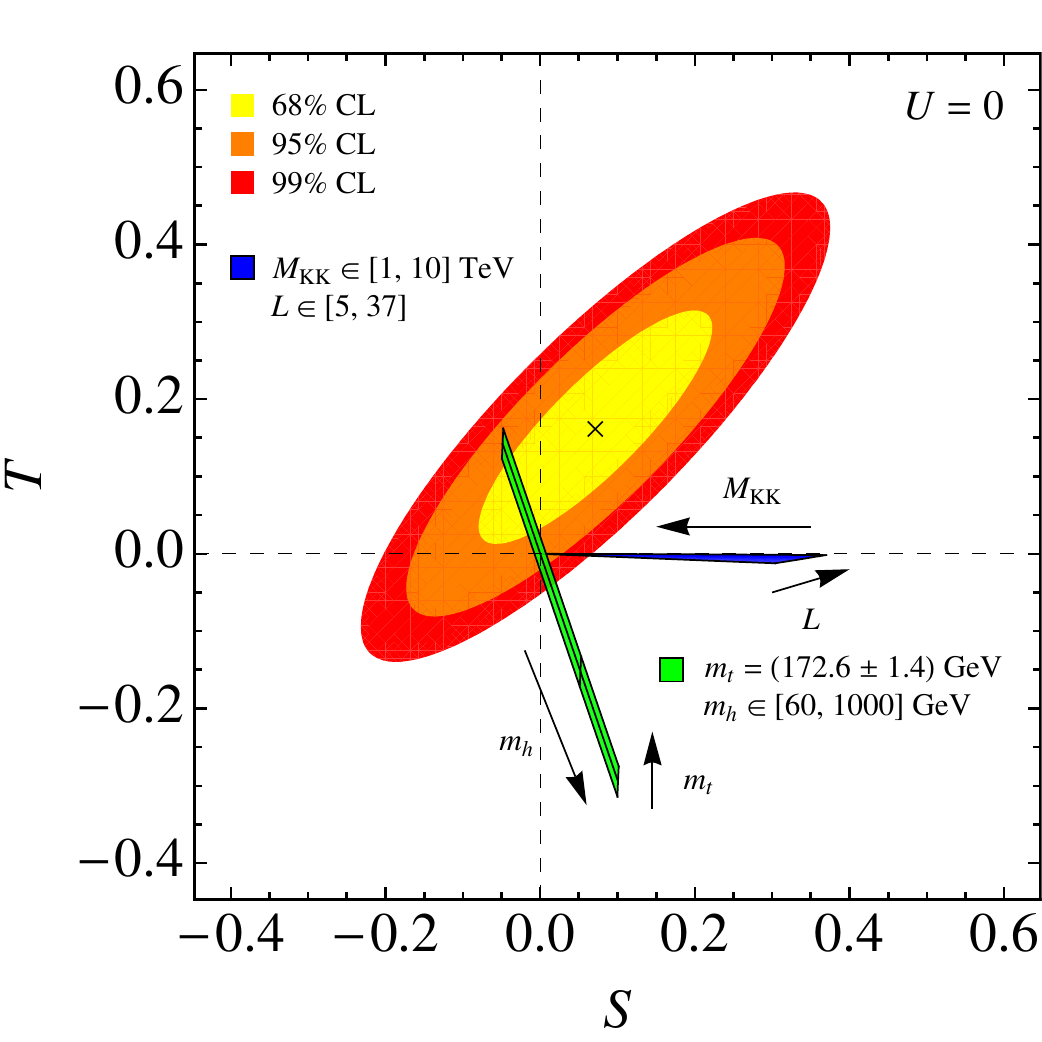}}
\vspace{-2mm}
\parbox{15.5cm}{\caption{\label{fig:STplots}
Regions of 68\%, 95\%, and 99\% probability in the $S$--$T$ plane. The green (medium gray) shaded stripes in both panels indicate the SM predictions for $m_t=(172.6\pm 1.4)$\,GeV and $m_h\in [60,1000]$\,GeV. The blue (dark gray) shaded area in the left/right panel represents the RS corrections without/with custodial protection for $\Mkk\in [1,10]$\,TeV and $L\in [5,37]$. See text for details.}}
\end{center}
\end{figure}

The experimental 68\% CL bounds on the $S$ and $T$ parameters, corrected to the present world average of the top-quark mass \cite{Group:2008nq}, and their correlation matrix are given by \cite{LEPEWWG:2005ema}
\beq\label{eq:STexp}
   \begin{array}{l}
    S = 0.07 \pm 0.10 \,, \\[0.5mm] 
    T = 0.16 \pm 0.10 \,,
   \end{array} \qquad 
   \rho = \begin{pmatrix} 1.00~ & 0.85 \\ 
                          0.85~ & 1.00 
          \end{pmatrix} .
\eeq
The regions of 68\%, 95\%, and 99\% probability in the $S$--$T$ plane are shown in the left panel of Figure~\ref{fig:STplots}. The light-shaded stripe indicates the SM predictions for different values of $m_h$ and $m_t$, while the dark-shaded area represents the RS corrections in (\ref{eq:STURS}) for different values of $\Mkk$ and $L$. In the global fit to the LEP and SLC measurements the parameter $U$ is set to zero, and the subtraction point corresponds to the SM reference values compiled in Appendix~\ref{app:masses} and $m_h=150$\,GeV. The $S$--$T$ error ellipses show that there are no large unexpected electroweak radiative corrections from physics beyond the SM, as the values of the $S$ and $T$ parameters are in agreement with zero.

Requiring that the RS contributions (\ref{eq:STURS}) satisfy the experimental constraint from $S$ and $T$ yields for the reference point (\ie, for the SM parameters given above)
\beq\label{eq:MKKboundslighthiggs}
   \Mkk > 4.0\,\mbox{TeV} \quad (99\%\;{\rm CL}) \,.
\eeq
Since the lightest KK modes have masses of around $2.45\Mkk$ in the
RS framework, the constraint from $T$ forces the first KK gauge-boson
excitations to be heavier than about 10\,TeV. Interestingly, one can show that the problem with $T$ persists in any 5D warped model with SM gauge symmetry in the bulk, in which the solution to the gauge hierarchy problem is associated with a moderately large volume factor \cite{Delgado:2007ne}. This strong constraint introduces a ``little hierarchy problem'' and calls for a cure.

\begin{figure}[!t]
\begin{center} 
\hspace{-2mm}
\mbox{\includegraphics[height=2.85in]{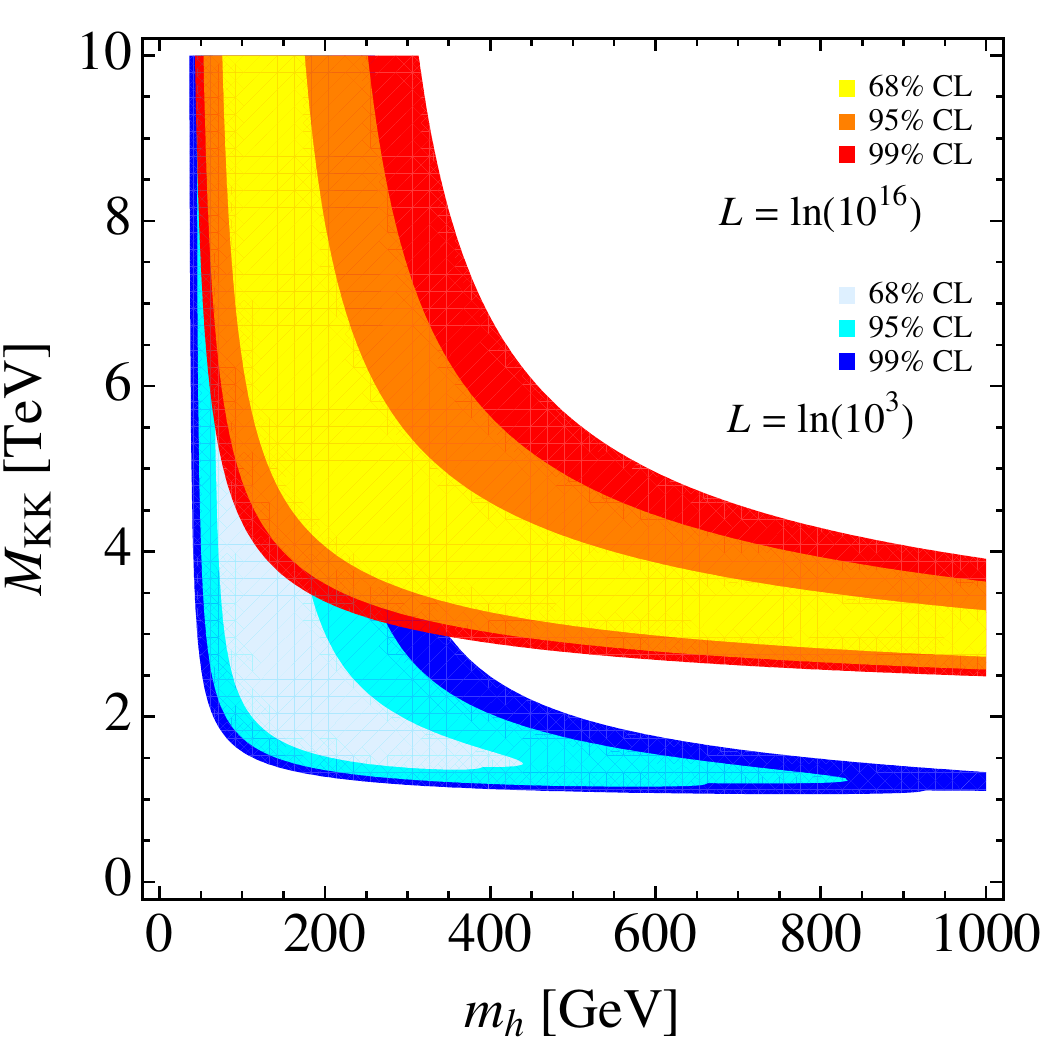}} 
\qquad 
\mbox{\includegraphics[height=2.85in]{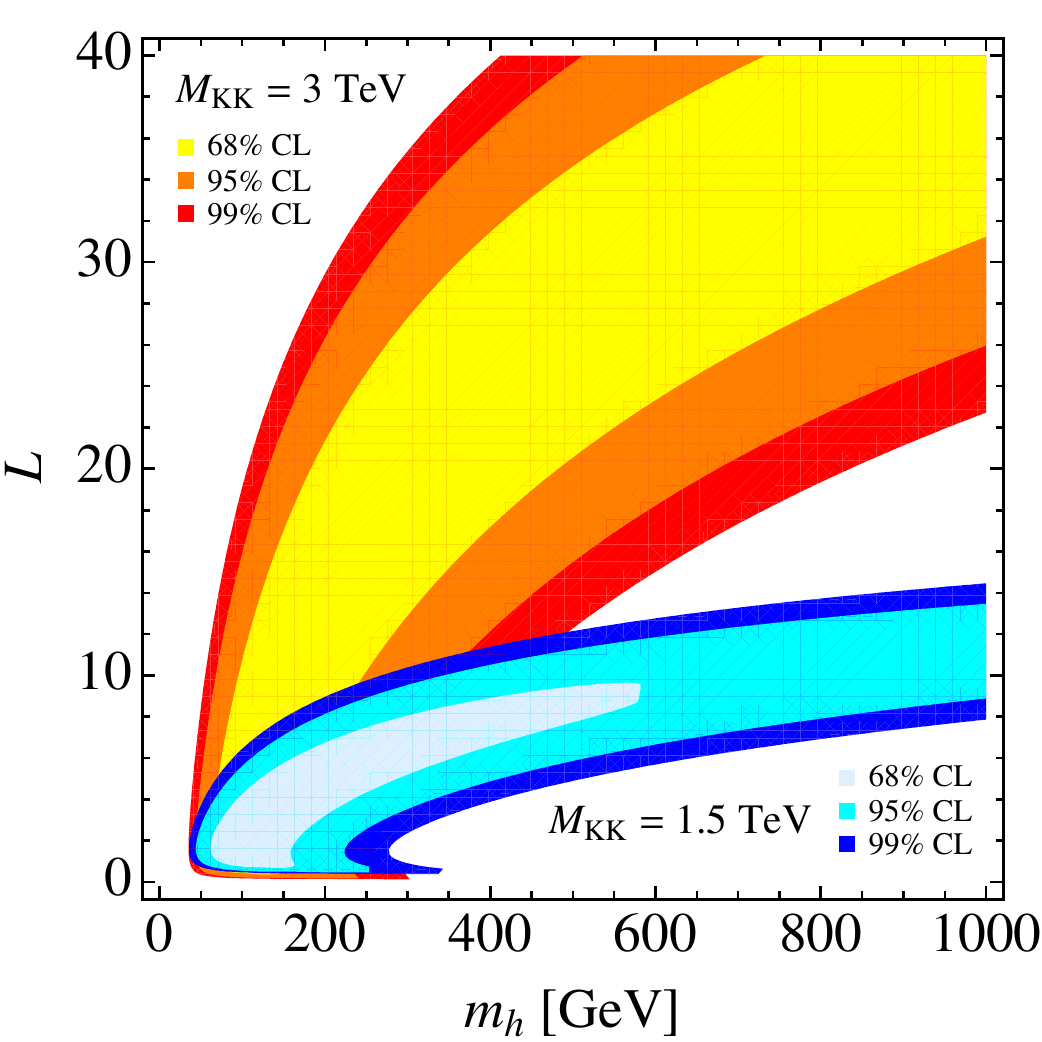}} 
\vspace{-2mm}
\parbox{15.5cm}{\caption{\label{fig:MHplots}
Regions of 68\%, 95\%, and 99\% probability in the $m_h$--$\Mkk$ (left panel) and $m_h$--$L$ (right panel) plane in the RS scenario without custodial protection. The upper (lower) area in the left and right panel corresponds to $L=\ln(10^{16})$ ($L=\ln(10^3)$) and $\Mkk=3$\,TeV ($\Mkk=1.5$\,TeV), respectively. See text for details.}}
\end{center}
\end{figure}

There are at least four possibilities that can mitigate the strong
constraint from $T$ in warped 5D scenarios. The first one consists in
canceling the large positive corrections (\ref{eq:STURS}) to the $T$
parameter by a large negative correction associated with a massive
Higgs boson \cite{Dobrescu:1997nm,Chivukula:1999az,Peskin:2001rw,Choudhury:2001hs,Barbieri:2006dq}. Keeping only the leading logarithmic loop effects in the SM, the shifts in $S$ and $T$ due to a Higgs-boson mass different from the reference value $m_h^{\rm ref}=150$\,GeV read \cite{Peskin:1991sw}
\beq\label{eq:SThiggs} 
   \Delta S = \frac{1}{6\pi}\,\ln\frac{m_h}{m_h^{\rm ref}} \,,
    \qquad 
   \Delta T = -\frac{3}{8\pi\cos^2\theta_W}\,
    \ln\frac{m_h}{m_h^{\rm ref}} \,,
\eeq
while $U$ remains unchanged. These relations imply that the bound on
the KK gauge-boson masses following from $T$ may be relaxed by taking
large values of $m_h$. A large Higgs-boson mass also induces a positive correction to the parameter $S$. However, since $\Delta T/\Delta S\approx -3$ the Higgs-boson mass can always be adjusted so that the bound on $\Mkk$ is alleviated. For example, taking $m_h=1$\,TeV,\footnote{Unitarity of longitudinal $W^\pm$-boson scattering requires the SM Higgs-boson mass to satisfy the bound $m_h\lesssim 870$\,GeV. In the RS model, unitarity can be satisfied for substantially higher values of $m_h$ \cite{Grzadkowski:2006nx}.}  
the shifts due to the Higgs boson amount to $\Delta S\approx 0.10$ and $\Delta T\approx -0.30$, and the lower bound (\ref{eq:MKKboundslighthiggs}) is relaxed to
\beq\label{eq:MKKboundsheavyhiggs}
   \Mkk > 2.6\,\mbox{TeV} \quad (99\% \; {\rm CL}) \,.
\eeq
This feature is illustrated by the upper sets of bands in the panels of Figure~\ref{fig:MHplots}, which show the regions of 68\%, 95\%, and 99\% probability in the $m_h$--$\Mkk$ and $m_h$--$L$ planes for $L=\ln(10^{16})$ (left plot) and $\Mkk=3$\,TeV (right plot). Since, as explained in Section~\ref{sec:before}, the Higgs-boson mass in warped models with the Higgs field residing on the IR-brane is naturally of order the KK rather than the electroweak scale, the possibility that the constraint from $T$ is tempered by the presence of a heavy Higgs boson should be considered seriously. In addition, a smaller value of the top-quark mass induces a small negative shift in $T$ without affecting $S$ and would thus help in relaxing the constraint from $T$ further. For example, the present total error on the top-quark mass of $\pm 1.4$\,GeV translates into a shift of $\Delta T\approx\pm 0.02$. A heavy Higgs boson in combination with a somewhat lighter top quark thus might allow for KK gauge-boson masses as low as 6\,TeV without invoking a symmetry that protects $T$ from unacceptably large positive corrections.

The second way to protect $T$ from vast corrections gives up on the
solution to the full gauge hierarchy problem by working in a volume-truncated RS background \cite{Davoudiasl:2008hx}. Taking for example $L=\ln(10^3)$ to address the hierarchy between the electroweak scale and $10^3$\,TeV, the RS bound (\ref{eq:MKKboundslighthiggs}) changes into the ``little RS'' limit
\beq\label{eq:MKKboundsLRS}
   \Mkk > 1.5\,\mbox{TeV} \quad (99\% \; {\rm CL}) \,.
\eeq
In the ``little RS'' scenario with $L=\ln(10^3)$, the lightest KK modes have masses of approximately $2.65\Mkk$, so that $T$ pushes the mass of the lowest-lying KK gauge-boson excitation to around 4\,TeV. This lower limit is weaker by about a factor 2.5 than the one obtained from (\ref{eq:MKKboundslighthiggs}). The bound (\ref{eq:MKKboundsLRS}) relaxes further for larger Higgs-boson mass. This feature is illustrated by the lower sets of bands in the panels of Figure~\ref{fig:MHplots}, which show the regions of 68\%, 95\%, and 99\% probability in the $m_h$--$\Mkk$ and $m_h$--$L$ planes for $L=\ln(10^3)$ (left plot) and $\Mkk=1.5$\,TeV (right plot). For example, using $m_h=500$\,GeV instead of the reference value of 150\,GeV turns the limit (\ref{eq:MKKboundsLRS}) into 1.1\,TeV.

A third possibility to obtain a consistent description of the experimental data, while allowing for masses of the first KK gauge-boson modes of the order of 5\,TeV, utilizes large brane-localized kinetic terms for the gauge fields \cite{Davoudiasl:2002ua,Carena:2002dz,Carena:2003fx}. Since such terms are needed as counterterms to cancel divergences appearing at the loop level \cite{Georgi:2000ks,Cheng:2002iz}, they are expected on general grounds to be present in any realistic orbifold theory. The bare contributions to the brane-localized kinetic terms encode the unknown UV physics at or above the cutoff scale. To retain the predictivity of the model, we simply assume that these bare contributions are small. As we concentrate on the leading contributions to the electroweak precision observables, we furthermore ignore possible effects of brane-localized kinetic terms appearing at the loop level. Even if these assumptions would be relaxed, the bound on the KK gauge-boson masses that derives from the constraints on the $S$ and $T$ parameters would remain anti-correlated with the mass of the Higgs boson. For example, Ref.~\cite{Carena:2003fx} finds that having a light KK spectrum would require a value of the Higgs-boson mass in the range of several hundred GeV.

The fourth cure for an excessive $T$ parameter is the custodial $SU(2)_R$ symmetry \cite{Agashe:2003zs}. In the context of 5D theories with warped background the custodial symmetry is promoted to a gauge symmetry. The hypercharge group is extended to $SU(2)_R\times U(1)_X$, such that the bulk gauge symmetry is $SU(3)_c\times SU(2)_L\times SU(2)_R\times U(1)_X$. The gauge group is then broken to $SU(3)_c\times SU(2)_L\times U(1)_Y$ on the UV brane. The tree-level $S$ and $T$ parameters in the RS scenario with this extended electroweak sector are given by \cite{Agashe:2003zs}
\beq\label{eq:STURScustodial}
   S = \frac{2\pi v^2}{\Mkk^2} \left( 1 - \frac{1}{L} \right) ,
    \qquad 
   T = - \frac{\pi v^2}{4\cos^2\theta_W\,\Mkk^2}\,\frac{1}{L} \,,
\eeq
while $U$ remains unaffected. Notice that the $S$ parameter is
given by the same expression as in the case without the custodial symmetry. For the $T$ parameter the custodial symmetry is at work, since $T$ turns out to be suppressed, rather than enhanced, by the logarithm $L$ of the warp factor. Thus the tree-level $T$ parameter ends up being tiny for the RS background with a large warp factor. The lower bound on the KK scale then follows from the experimental constraint on $S$. Numerically, one finds for the reference point
\beq\label{eq:MKKboundscustodial}
   \Mkk > 2.4 \,\mbox{TeV} \quad (99\% \; {\rm CL}) \,,
\eeq
which translates into a lower bound on the first KK gauge-boson mass
of about 6\,TeV. This limit is only marginally better that the bound (\ref{eq:MKKboundsheavyhiggs}) obtained in the original RS model without custodial symmetry, but with a heavy Higgs boson. Notice also that in the case of the RS scenario with extended electroweak sector, the existence of a heavy Higgs boson would spoil the global electroweak fit, since the corrections (\ref{eq:STURScustodial}) are generically too small to compensate for the negative shift $\Delta T$ due to a large value of $m_h$. This feature is illustrated in the right panel of Figure~\ref{fig:STplots}. On the other hand, it has been shown in \cite{Agashe:2003zs} that in this model fermionic loop corrections to $T$ are UV-finite and positive and allow to lower the KK scale for a sufficiently light Higgs boson. As a word of caution we should mention that, without a custodial bulk symmetry, the $T$ parameter could also be affected by unknown UV dynamics. It is thus possible that loop and cutoff effects might raise the bounds (\ref{eq:MKKboundsheavyhiggs}) and (\ref{eq:MKKboundsLRS}) to significantly higher values of the KK scale.

In \cite{Carena:2006bn, Carena:2007ua} one-loop corrections to the $T$
parameter arising from KK excitations that couple via the top Yukawa
coupling have been studied in the context of the custodially symmetric
$SU(2)_L\times SU(2)_R$ model. The interesting observation made by these authors is that the presence of quark bi-fundamentals under $SU(2)_L\times SU(2)_R$, introduced to protect the left-handed $Z^0 b\bar b$ coupling \cite{Agashe:2006at}, typically induces a negative shift in the $T$ parameter at the one-loop level, although a positive correction is possible in some regions of parameter space. It is apparent from the right plot in Figure~\ref{fig:STplots} that obtaining a negative contribution to $T$, together with a positive value of the $S$ parameter, would be rather problematic. In the RS scenario without $SU(2)_R$ one-loop corrections to $T$ have not been calculated. The precise impact of higher-order corrections on (\ref{eq:MKKboundslighthiggs}), (\ref{eq:MKKboundsheavyhiggs}), and (\ref{eq:MKKboundsLRS}) therefore remains unknown. 

\subsection{Flavor-Changing Interactions}
\label{sec:63}

We now turn our attention to the sources of flavor violation in the couplings of the weak gauge bosons to fermions. To this end we need to specify a large number of model parameters, namely the bulk mass parameters of the 5D fermion fields and the 5D Yukawa matrices. Their choice is restricted by the fact that one should reproduce the known values of the quark masses and CKM matrix elements within errors, but this information still leaves some freedom.

It will be useful for many considerations to have a default set of input parameters, which is consistent with all experimental constraints concerning the quark masses and CKM parameters. We use $\Mkk=1.5$\,TeV as the default KK scale, corresponding to masses of about 3.7\,TeV for the first KK excitations of the gauge bosons. We sometimes use twice that value, $\Mkk=3$\,TeV, as a second reference point. Unless otherwise noted, we set $L=\ln (10^{16})$ for the logarithm of the warp factor. Our default set of bulk mass parameters is\footnote{Here and below, results are given to at least three significant digits.}
\beq\label{eq:cparameter}
\begin{aligned}
   c_{Q_1} &= -0.579 \,, \qquad
    c_{Q_2} &= -0.517 \,, \qquad
    c_{Q_3} &= -0.473 \,, \\
   c_{u_1} &= -0.742 \,, \qquad
    c_{u_2} &= -0.558 \,, \qquad
    c_{u_3} &= +0.339 \,, \\
   c_{d_1} &= -0.711 \,, \qquad
    c_{d_2} &= -0.666 \,, \qquad
    c_{d_3} &= -0.553 \,.
\end{aligned}
\eeq
For the 4D Yukawa matrices in the normalization (\ref{Y4Ddef}) we take 
\begin{eqnarray}\label{eq:yukawas}
   \bm{Y}_u &=& \left(
    \begin{array}{rrr}   
      1.077-0.820 \hspace{0.5mm} i & 
      ~~0.102+1.177 \hspace{0.5mm} i~~ &  
      0.788-0.435 \hspace{0.5mm} i \\
      1.333-1.132 \hspace{0.5mm} i & 
      ~~0.084-0.755 \hspace{0.5mm} i~~ & 
      -0.394-0.628 \hspace{0.5mm} i \\
      -0.729+0.081 \hspace{0.5mm} i & 
      ~~0.682+0.276 \hspace{0.5mm} i~~ &  
      0.753-2.360 \hspace{0.5mm} i
    \end{array}
  \right) , \nonumber\\
  \bm{Y}_d &=& \left(
    \begin{array}{rrr}
      1.425-1.784 \hspace{0.5mm} i & 
      ~~0.358+0.464 \hspace{0.5mm} i~~ & 
      -0.301-0.128 \hspace{0.5mm} i \\
      0.055-0.602 \hspace{0.5mm} i & 
      ~-1.054-1.575 \hspace{0.5mm} i~~ & 
      -0.043+0.411 \hspace{0.5mm} i \\
      0.968+1.826 \hspace{0.5mm} i & 
      ~~0.697+0.277 \hspace{0.5mm} i~~ & 
      -0.853+0.646 \hspace{0.5mm} i
    \end{array}
    \right) .
\end{eqnarray}
These matrices have been obtained by random choice, subject to the constraints that the absolute value of each entry is between $1/3$ and $3$, and that the Wolfenstein parameters $\bar\rho$ and $\bar\eta$ in (\ref{eq:wolfenstein}) agree with experiment within errors. The bulk mass parameters $c_{A_i}$ have then be determined using the scaling relations described in Section~\ref{sec:quark}.

With these parameters, the exact values for the quark masses obtained from the eigenvalue equation (\ref{fermeigenvals}) are
\beq\label{eq:massesexact}
\begin{split}
   m_u &= 1.45\,\mbox{MeV} \,, \qquad
    m_c = 563\,\mbox{MeV} \,, \qquad
    \,\,m_t = 136\,\mbox{GeV} \,, \\
   m_d &= 2.98\,\mbox{MeV} \,, \qquad
    m_s = 49.6\,\mbox{MeV} \,, \qquad
    m_b = 2.23\,\mbox{GeV} \,.
\end{split}
\eeq
These values should be interpreted as the running quark masses in the $\overline{\rm MS}$ scheme evaluated at the scale $\mu=\Mkk$. After renormalization-group evolution, they agree with the mass values derived from low-energy measurements (see Appendix~\ref{app:masses}). Note that essentially the same values are obtained using relations (\ref{eq:singular}) and (\ref{eq:lambdaud}) valid in the ZMA. The only exception is the top-quark mass, which comes out 5.5\,GeV too large in this approximation.

In parts of our analysis below, specifically in Sections~\ref{sec:631}, \ref{sec:Zbb}, \ref{sec:tcZ}, and \ref{sec:tch}, we will perform a scan over the parameter space of the RS model. Whenever this is done, we generate 3000 randomly chosen points using uniform initial distributions for the input parameters. The parameter ranges are $\Mkk\in [1,10]$\,TeV for the KK scale and $|(Y_{u,d})_{ij}|\in [1/3,3]$ for the Yukawa couplings. We further require in a somewhat {\it ad hoc\/} way that $|F(c_{A_i})|\le\sqrt{2}$, which means $c_{A_i}<1/2$. The bulk mass parameters are then chosen such that all points reproduce the quark masses and CKM parameters with a global $\chi^2/{\rm dof}$ of better than 11.5/10 (corresponding to 68\% CL). This large set of points provides a reasonable range of predictions that can be obtained for a given observable. Only for a subset of the 3000 scatter points the RS predictions for the $Z^0 b\bar b$ couplings are consistent at the 99\% CL with experimental measurements. (A detailed discussion of the $Z^0\to b\bar b$ ``pseudo observables'' is deferred to Section~\ref{sec:Zbb}.) We will generally show two plots for a given observable, one made with all 3000 parameter points, and one restricted to those points that are compatible with the measured $Z^0 b\bar b$ couplings. In the latter case, we add further scatter points that pass all constraints, so that the number of shown parameter points also amounts to 3000 in total.

\subsubsection{Mixing Matrices in the Charged-Current Sector}
\label{sec:631}

We next consider the results for the flavor-mixing matrices, starting with the CKM matrix. In all cases we adjust the phases of the SM quark fields according to the standard CKM phase convention \cite{Yao:2006px}, which is defined by the requirements that the matrix elements $V_{ud}$, $V_{us}$, $V_{cb}$, $V_{tb}$ are real, and that
\beq
   \mbox{Im}\,V_{cs} = \frac{V_{us} V_{cb}}{V_{ud}^2 + V_{us}^2}\,
   \mbox{Im}\,V_{ub} \,.
\eeq
These five conditions fix the phase differences between the six quark fields uniquely. 

With our default parameters, the exact expression for the left-handed charged-current mixing matrix $\bm{V}_L$ defined in (\ref{VLVRdef}) is
\beq\label{eq:CKMexact}
   \bm{V}_L = \left(
    \begin{array}{ccc}
     0.975 & 0.225 & 0.00322\,e^{-i\,66.6^\circ} \\
    -0.225\,e^{i\,0.02^\circ} & 0.973\,e^{-i\,0.002^\circ} & 0.0417 \\
     0.00880\,e^{-i\,20.5^\circ} & ~~ -0.0400\,e^{i\,1.5^\circ} ~~
      & 0.996
   \end{array}
   \right) .
\eeq 
For the purposes of this work, we define the CKM matrix as $\bm{V}_{\rm CKM}\equiv\bm{V}_L$, \ie, as the matrix governing the charged-current couplings of the lightest $W^\pm$ bosons in units of $g/\sqrt2$, see (\ref{Wff}).\footnote{An alternative, more physical definition is based on the effective four-fermion interactions induced by the exchange of the entire tower of $W^\pm$ bosons and their KK excitations \cite{inprep}.} From this matrix we extract for the Wolfenstein parameters the values
\beq\label{eq:exactwolf}
   \lambda = 0.225 \,, \qquad
   A = 0.822 \,, \qquad
   \bar\rho = 0.132 \,, \qquad
   \bar\eta = 0.306 \,.
\eeq
They are all in good agreement with experiment (see Appendix~\ref{app:masses}). Unlike the CKM matrix in the SM, the left-handed mixing matrix in the RS model is not a unitary matrix. As two measures of this effect we consider the deviation of the sum of the squares of the matrix elements in the first row from 1, and the lack of closure of the unitarity triangle \cite{Bjorken:1988ni}. A much more detailed analysis of new physics effects on determinations of CKM and unitarity-triangle parameters will be presented in \cite{inprep}. With our default parameters, we obtain
\beq\label{eq:unitarityviolation}
\begin{split}
   1 - \left( |V_{ud}|^2 + |V_{us}|^2 + |V_{ub}|^2 \right)
   &= - 0.00048 \,, \\
   1 + \frac{V_{ud} V_{ub}^*}{V_{cd} V_{cb}^*}
    + \frac{V_{td} V_{tb}^*}{V_{cd} V_{cb}^*} 
   &= - 0.0068 + 0.0209 \, i \,.
\end{split}
\eeq
The first number should be compared with the one following from a global fit performed in \cite{Antonelli:2008jg}, which yields
\beq\label{eq:flavianet}
   1 - \left( |V_{ud}|^2 + |V_{us}|^2 + |V_{ub}|^2 \right)
   = 0.00022\pm 0.00051_{V_{ud}}\pm 0.00041_{V_{us}} \,.
\eeq
The unitarity violation predicted in the RS model with our default parameters is at the same level as the uncertainties in (\ref{eq:flavianet}). An improvement of the determination of both $V_{ud}$ and $V_{us}$ might thus allow to detect the non-unitarity of the CKM matrix induced in the RS framework. The second prediction in (\ref{eq:unitarityviolation}) can be compared with the current precision on the values of the Wolfenstein parameters, which are $\bar\rho=0.141\,_{-0.017}^{+0.029}$ and $\bar\eta=0.343\pm 0.016$ according to the CKMfitter Group \cite{Charles:2004jd}, and $\bar\rho=0.147\pm 0.029$ and $\bar\eta=0.342\pm 0.016$ according to the UT$_{fit}$ Collaboration \cite{Bona:2006ah}. Again, the predicted unitarity violation is of the same order as the present errors.

Our default result for the right-handed mixing matrix $\bm{V}_R$ in (\ref{VLVRdef}) is given in Appendix~\ref{app:results}. Its entries are extremely small compared to the corresponding entries of the CKM matrix due to the suppression proportional to two powers of light quark masses, which is evident from (\ref{VRres}). The only two elements exceeding the level of $10^{-5}$ in magnitude are $(V_R)_{32}=-1.17\cdot 10^{-5}\,e^{-i\,33.4^\circ}$ and $(V_R)_{33}=1.20\cdot 10^{-3}$. However, these couplings are still too small to give rise to any observable effect. Consider the right-handed $Wtb$ coupling as an example. To leading power in hierarchies only the profile $F(c_{Q_3})$ and certain combinations of $c_{Q_1}$, $c_{Q_2}$, and the elements of the up- and down-type Yukawa matrices enter in the formula for this quantity. Explicitly, we find in the ZMA
\beq\label{eq:VRexplicitRS}
\begin{split}
   v_R\equiv (V_R)_{33}
   &\to \frac{m_b m_t}{\Mkk^2}\,\Bigg[ \frac{1}{1-2c_{Q_3}} 
    \left( \frac{1}{F^2(c_{Q_3})} - 1 
    + \frac{F^2(c_{Q_3})}{3+2c_{Q_3}} \right) \\
   &\quad\mbox{}+ \sum_{i=1,2} 
    \frac{(Y_u)_{i3}^\ast (Y_d)_{i3}}{(Y_u)_{33}^\ast (Y_d)_{33}}\, 
    \frac{1}{1-2c_{Q_i}}\,\frac{1}{F^2(c_{Q_3})} \Bigg] \,.
\end{split}
\eeq
Interestingly, the inclusive $B\to X_s\gamma$ decay provides a stringent bound on the potential size of the anomalous $W t b$ coupling proportional to $v_R$. Due to the chiral $m_t/m_b$ enhancement present in $B\to X_s\gamma$ \cite{Cho:1993zb,Fujikawa:1993zu}, the bound on this specific effective coupling turns out to be more than an order of magnitude stronger than the limit expected from future measurements of top-quark production and decay at the LHC
\cite{AguilarSaavedra:2007rs}. A recent careful analysis arrives at \cite{Grzadkowski:2008mf}
\beq\label{eq:vRbound} 
   v_R\in [-0.0007, 0.0025] \quad \mbox{(95\% CL)} \,. 
\eeq

\begin{figure}[!t]
\begin{center}
\hspace{-2mm}
\mbox{\includegraphics[height=2.85in]{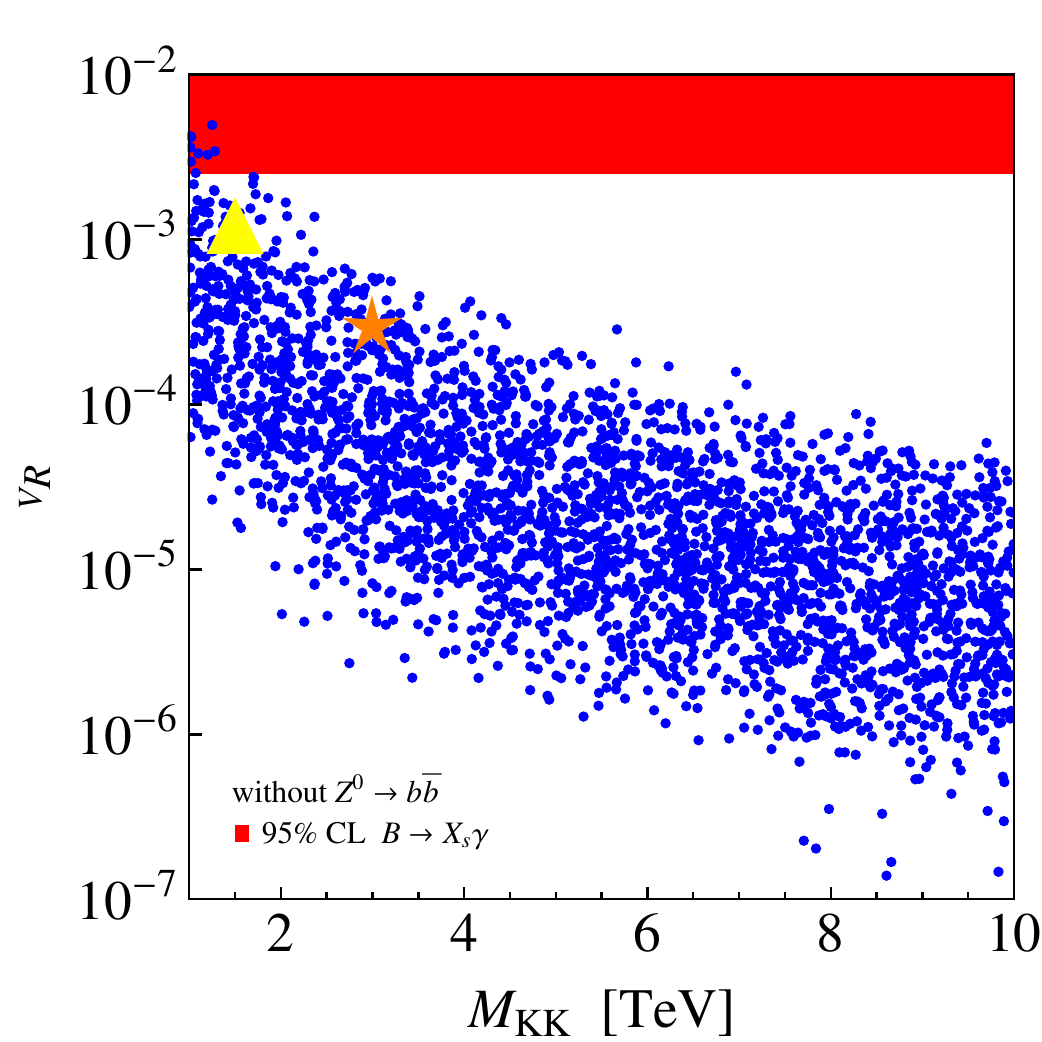}}
\qquad 
\mbox{\includegraphics[height=2.85in]{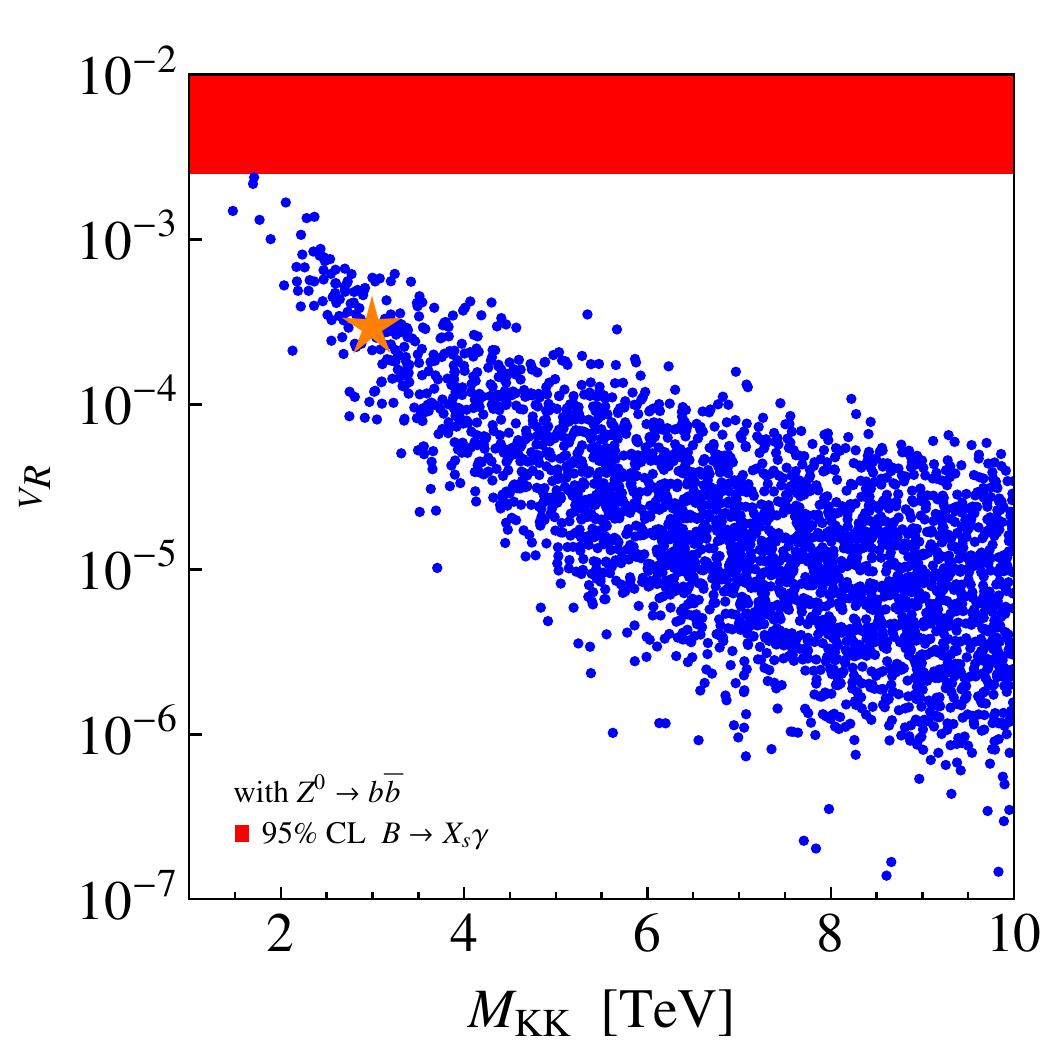}}
\vspace{-2mm}
\parbox{15.5cm}{\caption{\label{fig:vRplots}
Coefficient $v_R$ as a function of $\Mkk$. The red (dark gray) band represents the parameter region disfavored by $B\to X_s\gamma$ at the 95\% CL. All scatter points reproduce the correct quark masses and CKM parameters within errors. In the left (right) plot, points that violate the constraints from the $Z^0\to b\bar b$ ``pseudo observables'' are included (excluded). The values of $v_R$ for our RS reference points with $\Mkk=1.5$\,TeV and $\Mkk=3$\,TeV are indicated by the yellow (light gray) triangle and the orange (medium gray) stars. See text for details.}}
\end{center}
\end{figure}

In Figure~\ref{fig:vRplots} we show the RS predictions for $v_R$ as a function of $\Mkk$ for 3000 randomly generated parameter points, see Section~\ref{sec:63}. The range (\ref{eq:vRbound}) disfavored by $B\to X_s\gamma$ is indicated by the horizontal band. In the right plot we have excluded scatter points that do not lead to an agreement with the measured $Z^0 b\bar b$ couplings at the 99\% CL. The plots illustrate that although values of $v_R$ at the level of $10^{-3}$ are possible for light KK masses, the RS corrections to $v_R$ are generically too small to lead to an observable effect in $B\to X_s\gamma$. The possibility that an anomalous right-handed $W t b$ coupling changes the top-quark production and decay in a way that would be detectable at the LHC thus seems highly unlikely in the minimal RS framework. 

Note that the correlation between the $Wtb$ and $Z^0 b \bar b$ couplings is in general different in warped models with extended electroweak sector, because the custodial symmetry cannot simultaneously protect the $Wtb$ and $Z^0 b\bar b$ couplings \cite{Agashe:2006at}. We therefore expect that, depending on the exact realization of the fermionic sector, the corrections to the $Wtb$ couplings could be more pronounced, which might allow for an indirect detection of the anomalous right-handed $Wtb$ coupling through a deviation in the $B\to X_s\gamma$ branching ratio.

\subsubsection{Mixing Matrices in the Neutral-Current Sector}
\label{sec:632}

The largest flavor-changing effects in the neutral-current sector arise from the $Z^0$-boson couplings to SM fermions. They are encoded in the left- and right-handed couplings $\bm{g}_{L,R}^f$ given in (\ref{gLR}). The dominant effects arise from the matrices $\bm{\Delta}_{Q,q}$ (whose contributions are enhanced by a factor $L$) and $\bm{\delta}_{Q,q}$. We now list the results for these matrices obtained with our default parameters. Note that the $\bm{\Delta}_A$ terms are multiplied by a factor $m_Z^2 L/(2\Mkk^2)\approx 0.07\approx 1/15$ relative to the $\bm{\delta}_A$ terms. This needs to be taken into account when comparing the numerical values of the various matrices.\\

\noindent
Left-handed down-quark sector:
\beq
\begin{split}
   \bm{\Delta}_D &= 10^{-3} \left(
    \begin{array}{ccc}
     0.290 & 0.235\,e^{-i\,3.6^\circ} & -0.788\,e^{i\,71.8^\circ} \\
     0.235\,e^{i\,3.6^\circ} & 7.864 & -4.305\,e^{-i\,8.6^\circ} \\
     -0.788\,e^{-i\,71.8^\circ} & ~~ -4.305\,e^{i\,8.6^\circ} ~~
      & 30.93
    \end{array} \right) , \\
   \bm{\delta}_D &= 10^{-4} \left(
    \begin{array}{ccc}
     0.214 & 0.407\,e^{i\,60.2^\circ} & -1.736\,e^{i\,82.2^\circ} \\
     0.407\,e^{-i\,60.2^\circ} & 3.778 & -5.309\,e^{i\,43.0^\circ} \\
     -1.736\,e^{-i\,82.2^\circ} & ~~ -5.309\,e^{-i\,43.0^\circ} ~~
      & 20.91
    \end{array} \right) .
\end{split}
\eeq
Right-handed down-quark sector:
\beq
\begin{split}
   \bm{\Delta}_d &= 10^{-4} \left(
    \begin{array}{ccc}
     0.00168 & ~~ 0.00176\,e^{-i\,29.2^\circ} ~~
      & 0.112\,e^{i\,72.5^\circ} \\
     0.00176\,e^{i\,29.2^\circ} & 0.0136 & 0.218\,e^{i\,70.6^\circ} \\
     0.112\,e^{-i\,72.5^\circ} & 0.218\,e^{-i\,70.6^\circ} & 11.82
    \end{array} \right) , \\
   \bm{\delta}_d &= 10^{-6} \left(
    \begin{array}{ccc}
     0.00390 & ~~ -0.00226\,e^{-i\,31.6^\circ} ~~
      & 0.0754\,e^{i\,80.4^\circ} \\
     -0.00226\,e^{i\,31.6^\circ} & 0.0393
      & 0.267\,e^{-i\,19.3^\circ} \\
     0.0754\,e^{-i\,80.4^\circ} & 0.267\,e^{i\,19.3^\circ} & 20.40
    \end{array} \right) .
\end{split}
\eeq
Left-handed up-quark sector:
\beq
\begin{split}
   \bm{\Delta}_U &= 10^{-3} \left(
    \begin{array}{ccc}
     0.775 & 1.823\,e^{-i\,1.5^\circ} & -1.337\,e^{i\,23.9^\circ} \\
     1.823\,e^{i\,1.5^\circ} & 7.059 & -3.177\,e^{-i\,16.7^\circ} \\
     -1.337\,e^{-i\,23.9^\circ} & ~~ -3.177\,e^{i\,16.7^\circ} ~~
      & 29.23
    \end{array} \right) , \\
   \bm{\delta}_U &= 10^{-4} \left(
    \begin{array}{ccc}
     0.419 & 1.006\,e^{-i\,3.9^\circ} & -0.733\,e^{i\,0.7^\circ} \\
     1.006\,e^{i\,3.9^\circ} & 2.812 & -1.918\,e^{i\,31.2^\circ} \\
     -0.733\,e^{-i\,0.7^\circ} & ~~ -1.918\,e^{-i\,31.2^\circ} ~~
      & 15.46
    \end{array} \right) .
\end{split}
\eeq
Right-handed up-quark sector:
\beq
\begin{split}
   \bm{\Delta}_u &= 10^{-4} \left(
    \begin{array}{ccc}
     0.000421 & ~~ -0.0580\,e^{-i\,3.1^\circ} ~~
      & 0.269\,e^{-i\,67.8^\circ} \\
     -0.0580\,e^{i\,3.1^\circ} & 9.382
      & -40.66\,e^{-i\,82.5^\circ} \\
     0.269\,e^{i\,67.8^\circ} & -40.66\,e^{i\,82.5^\circ}
      & 4596.8
    \end{array} \right) , \\
   \bm{\delta}_u &= 10^{-6} \left(
    \begin{array}{ccc}
     0.000867 & ~~ -0.0857\,e^{-i\,4.7^\circ} ~~
      & 2.593\,e^{i\,61.7^\circ} \\
     -0.0857\,e^{i\,4.7^\circ} & 13.76
      & 276.9\,e^{-i\,85.0^\circ} \\
     2.593\,e^{-i\,61.7^\circ} & 276.9\,e^{i\,85.0^\circ}
      & 75086
    \end{array} \right) .
\end{split}
\eeq
The mixing matrices $\bm{\Delta}'_A$ take similar values (both in magnitudes and phases) as the corresponding matrices $\bm{\Delta}_A$. They are collected in Appendix~\ref{app:results}. Also given there are our results for the matrices $\bm{\varepsilon}_A^{(\prime)}$, which are more than an order of magnitude smaller in magnitude than the corresponding matrices $\bm{\Delta}_A^{(\prime)}$. Their contributions to the $Z^0$-boson couplings in (\ref{gLR}) are negligible. In fact, these contributions are formally of $\ord(v^4/\Mkk^4)$.

It is instructive to compare the above exact results for the various mixing matrices with the approximations (\ref{ZMA1}) and (\ref{ZMA2}) obtained using the ZMA. We find that in all cases the ZMA reproduces the magnitudes and phases of the matrix elements with very good accuracy. Deviations typically arise at the percent level.

While the various mixing matrices presented above refer to a particular point in parameter space, they are nevertheless useful in order to obtain a picture of ``typical'' flavor-changing effects that can be expected in the RS model. The input parameters have been chosen such that the resulting values of the fermion masses and CKM parameters are in good agreement with the experimental values of these quantities. Apart from $\ord(1)$ effects on the various entries, the freedom one has in varying these matrices is thus restricted to the reparametrization transformations (\ref{RPI1Delta}) and (\ref{RPI2Delta}). Finally, the overall scale of all flavor-changing effects in the neutral-current sector can be tuned by changing the KK scale, since these effects are proportional to $v^2/\Mkk^2$. 

\subsubsection{Masses and Mixings of KK Fermions}

\begin{figure}[!t]
\begin{center}
\includegraphics[height=2.85in]{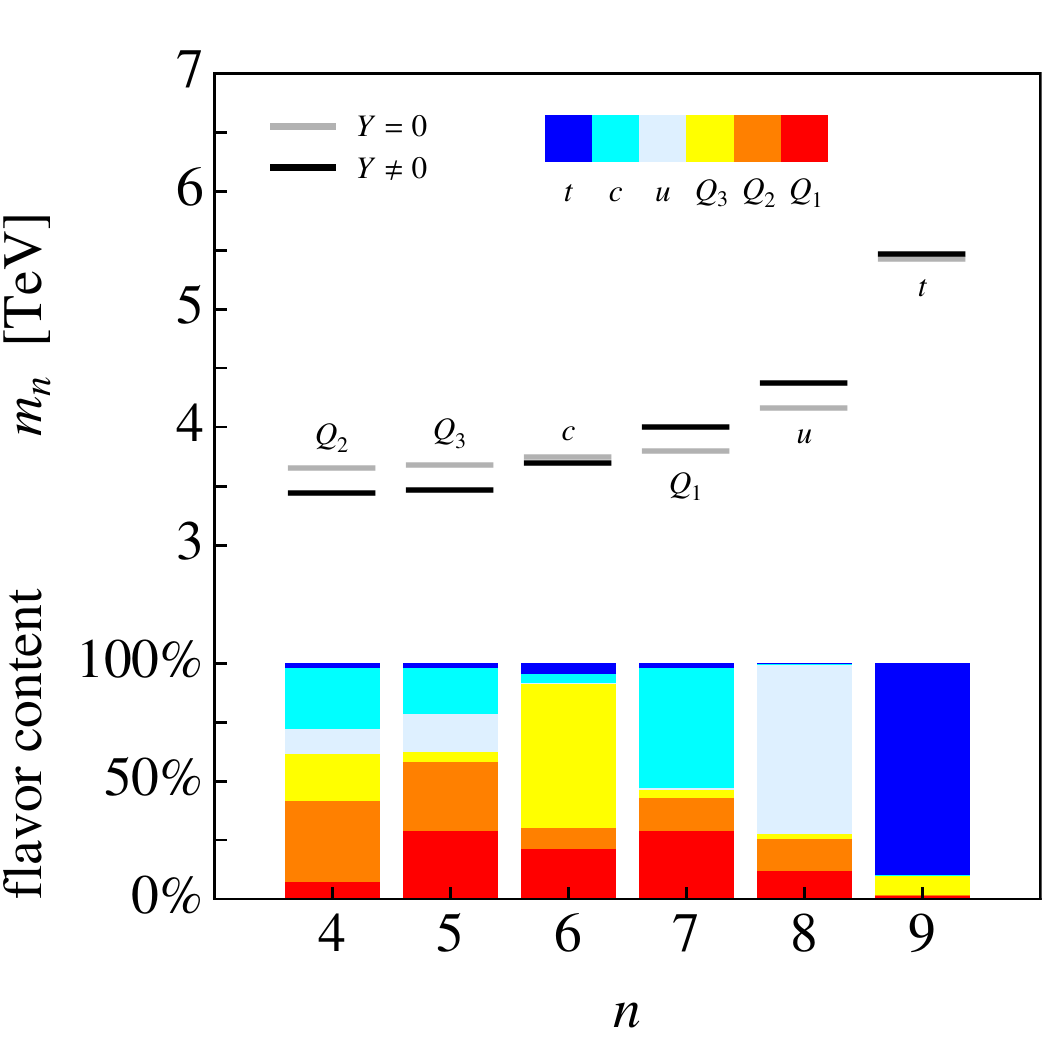} 
\qquad
\includegraphics[height=2.85in]{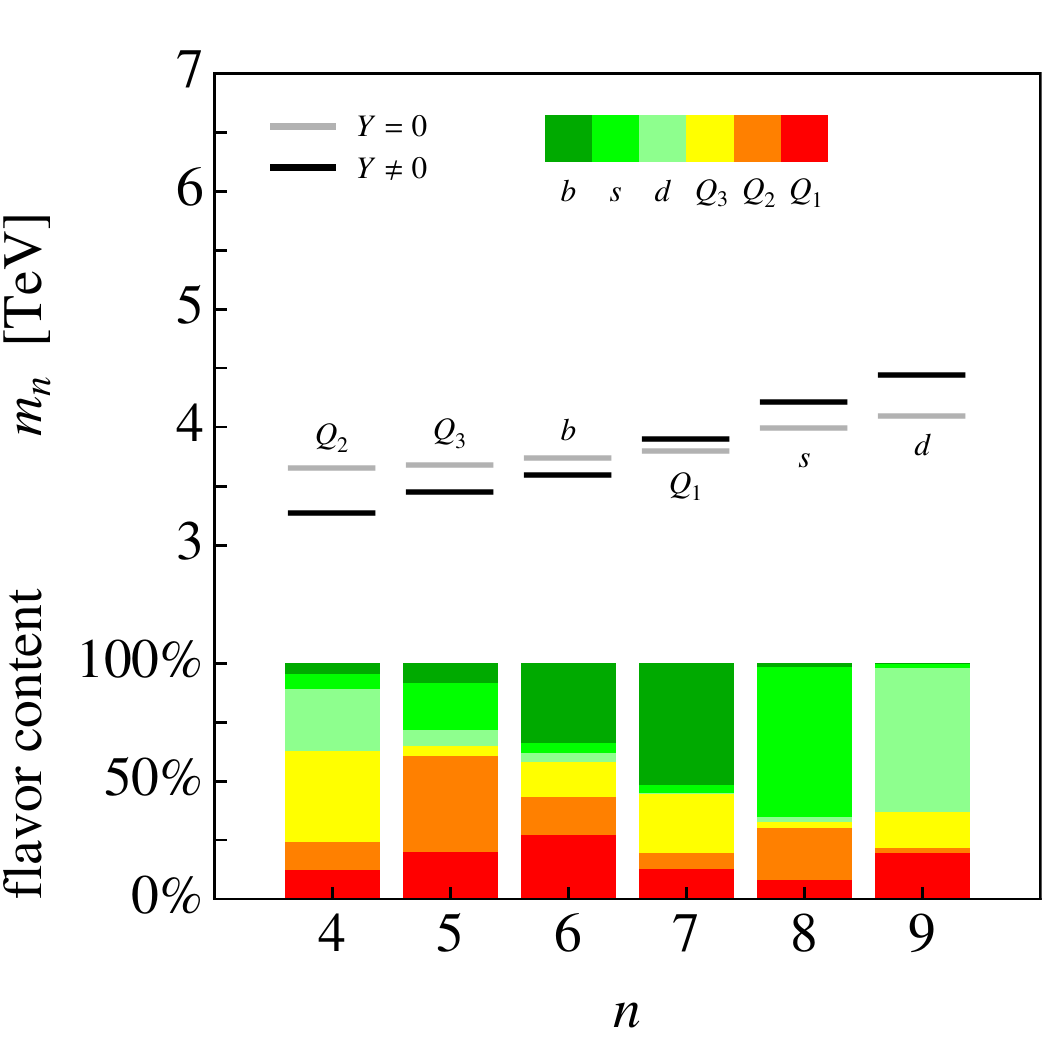}
\vspace{-4mm}
\parbox{15.5cm}{\caption{\label{fig:up}
Mass spectrum of the first KK excitations of the up- (left) and down-type (right) quarks. Black lines show the exact masses, while gray lines show the masses obtained by switching off the Yukawa couplings. The mixings of the mass eigenstates are visualized by the bar charts at the bottom of each panel. See text for details.}}
\end{center}
\end{figure}

The equations derived in Section~\ref{sec:fermions} allow us to study also the mass spectrum and mixings of the KK excitations of the SM fermions. These mixings can give rise to flavor-violating transitions among the light SM quarks when inserted into loop diagrams. As an example, we study the first level of KK quarks numerically using the default parameters given earlier in this section and assuming $\Mkk=1.5$\,TeV. Since there are vector-like excitations for each chiral SM fermion, we obtain six states each in the up- and down-type quark sectors at the first KK level. Figure~\ref{fig:up} shows the exact mass spectrum of these states together with the spectrum one would obtain without Yukawa couplings. The ``undisturbed states'' calculated with $\bm{Y}_{u,d}=0$ correspond to pure $SU(2)_L$ doublets and singlets, labeled by $Q_1$, $Q_2$, $Q_3$ and $u$, $c$, $t$ or $d$, $s$, $b$, respectively. Introducing the Yukawa couplings leads to mixings between fields with different flavor and $SU(2)_L$ quantum numbers, which are visualized by the bar charts at the bottom of each panel. The area of each colored region is proportional to the square of the absolute value of the corresponding entry in the mixing vectors $a_{4-9}^{(U,D)}$ and $a_{4-9}^{(u,d)}$ (altogether 36 numbers), which appear in the KK decomposition (\ref{KKdecomp}) of the fermion fields. 

As the KK scale is much larger than the Higgs vacuum expectation value, one might expect the flavor mixings between KK fermions to be small. To the contrary, however, one finds very large mixing effects especially in the down-type quark sector due to the near degeneracy of the 5D bulk masses of the corresponding fermion fields. The mass splittings of the undisturbed KK states are typical of order 100\,GeV, which is not large compared to $v$. Thus the Yukawa couplings generically induce $\ord(1)$ mixings among the KK excitations of the same KK level.

The generation mixing of KK modes provides an important example where the approach of treating the Yukawa couplings as a small perturbation is inadequate. On the other hand, we have checked that good approximations to the exact results for the masses and mixings of KK fermions can be obtained by working with the undisturbed states, treating the Yukawa couplings as interactions, and diagonalizing the resulting mass matrices on a truncated basis of KK states \cite{Goertz:2008vr}.

\subsection{\boldmath$Z^0 b\bar b$ Couplings}
\label{sec:Zbb}

From (\ref{Zff}) and (\ref{gLR}), it follows that the flavor-diagonal $Z^0$-boson couplings to down-type quarks can be written in the form
\beq
\begin{split}
   {\cal L}_{\rm 4D}
   &\ni \left( 4\sqrt2 G_F m_Z^2 \right)^{1/2} 
    \left[ 1 + \frac{m_Z^2}{4\Mkk^2}
    \left( L\sin^2\theta_W - \frac{1}{2L} \right) 
    + \ord\left( \frac{m_Z^4}{\Mkk^4} \right) \right] Z_\mu^0 \\
   &\quad\times
    \left[ \big( g_L^d \big)_{ii}\,\bar d_{L,i}\gamma^\mu d_{L,i}
    + \big( g_R^d \big)_{ii}\,\bar d_{R,i}\gamma^\mu d_{R,i} 
    \right] ,
\end{split}
\eeq
where $i=1,2,3$. For the light down and strange quarks the left- and right-handed couplings are, to excellent approximation, given by the SM expressions, $(g_L^d)_{11}\approx-1/2+\sin^2\theta_W/3$ and $(g_R^d)_{11}\approx\sin^2\theta_W/3$. For the bottom quark, the non-universal corrections in (\ref{gLR}) can be significant. Using the approximate expressions (\ref{ZMA1}) and (\ref{ZMA2}) valid in the ZMA, we can derive compact analytical expressions for these corrections. From the explicit results for the mixing matrices collected in Appendix~\ref{app:textures}, we see that to leading power in hierarchies only the two profiles $F(c_{Q_3})$ and $F(c_{d_3})$ and simple combinations of $c_{Q_{1,2}}$, $c_{d_{1,2}}$, and the elements of the down-type Yukawa matrix appear. Denoting $c_{b_L}\equiv c_{Q_3}$ and $c_{b_R}\equiv c_{d_3}$, we obtain
\beq\label{eq:ZbbRS}
\begin{split}
   g_L^b\equiv \big( g_L^d \big)_{33}
   &\to \left( - \frac12 + \frac{\sin^2\theta_W}{3} \right)
    \left[ 1 - \frac{m_Z^2}{2\Mkk^2}\,
    \frac{F^2(c_{b_L})}{3+2c_{b_L}} \left(
    L - \frac{5+2c_{b_L}}{2(3+2c_{b_L})} \right) \right] \\
   &\hspace{-12.5mm} \mbox{}+ \frac{m_b^2}{2\Mkk^2}\,
   \left [ \frac{1}{1-2c_{b_R}} 
    \left( \frac{1}{F^2(c_{b_R})} - 1 
    + \frac{F^2(c_{b_R})}{3+2c_{b_R}} \right) + 
    \sum_{i=1,2} \frac{|(Y_d)_{3i}|^2}{|(Y_d)_{33}|^2} 
    \frac{1}{1 - 2 c_{d_i}} \frac{1}{F^2(c_{b_R})} \right] , 
\end{split}
\eeq
\[
\begin{split}
   g_R^b\equiv \big( g_R^d \big)_{33} 
   &\to \frac{\sin^2\theta_W}{3}
    \left[ 1 - \frac{m_Z^2}{2\Mkk^2}\,
    \frac{F^2(c_{b_R})}{3+2c_{b_R}} \left(
    L - \frac{5+2c_{b_R}}{2(3+2c_{b_R})} \right) \right] \\
   &\hspace{-12.5mm} \mbox{}- \frac{m_b^2}{2\Mkk^2}\,
    \left [ \frac{1}{1-2c_{b_L}} 
    \left( \frac{1}{F^2(c_{b_L})} - 1 
    + \frac{F^2(c_{b_L})}{3+2c_{b_L}} \right) + 
    \sum_{i=1,2} \frac{|(Y_d)_{i3}|^2}{|(Y_d)_{33}|^2} 
    \frac{1}{1 - 2 c_{Q_i}} \frac{1}{F^2(c_{b_L})} \right] . \end{split}
\]
Note that the non-universal corrections always reduce the couplings with respect to their SM values in magnitude. Using the freedom of reparametrization invariance, one can rescale $F(c_{b_L})$, $F(c_{b_R})$, and the Yukawa couplings in such a way that the values of the quark masses and the CKM mixing angles remain unaffected, see (\ref{RPI1}) and (\ref{RPI2a}). This has the effect of redistributing contributions between the left-handed and right-handed couplings. In practice, however, the value of $F(c_{b_L})$ cannot be made too small if one wants to reproduce the large top-quark mass with reasonable Yukawa couplings.

The ratio of the width of the $Z^0$-boson decay into bottom quarks and
the total hadronic width, $R_b^0$, the bottom quark left-right asymmetry parameter, $A_b$, and the forward-backward asymmetry for
bottom quarks, $A_{\rm FB}^{0,b}$, are given in terms of the left- and
right-handed bottom quark couplings as \cite{Field:1997gz}
\begin{eqnarray}\label{eq:bPOtheory}
   R_b^0 &=& \left [ 1 
    + \frac{4\sum_{q=u,d} \left[ (g_L^q)^2 + (g_R^q)^2\right]}%
           {\eta_{\rm QCD}\,\eta_{\rm QED} \left[ (1-6z_b)
            (g_L^b-g_R^b)^2 + (g_L^b+g_R^b)^2 \right]} 
    \right]^{-1} , \nonumber\\
   A_b &=& \frac{2\sqrt{1-4z_b}\,\,
    {\displaystyle \frac{g_L^b+g_R^b}{g_L^b-g_R^b}}}%
    {1-4z_b+(1+2z_b) {\displaystyle \left( 
     \frac{g_L^b+g_R^b}{g_L^b-g_R^b} \right)^2}} \,,
    \qquad 
   A_{\rm FB}^{0,b} = \frac34\,A_e\,A_b \,,
\end{eqnarray} 
where $\eta_{\rm QCD}=0.9954$ and $\eta_{\rm QED}=0.9997$ are QCD and QED radiative correction factors. The parameter $z_b\equiv m_b^2(m_Z)/m_Z^2=0.997\cdot 10^{-3}$ describes the effects of the non-zero bottom
quark mass. To an excellent approximation we can neglect the RS
contributions to the left- and right-handed couplings of the light
quarks, $g_{L,R}^q$, and to the asymmetry parameter of the electron, $A_e$, and fix these quantities to their SM values. In what follows we
will employ $g_L^u=0.34674$, $g_R^u=-0.15470$, $g_L^d=-0.42434$, $g_R^d=0.077345$ \cite{LEPEWWG:2005ema}, and $A_e=0.1462$ \cite{Bardin:1999yd, Arbuzov:2005ma}. The numerical values quoted
above correspond to the SM reference parameters collected in Appendix~\ref{app:masses}.

Inserting the predicted SM values $g_L^b=-0.42114$ and $g_R^b=0.077420$ \cite{LEPEWWG:2005ema} for the left- and right-handed bottom-quark couplings into the relations (\ref{eq:bPOtheory}), we obtain for the central values of the quantities in question
\beq\label{eq:bPOSM}
   R_b^0 = 0.21579 \,, \qquad 
   A_b = 0.935 \,, \qquad 
   A_{\rm FB}^{0,b} = 0.1025 \,.
\eeq
These SM expectations should be compared to the experimentally extracted values for the three ``pseudo observables''. They are \cite{LEPEWWG:2005ema}
\beq\label{eq:bPOsexp}
   \begin{array}{l}
    R_b^0 = 0.21629\pm 0.00066 \,, \\[0.25mm] 
    A_b = 0.923\pm 0.020 \,, \\[1mm]
    A_{\rm FB}^{0,b} = 0.0992\pm 0.0016 \,, 
   \end{array}
    \qquad 
   \rho = \begin{pmatrix}
    1.00 \, & \, -0.08 & \, -0.10 \\ 
    \, -0.08 & \, 1.00 & \, 0.06 \\ 
    -0.10 \, & \,  0.06 & \, 1.00  
   \end{pmatrix} ,
\eeq
where $\rho$ is the correlation matrix. We see that while the $R_b^0$ and $A_b$ measurements agree within $+0.8\sigma$ and $-0.6\sigma$ with their SM predictions, the $A_{\rm FB}^{0,b}$ measurement is almost $-2.1\sigma$ away from its SM value.\footnote{For $m_h=115$\,GeV the discrepancy in $A_{\rm FB}^{0,b}$ would amount to around $-2.5\sigma$.} 
Whether this is an experimental problem, a statistical fluctuation or an effect of new physics in the bottom-quark couplings is up to date unresolved. In fact, the relative experimental error is much larger in $A_{\rm FB}^{0,b}$ than in $R_b^0$ and $A_b$, where no anomalies are observed. Furthermore, the value of the left-handed bottom-quark coupling, which is essentially determined by the measurement of $R_b^0\propto (g_L^b)^2+(g_R^b)^2$ due to the smallness of $g_R^b$, shows no discrepancy. The data therefore invite an explanation in terms of a possible deviation of the right-handed bottom-quark coupling from its SM value. This would affect $A_b$ and $A_{\rm FB}^{0,b}$, which both depend linearly on the ratio $g_R^b/g_L^b$, more strongly than $R_b^0$. 

\begin{figure}[!t]
\begin{center} 
\hspace{-2mm}
\mbox{\includegraphics[height=2.85in]{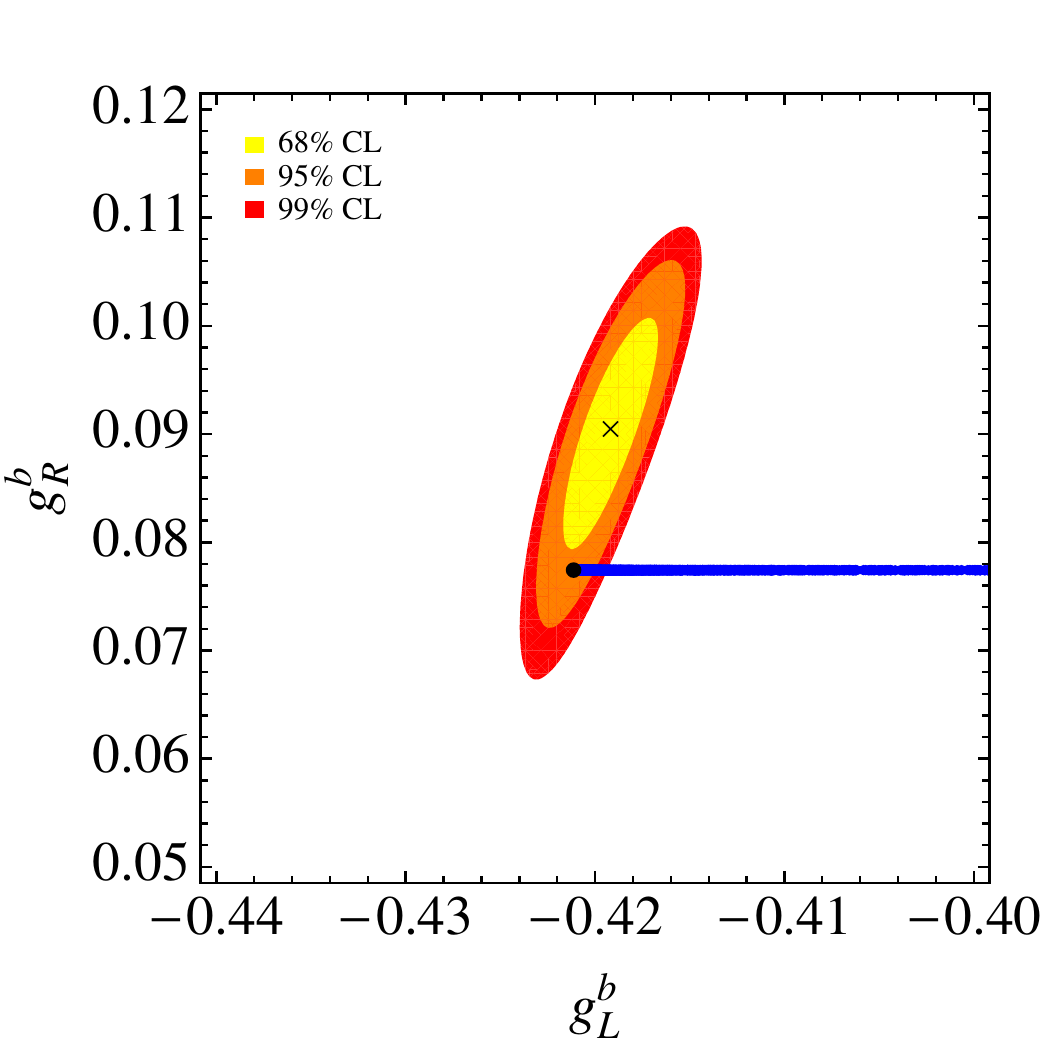}} 
\qquad 
\mbox{\includegraphics[height=2.85in]{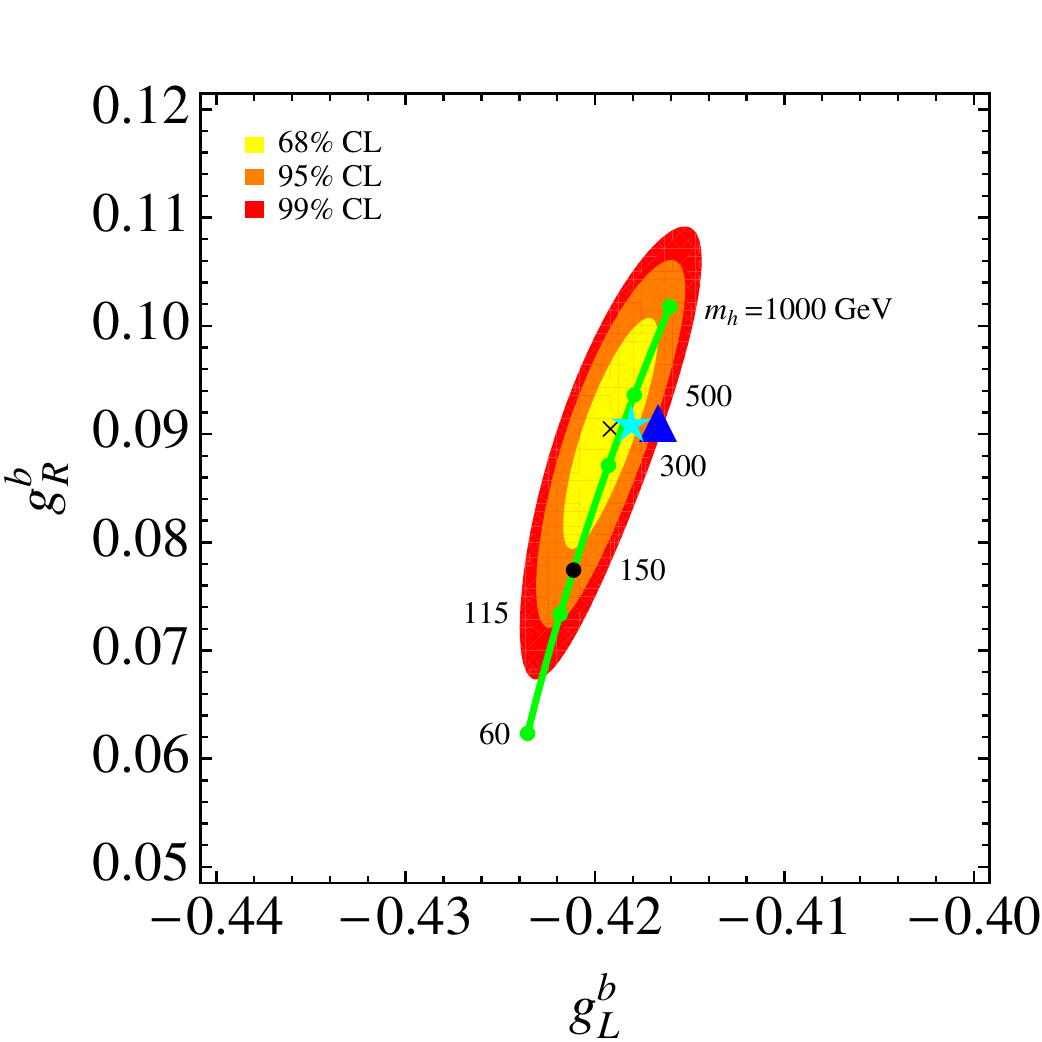}} 
\vspace{-2mm}
\parbox{15.5cm}{\caption{\label{fig:gLgRplots}
Regions of 68\% , 95\% , and 99\% probability in the $g_L^b$--$g_R^b$ plane. The horizontal stripe in the left plot consists of a large number of points in parameter space. The triangle and star in the right panel correspond to the results for our reference RS point with $\Mkk=1.5$\,TeV and $\Mkk=3$\,TeV, assuming a Higgs-boson mass of $m_h=400$\,GeV. The black dot is the SM expectation for the reference point, and the green (medium gray) line in the right panel indicates the SM predictions for $m_h\in [60,1000]$\,GeV. See text for details.}}
\end{center}
\end{figure}

The results of our fit to the $Z^0\to b\bar b$ ``pseudo observables'' are shown in Figure~\ref{fig:gLgRplots}. In the left panel we have superimposed the RS predictions obtained for 3000 randomly generated points in parameter space. It is evident that the RS prediction for $g_L^b$ is always larger than the SM reference value indicated by the black dot, while $g_R^b$ is essentially unaffected.\footnote{The tiny RS shifts in $g_R^b$ are always negative.} 
This implies that the RS corrections to the $Z^0 b\bar b$ couplings necessarily shift the values $g_{L,R}^b$ further away from the best fit $g_L^b=-0.41918$ and $g_R^b=0.090677$. The RS corrections (\ref{eq:ZbbRS}) thus cannot account for the positive shift in $g_R^b$ needed to explain the anomaly in $A_{\rm FB}^{0,b}$. 

\begin{figure}[!t]
\begin{center}
\hspace{-2mm}
\mbox{\includegraphics[height=2.85in]{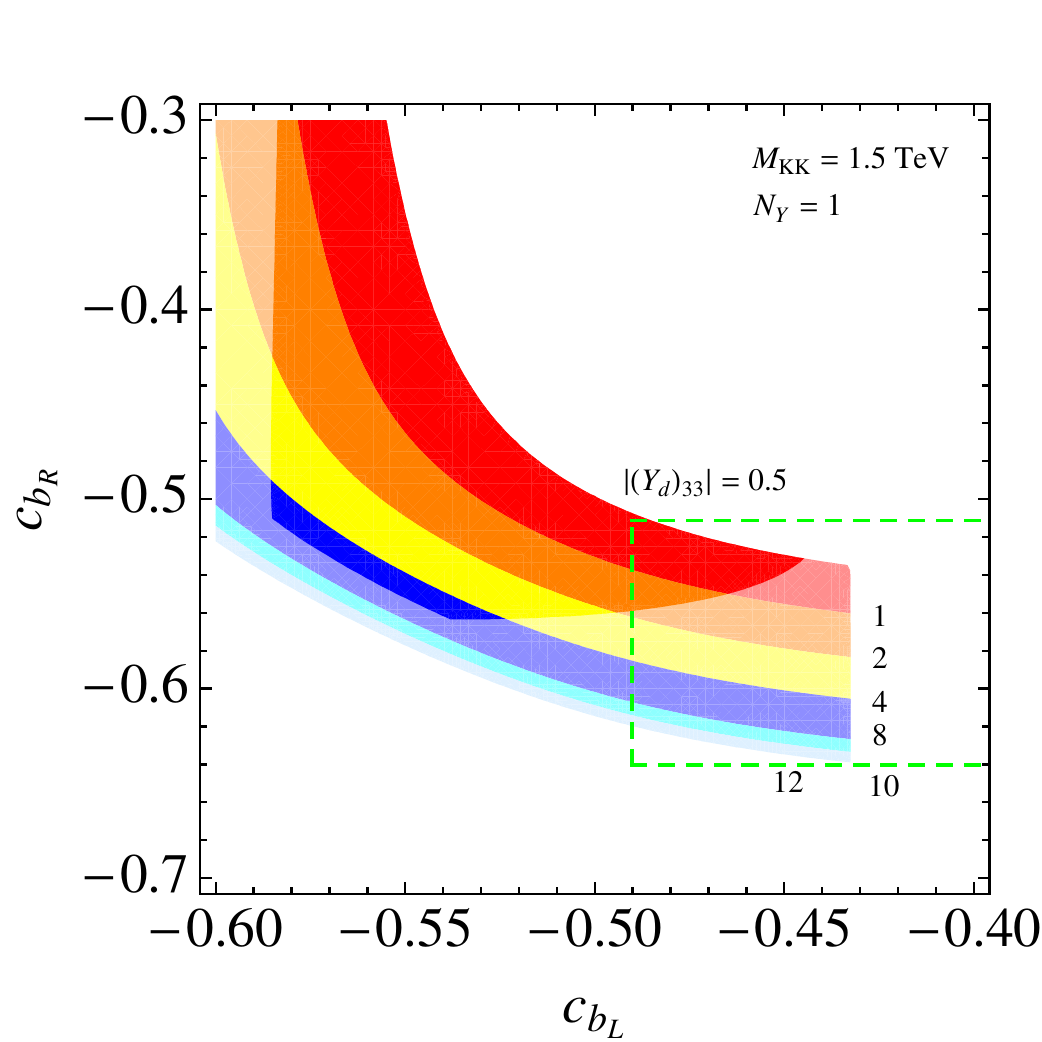}}
\qquad 
\raisebox{0.025cm}{
\mbox{\includegraphics[height=2.85in]{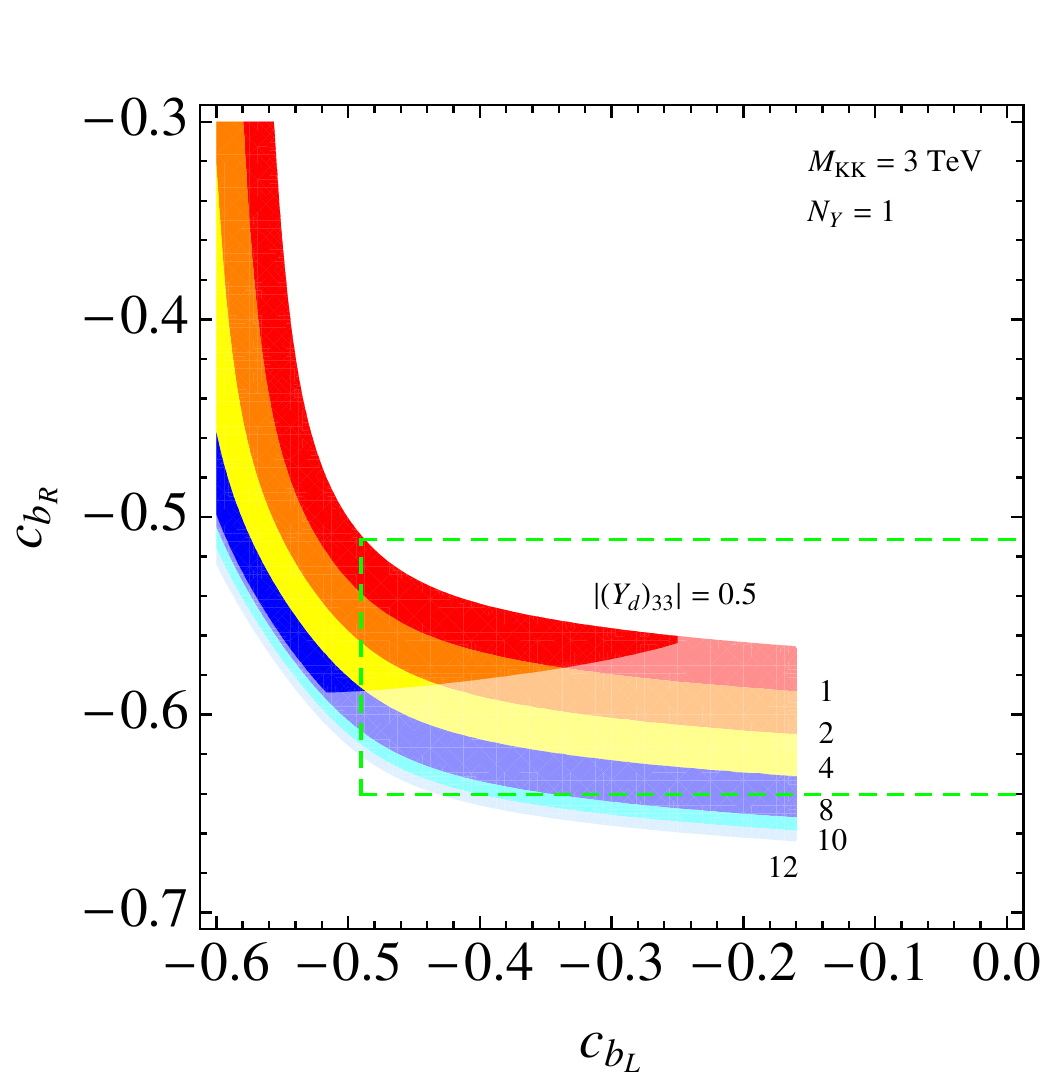}}}
\vspace{-2mm}
\parbox{15.5cm}{\caption{\label{fig:cLcRplots}
Regions of 99\% probability in the $c_{b_L}$--$c_{b_R}$ plane for $\Mkk=1.5$\,TeV (left) and $\Mkk=3$\,TeV (right). We set $c_{Q_{1,2}}=c_{d_{1,2}}=-1/2$ and $N_Y\equiv |(Y_d)_{ij}|/|(Y_d)_{33}|=1$. The colored contours indicate the value of $|(Y_d)_{33}|$ necessary to reproduce the value of the bottom-quark mass. The RS results with (without) the $m_b^2/\Mkk^2$ terms are indicated by bright (faint) colors. Only the parameter region inside the dashed rectangles leads to consistent values for the quark masses and mixings. See text for details.}}
\end{center}
\end{figure}

The apparent large positive corrections to $g_L^b$ imply that the $R_b^0$, $A_b$, and $A_{\rm FB}^{0,b}$ measurements impose stringent constraints on the parameter space of the RS scenario. The distribution of points depends strongly on the bulk mass parameters, while the exact values of the KK scale and the elements of the down-type Yukawa matrices have only a minor impact on the overall picture. Notice that the dependence of $g_{L,R}^b$ on $c_{b_{L,R}}$ is much more pronounced than the one on $c_{Q_{1,2}}$ and $c_{d_{1,2}}$, because the parameters $c_{b_{L,R}}$ enter (\ref{eq:ZbbRS}) through their corresponding zero-mode profiles $F(c_{b_{L,R}})$. As a result, the allowed values of $c_{b_{L,R}}$ are strongly constrained by the $Z^0\to b\bar b$ ``pseudo observables'', while the bulk mass parameters $c_{Q_{1,2}}$ and $c_{d_{1,2}}$ are only weakly bounded. The former feature is illustrated in Figure~\ref{fig:cLcRplots}, which shows the regions of 99\% probability in the $c_{b_L}$--$c_{b_R}$ plane for $\Mkk=1.5$\,TeV (left) and $\Mkk=3$\,TeV (right). The colored contours in both plots indicate the magnitude $|(Y_d)_{33}|$ necessary to achieve the correct value of the bottom-quark mass. We find that under the restriction $|(Y_d)_{33}|<12$ the allowed $c_{b_{L,R}}$ parameters all lie in the intervals $c_{b_L}\in [-0.59,-0.45]$ ($c_{b_L}\in [-0.62,-0.24]$) and $c_{b_R}\in [-0.57,0.30]$ ($c_{b_R}>-0.60$) for $\Mkk=1.5$\,TeV ($\Mkk=3$\,TeV). Notice that increasing the magnitude $|(Y_d)_{33}|$ decreases the available parameter space. Requiring a consistent fit of all quark masses and CKM parameters restricts the allowed region in the $c_{b_L}$--$c_{b_R}$ plane even further. This feature is illustrated by the dashed rectangles in Figure~\ref{fig:cLcRplots}, which represent the allowed $c_{b_{L,R}}$ ranges that lead to a global $\chi^2/{\rm dof}$ of better than 11.5/10 for $|(Y_{u,d})_{ij}|$ restricted to the range $[1/3,3]$. The strongest constraint is provided by the top-quark mass. We recall in this context that we require $c_{u_3}<1/2$. 

In this figure, the RS results with (without) the $m_b^2/\Mkk^2$ terms are indicated by bright (faint) colors. Comparing the allowed regions makes clear that the $m_b^2/\Mkk^2$ terms in (\ref{eq:ZbbRS}), which result from the matrix elements $(\delta_D)_{33}$ and $(\delta_d)_{33}$ in (\ref{gLR}), are numerically important, especially if $|(Y_d)_{33}|$ is allowed to take values larger than a few. Neglecting these contributions is in general not justified. Notice also that the terms in (\ref{eq:ZbbRS}) proportional to $m_Z^2/\Mkk^2$ depend linearly on $L$, whereas the terms proportional to $m_b^2/\Mkk^2$ are independent of the logarithm of the warp factor. These latter terms thus cannot be removed by truncating the volume of the RS background. In turn, they are typically dominant for moderate and small values of $L$.

After a proper adjustment of bulk mass parameters, the constraint on $\Mkk$ coming from the combined $R_b^0$, $A_b$, and $A_{\rm FB}^{0,b}$ measurements is in general weaker than the one stemming from the electroweak precision measurements, encoded in $S$ and $T$. For example, for the RS reference input values spelled out in Section~\ref{sec:63}, we find
\beq\label{eq:bPOMkkbound} 
   \Mkk > 1.6\,\mbox{TeV} \quad (99\% \; {\rm CL}) \,.
\eeq 

Like in the case of the $S$ and $T$ parameters, it turns out to be
instructive to study the Higgs mass dependence of the observables in
question. The leading logarithmic Higgs-mass corrections to the $Z^0\to b\bar b$ ``pseudo observables'' are well approximated by
\begin{eqnarray}\label{eq:bPOmhdep}
   \Delta R_b^0 &=& 3.3\cdot 10^{-5}\,\ln\frac{m_h}{m_h^{\rm ref}}
    \,, \nonumber\\
   \Delta A_b &=& -2.7\cdot 10^{-4}\,\ln\frac{m_h}{m_h^{\rm ref}}
    \,, \\ 
   \Delta A_{\rm FB}^{0,b} &=& -2.7\cdot 10^{-3}\,
    \ln\frac{m_h}{m_h^{\rm ref}} \,. \nonumber
\end{eqnarray}
These relations have been derived with the help of {\tt ZFITTER} \cite{Bardin:1999yd,Arbuzov:2005ma}.\footnote{The default flags of {\tt ZFITTER} version 6.42 are used, except for setting {\tt ALEM=2} to take into account the externally supplied value of $\Delta\alpha^{(5)}_{\rm had}(m_Z)$.} 
Interestingly, the shifts in $R_b^0$ and $A_b$ are only moderate compared to the experimental errors, while the shift in $A_{\rm FB}^{0,b}$ is rather pronounced, due to the strong Higgs-mass dependence of $A_e$, and negative. These features allow to significantly improve the quality of the fit to the $Z^0\to b\bar b$ ``pseudo observables'' by taking large values of $m_h$. For example, a Higgs-boson mass of $m_h=400$\,GeV would bring the predictions of $g_{L,R}^b$ very close to the best fit value. This feature is illustrated by the triangle and star in the right panel of Figure~\ref{fig:gLgRplots}, which correspond to the predictions (\ref{eq:ZbbRS}) for our two RS reference points. Numerically, we obtain $g_L^b=-0.416658$ and $g_R^b=0.090813$ ($g_L^b=-0.418055$ and $g_R^b=0.090826$) for $\Mkk=1.5$\,TeV ($\Mkk=3$\,TeV). The line in the same plot indicates the SM predictions for $m_h\in[60,1000]$\,GeV. Warped models with the Higgs field localized in the IR might thus indirectly allow for an explanation of the $A_{\rm FB}^{0,b}$ anomaly, since in these setups the Higgs boson is expected to be heavy, which leads to a good agreement between $Z^0\to b\bar b$ data and theory. Needless to say, even for a heavy Higgs boson the tight constraints on $c_{b_{L,R}}$ arising from the measurements of $R_b^0$, $A_b$, and $A_{\rm FB}^{0,b}$ persist.

The strong bulk-mass parameter dependence of the RS corrections to the $Z^0 b\bar b$ vertex can be greatly reduced \cite{Carena:2006bn} by using an embedding of the SM fermions into the custodially symmetric $SU(2)_L\times SU(2)_R$ model, under which the left-handed bottom quark is symmetric under the exchange of $SU(2)_L$ and $SU(2)_R$ \cite{Agashe:2006at}. The simplest implementation of this $Z_2$ symmetry is realized by choosing the left-handed quarks to be bi-fundamentals under $SU(2)_L\times SU(2)_R$, while the right-handed top quark is chosen to be a singlet. In these variations of the original RS scenario, mixing between the $Z^0$ boson and the KK excitations allows to explain the $A_{\rm FB}^{0,b}$ anomaly without affecting the agreement of the other precision electroweak observables with experimental data for moderate KK masses of the order of 3\,TeV \cite{Djouadi:2006rk}.

\subsection{Rare Decay \boldmath$t\to cZ^0$\unboldmath}
\label{sec:tcZ}

A particularly interesting class of FCNC processes in the RS model involves the flavor-violating couplings of the top quark. As the heaviest fermion in the SM, the top quark is located closest to the IR brane and hence couples most strongly to the KK excitations of the gauge bosons. Sizable flavor-changing effects involving the top quark are hence expected in the RS scenario. Because flavor transitions in the up-type quark sector are far less constrained by kaon and $B$-meson physics than those in the down-type quark sector, the presence of non-negligible anomalous flavor-changing top-quark couplings is not ruled out experimentally. This makes searches for radiative and rare $\Delta F=1$ processes involving the top quark unique probes of the RS framework.

The flavor-changing couplings of quarks to the $Z^0$ boson in (\ref{Zff}) allow the top quark to decay via the process $t\to c Z^0$. The branching ratio is given to excellent approximation by
\beq
\begin{split}
   {\cal B}(t\to c Z^0) 
   &= \frac{2\left( 1-r_Z^2 \right)^2 \left(1+2r_Z^2 \right)}%
           {\left( 1-r_W^2 \right)^2 \left( 1+2r_W^2 \right)} \\
   &\quad\times
    \left\{ \left| \left( g_L^u \right)_{23} \right|^2
    + \left| \left( g_R^u \right)_{23} \right|^2 
    - \frac{12r_c r_Z^2}%
           {\left( 1-r_Z^2 \right) \left( 1+2r_Z^2 \right)}\,
    \mbox{Re}\big[ \left( g_L^u \right)_{23}^*
     \left( g_R^u \right)_{23} \big] \right\} \\
   &\approx 1.842\,\Big[ \left| \left( g_L^u \right)_{23} \right|^2
    + \left| \left( g_R^u \right)_{23} \right|^2 \Big]
    - 0.048\,\mbox{Re}\big[ \left( g_L^u \right)_{23}^*
     \left( g_R^u \right)_{23} \big] \,,
\end{split}
\eeq
where $r_i\equiv m_i^{\rm pole}/m_t^{\rm pole}$, and for simplicity we have only kept terms up to first order in $v^2/\Mkk^2$ and the charm-quark mass ratio $r_c\approx 8.7\cdot 10^{-3}$. The flavor-changing couplings are given by
\beq
\begin{split}
   \left( g_L^u \right)_{23} 
   &= - \frac{m_Z^2}{2\Mkk^2}
    \left( \frac12 - \frac23\sin^2\theta_W \right)
    \Big[ L \left( \Delta_U \right)_{23}
    - \left( \Delta'_U \right)_{23} \Big] 
    - \frac12 \left( \delta_U \right)_{23} \,, \\
   \left( g_R^u \right)_{23} 
   &= \frac{m_Z^2}{2\Mkk^2}\,\frac23\sin^2\theta_W\,
    \Big[ L \left( \Delta_u \right)_{23} 
    - \left( \Delta'_u \right)_{23} \Big] 
    + \frac12 \left( \delta_u \right)_{23}  \, .
\end{split}
\eeq
The branching ratio and the flavor-changing couplings relevant for the
$t\to u Z^0$ decay are obtained from the above expressions by replacing subscripts $23$ by $13$ and $m_c$ by $m_u$. Due to the RS-GIM mechanism this decay is, however, strongly suppressed compared to $t\to c Z^0$.

Using the ZMA expressions (\ref{ZMA1}) and (\ref{ZMA2}), it is
straightforward to derive that to leading power in hierarchies the
flavor-changing couplings inducing $t\to c Z^0$ decay take the
form
\begin{eqnarray}\label{eq:tcZRS}
\begin{split}
   \big( g_L^u \big)_{23}
   &\to \frac{m_Z^2}{2\Mkk^2} 
    \left( \frac12 - \frac23 \sin^2\theta_W \right) 
    \frac{F(c_{Q_2}) F(c_{Q_3})}{3+2c_{Q_3}}\,
    \frac{(Y_u)_{23}}{(Y_u)_{33}} 
    \left( L - \frac{5+2c_{Q_3}}{2(3+2c_{Q_3})} \right) \\
   &\quad\mbox{}- e^{-i(\phi_2-\phi_3)}\, 
    \frac{m_c m_t}{2\Mkk^2}\,
    \frac{1}{F(c_{u_2}) F(c_{u_3})} \left[ \frac{1}{1-2c_{u_1}}\,
    \frac{(Y_u)_{31}^\ast (M_u)_{12}}{(Y_u)_{33}^\ast (M_u)_{11}} 
    + \frac{1}{1-2c_{u_2}}\,\frac{(Y_u)_{32}^\ast}{(Y_u)_{33}^\ast} 
     \right] ,  \hspace{6.5mm} \\
   \big( g_R^u \big)_{23} 
   &\to - e^{-i(\phi_2-\phi_3)}\, 
    \frac{m_Z^2}{2\Mkk^2}\,\frac23 \sin^2\theta_W\,
    \frac{F(c_{u_2}) F(c_{u_3})}{3+2c_{u_3}}\, 
    \frac{(Y_u)_{32}^\ast}{(Y_u)_{33}^\ast} 
    \left( L - \frac{5+2c_{u_3}}{2(3+2c_{u_3})} \right) \\
   &\quad\mbox{}+ \frac{m_c m_t}{2\Mkk^2}\,
   \frac{1}{F(c_{Q_2}) F(c_{Q_3})} \left[ \frac{1}{1-2c_{Q_1}}\,
   \frac{(Y_u)_{13} (M_u)_{21}^\ast}{(Y_u)_{33} (M_u)_{11}^\ast} 
   + \frac{1}{1-2c_{Q_2}}\,\frac{(Y_u)_{23}}{(Y_u)_{33}} 
   \right] . \hspace{6.5mm}
\end{split}
\end{eqnarray}
Here $(M_u)_{ij}$ again denotes the minor of the up-type Yukawa matrix $\bm{Y}_u$, and the definitions of the phase factors $e^{i\phi_j}$ can be found in (\ref{eq:expphij}).

\begin{figure}[!t]
\begin{center} 
\hspace{-2mm}
\mbox{\includegraphics[height=2.85in]{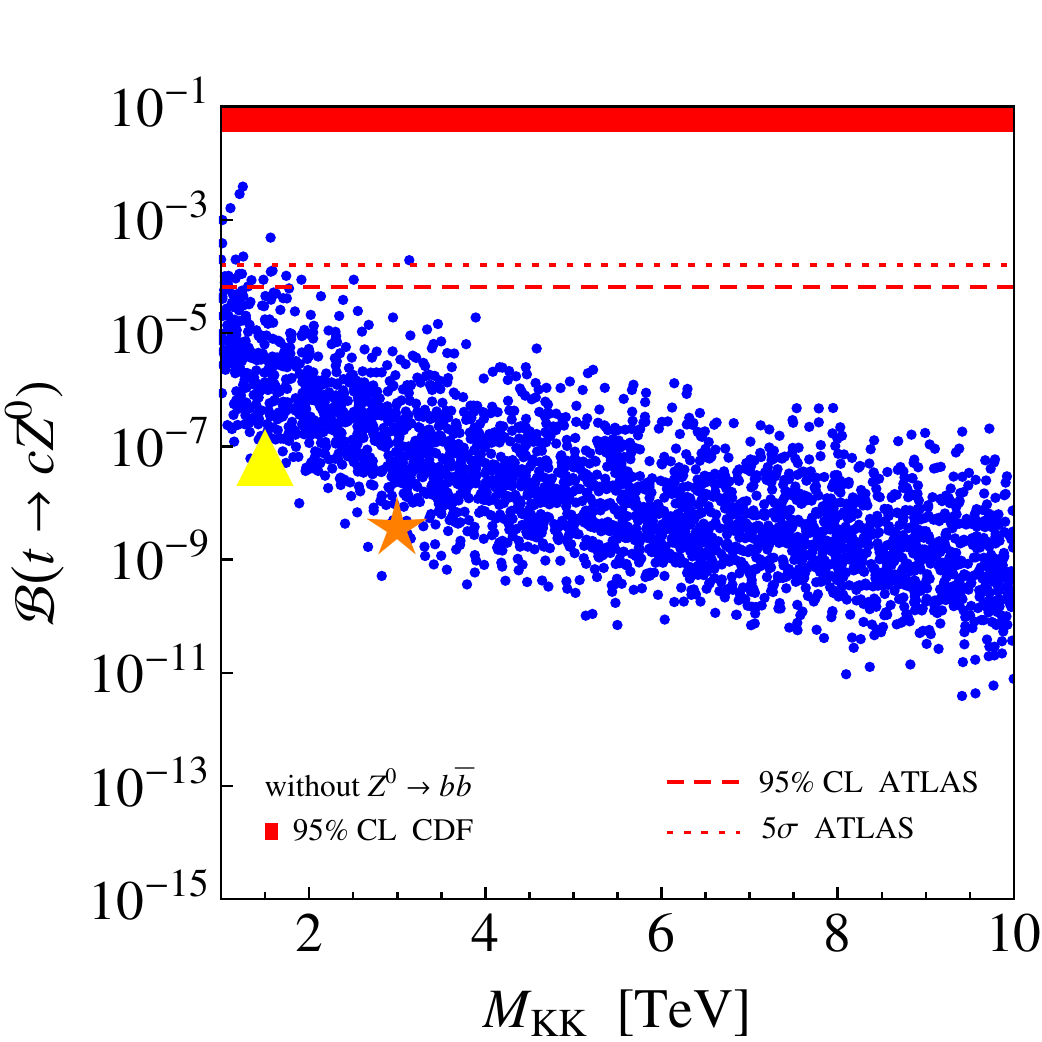}} 
\qquad 
\mbox{\includegraphics[height=2.85in]{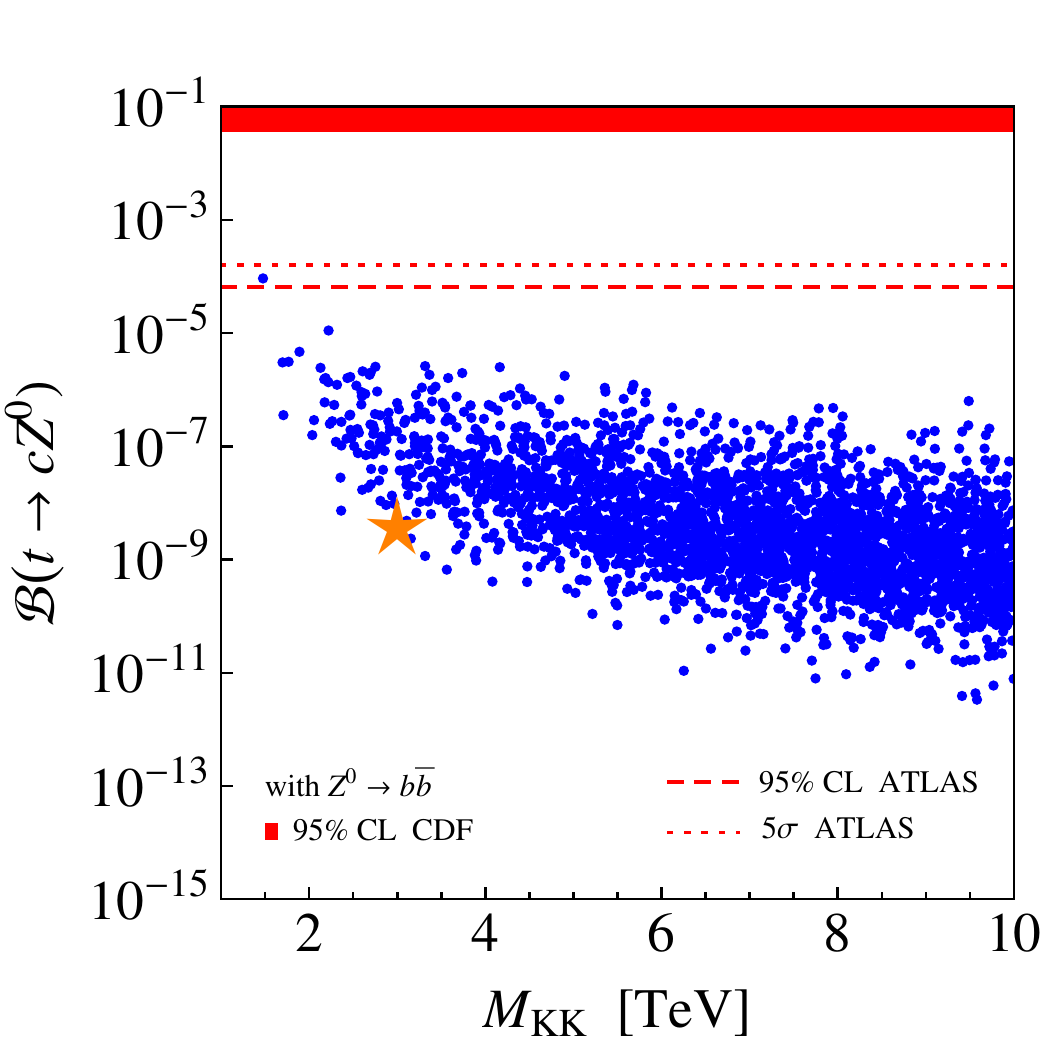}} 
\vspace{-2mm}
\parbox{15.5cm}{\caption{\label{fig:tcZplots}
Branching ratio of the rare decay $t\to c Z^0$ as a function of $\Mkk$ in the RS model. The red (dark gray) band is disfavored at 95\% CL by the CDF search for $t\to u(c) Z^0$. The dashed and dotted red (dark gray) lines indicate the expected sensitivities of ATLAS for 100\,fb$^{-1}$ integrated luminosity. All scatter points reproduce the correct quark masses, mixing angles, and the CKM phase. In the left (right) plot points that violate the constraints from the $Z^0\to b\bar b$ ``pseudo observables'' are included (excluded). The yellow (light gray) triangle and the orange (medium gray) stars represent the results for our RS reference points. See text for details.}}
\end{center}
\end{figure}

The RS predictions for ${\cal B}(t\to c Z^0)$ as a function of $\Mkk$ are shown in Figure~\ref{fig:tcZplots}. Each plot contains 3000 randomly generated parameter points. The values of ${\cal B}(t\to c Z^0)=6.1\cdot 10^{-8}$ and ${\cal B}(t\to c Z^0)=3.8\cdot 10^{-9}$ corresponding to our RS reference point for $\Mkk=1.5$\,TeV and $\Mkk=3$\,TeV are displayed by the triangle and stars. Notice that for $\Mkk=1.5$\,TeV our RS reference point barely fails to satisfy the constraints from the $Z^0\to b\bar b$ ``pseudo observables''. The presently most precise experimental upper bound on FCNC $t\to u(c) Z^0$ decays stems from the CDF experiment and amounts to ${\cal B}(t\to u(c) Z^0)<3.7\%$ at 95\% CL \cite{tcZ:2008aaa}. It is shown as a band. Notice that the recent CDF bound supersedes the 95\% CL upper limit ${\cal B}(t\to u(c) Z^0)<13.7\%$ \cite{Achard:2002vv} set by the L3 experiment from the non-observation of FCNC single top-quark production. At the LHC, rare FCNC top-quark transitions can be searched for in top-quark production and decays. The best identification will be reached for $t\to u(c) Z^0\to u(c) l^+ l^-$ and $t\to u(c)\gamma$. Simulation studies have been performed by both the ATLAS \cite{Carvalho:2007yi} and the CMS \cite{Ball:2007zza} Collaboration. The minimum of ${\cal B}(t\to c Z^0)$ allowing for a signal discovery with $5\sigma$ significance with 100\,fb$^{-1}$ integrated luminosity is expected to be $1.6\cdot 10^{-4}$ at ATLAS. In the absence of a signal, the expected limit at 95\% CL is $6.5\cdot 10^{-5}$. These sensitivities are indicated by the dashed and dotted lines in the panels.

\begin{figure}[!t]
\begin{center} 
\hspace{-2mm}
\mbox{\includegraphics[height=2.85in]{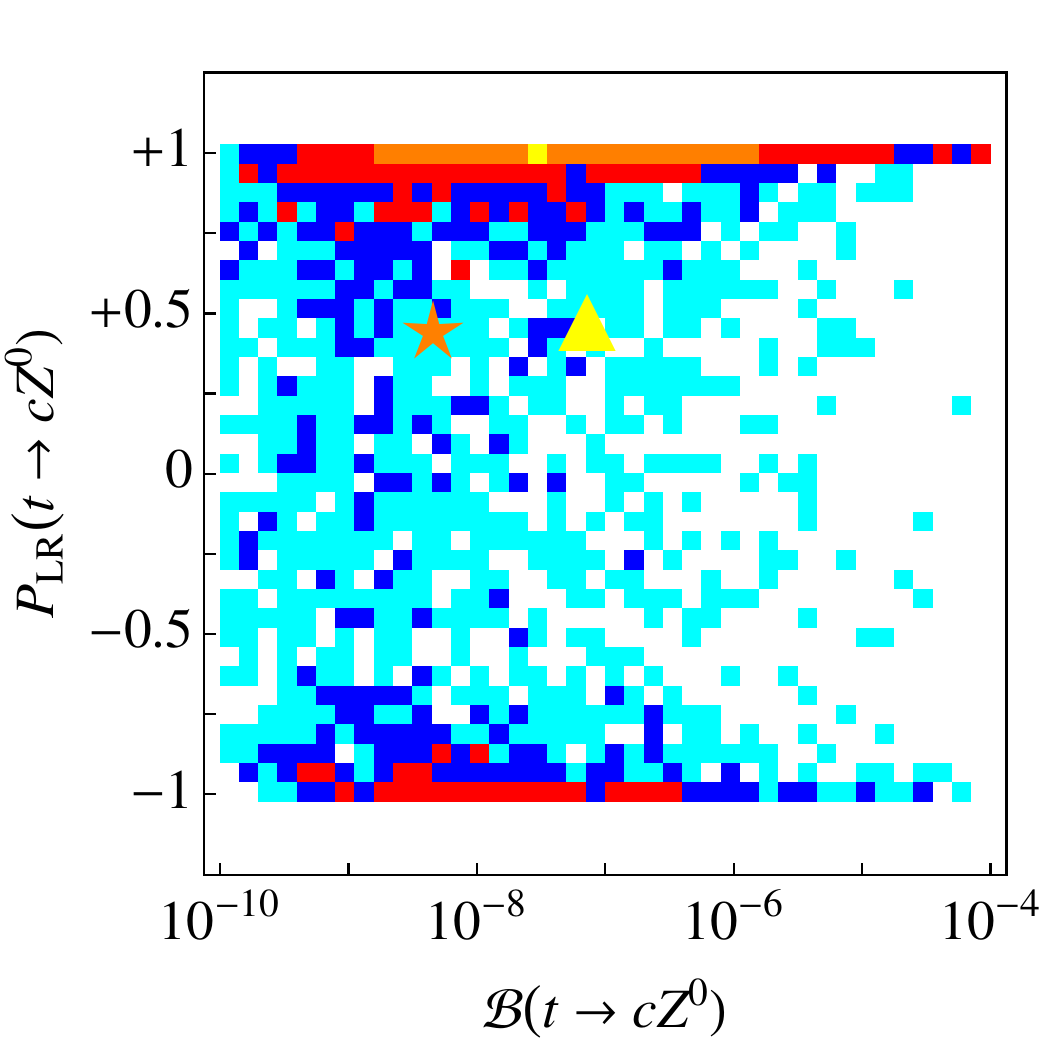}} 
\qquad 
\mbox{\includegraphics[height=2.85in]{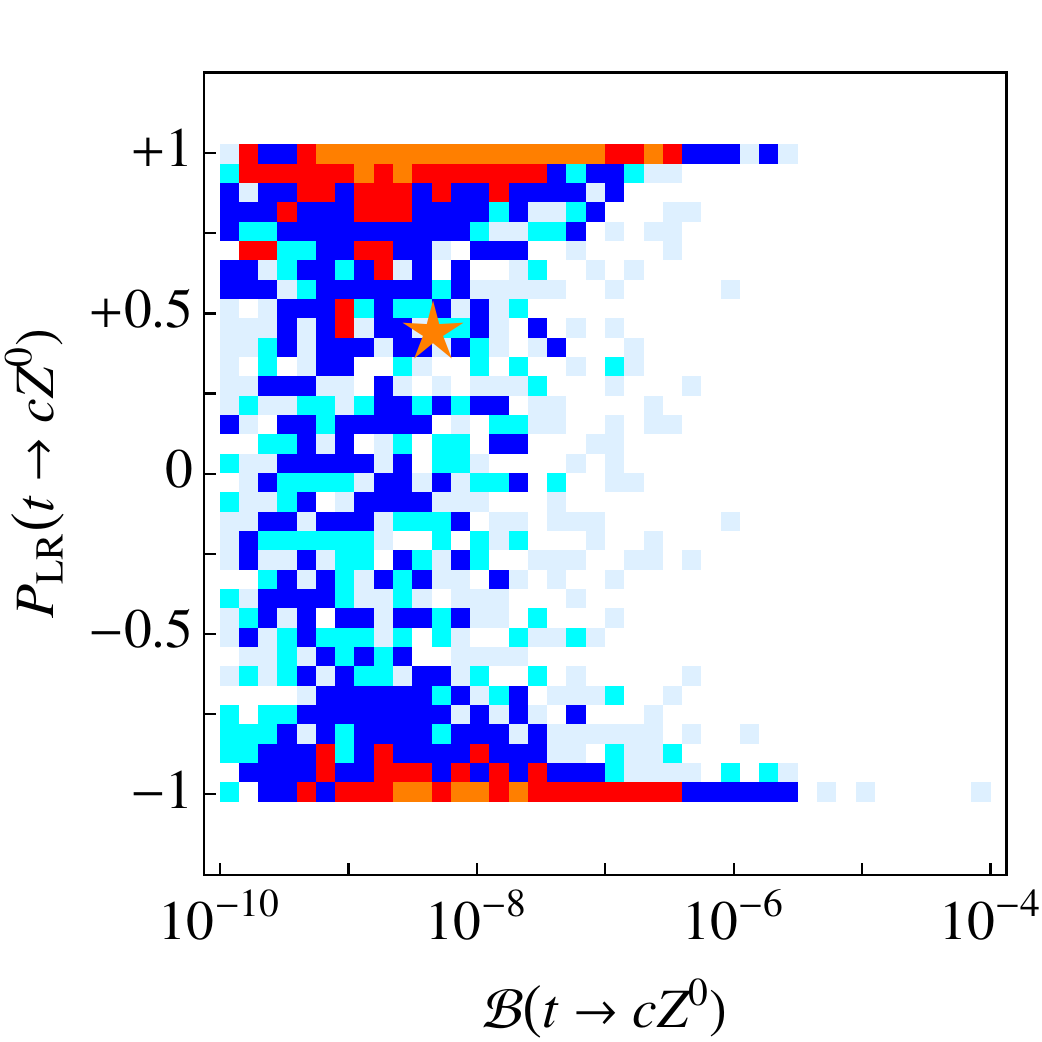}} 
\vspace{-2mm}
\parbox{15.5cm}{\caption{\label{fig:PLRplots}
Left-right polarization asymmetry of $t\to c Z^0$ as a function of the branching ratio of $t\to c Z^0$. The results before and after removing the points that violate the constraints from the $Z^0\to b\bar b$ ``pseudo observables'' are shown in the left and right panel. Bins with the lowest density of points are colored light blue (light gray), while those with the highest density of points are colored yellow (light gray). The yellow (light gray) triangle and the orange (medium gray) stars represent the results for our RS reference points. See text for details.}}
\end{center}
\end{figure}

The distribution of scatter points in the plots of Figure~\ref{fig:tcZplots} demonstrates that RS realizations which lead to values of ${\cal B}(t\to c Z^0)$ above $10^{-6}$ are typically in
conflict with the constraints from $R_b^0$, $A_b$, and $A_{\rm FB}^{0,b}$. This correlation arises from the presence of $F(c_{Q_3})$ in (\ref{eq:ZbbRS}) and (\ref{eq:tcZRS}). Making the profile $F (c_{Q_3})$ bigger by locating $(t_L,b_L)$ closer to the IR brane enhances ${\cal B}(t\to c Z^0)$, but at the same time this leads to a positive shift in $g_L^b$, which worsens the quality of the fit in
the $Z^0\to b\bar b$ sector. To elucidate this feature, we depict
in Figure~\ref{fig:PLRplots} the left-right polarization asymmetry
\beq\label{eq:tcZasymmetry} 
   P_{\rm LR}(t\to c Z^0) 
   = \frac{\Gamma(t_L\to c_L Z^0) - \Gamma(t_R\to c_R Z^0)}%
          {\Gamma(t_L\to c_L Z^0) + \Gamma(t_R\to c_R Z^0)} 
   = \frac{|(g_L^u)_{23}|^2 - |(g_R^u)_{23}|^2}%
          {|(g_L^u)_{23}|^2 + |(g_R^u)_{23}|^2}
\eeq 
as a function of ${\cal B}(t\to c Z^0)$. The left and right panels represent the RS results before and after removing the points that fail to reproduce $R_b^0$, $A_b$, and $A_{\rm FB}^{0,b}$ within $99 \%$ probability. The density of points\footnote{It is not possible to attach a statistically rigorous meaning to the shown distributions, since they depend on the way in which the parameter space is sampled and the results are binned.}  
in each bin is indicated by the color shading. The triangle and stars indicate the values $P_{LR}(t\to c Z^0)=0.44$ and $P_{LR}(t\to c Z^0)=0.41$ corresponding to our RS reference point with $\Mkk=1.5$\,TeV and $\Mkk=3$\,TeV. The distribution of points, with the largest concentration arising close to $P_{LR}(t\to c Z^0)=\pm 1$, can be understood by recalling that only the product $|F(c_{u_3})\,F(c_{Q_3})|$ is constrained by the requirement to reproduce the observed top-quark mass. From the structure of (\ref{eq:tcZRS}) and the fact that the contributions proportional to $m_Z^2$ and $m_c m_t$ are typical of the same size, it then follows that the magnitudes of $(g_{L,R}^u)_{23}$ are strongly anti-correlated. Cases in which $|(g_L^u)_{23}|\approx |(g_R^u)_{23}|$, corresponding to a left-right polarization asymmetry close to zero, thus require a tuning of the elements of the up-type Yukawa matrix. Notice also that on average the coupling $(g_L^u)_{23}$ tends to be larger in magnitude than $(g_R^u)_{23}$ due to the ratio of prefactors $(1/2-2/3\,\sin^2\theta_W)/(-2/3\,\sin^2\theta_W)\approx -2.2$ appearing in (\ref{eq:tcZRS}). This explains why the concentration of points in the left plot is most dense around $P_{LR}(t\to c Z^0)=+1$. 

The constraints from $R_b^0$, $A_b$, and $A_{\rm FB}^{0,b}$ restrict more strongly the magnitude of the left-handed than that of the right-handed coupling and so tend to exclude points close to $P_{LR}(t\to c Z^0)=+1$ and branching ratios above $10^{-6}$. This reduces the density of points corresponding to almost purely left-handed $Z^0 tc$ interactions, as can be seen from the right plot. In consequence, for the phenomenologically interesting range of ${\cal B}(t\to c Z^0)$ above $10^{-6}$, the right-handed coupling more frequently gives the dominant contribution to the $t\to c Z^0$ decay width. The same conclusion has been reached in \cite{Agashe:2006wa}, although the correlation between the $t\to c Z^0$ and the $Z^0\to b\bar b$ transition has not been studied there. The constraints from $R_b^0$, $A_b$, and $A_{\rm FB}^{0,b}$ have also not been considered in \cite{Chang:2008zx}, which finds that ${\cal B}(t\to c Z^0)$ is dominated by the left-handed coupling. The wide spread of points in our plots suggests however that the RS framework {\it per se\/} does not lead to a firm prediction of the chirality of the $Z^0  tc$ interactions. Since the RS predictions for ${\cal B}(t\to c Z^0)$ are typically below the expected LHC sensitivity, the prospects of angular analyses \cite{Hubaut:2005er} that would allow to determine the chirality of the $Z^0  tc$ interactions seem in any case quite challenging.

Important constraints on the structure and size of flavor-changing top-quark couplings also arise from $B$ physics. In particular, the constraints from $B\to X_s\gamma$ and $B\to X_s\,l^+ l^-$ decay disfavor flavor-changing top-quark interactions at a level observable at the LHC due to most of the operators containing left-handed up- or charm-quark fields \cite{Fox:2007in}. It is easy to understand that in the minimal RS framework the correlation between the $Z^0 s\bar b$ couplings $(g_{L,R}^d)_{23}$ affecting $B\to X_s\,l^+ l^-$ and $(g_{L,R}^u)_{23}$ entering $t\to c Z^0$ provides the strongest constraint on the potential size of ${\cal B}(t\to c Z^0)$.\footnote{In our analysis we neglect semileptonic four-fermion operators like $(\bar b s) (\bar l l)$ arising from integrating out the KK modes of the $Z^0$ boson. This is justified, since these contributions are suppressed by the logarithm of the warp factor with respect to the $Z^0 s\bar b$ couplings.} 
Corrections due to anomalous $Z^0 t\bar t$ and $W t b$ couplings enter $B\to X_s\,l^+ l^-$ first at the loop level and are therefore subleading. Within experimental and theoretical errors, the $Z^0 s\bar b$ couplings $(g_{L,R}^d)_{23}$ are constrained at the level of a few $10^{-4}$ by $B\to X_s\,l^+ l^-$ decay. We have verified that the vast majority of points that satisfy the $Z^0 \to b \bar b$ constraints lead to values of ${\cal B}(B\to X_s\,l^+ l^-)$ that lie safely within the allowed range. This feature is expected in models in which the modification of the flavor structure is connected to the third generation \cite{Haisch:2007ia}. Our analysis shows that values of the $t\to c Z^0$ branching ratio of up to $10^{-5}$ are not in conflict with any currently available information on $Z^0$-penguin transitions.

The branching ratio of $t\to u Z^0$ is typically suppressed by two orders of magnitude compared to the one of $t\to c Z^0$. This suppression factor is readily understood from the scaling $|F(c_{Q_1})|/|F(c_{Q_2})|\sim\lambda$ of the fermion profiles and the smallness of the up-quark mass, which renders effects due to mixing of zero and KK fermion modes negligibly small. The structure of (\ref{eq:tcZRS}) then implies that generically $|(g_L^u)_{13}|\gg |(g_R^u)_{13}|$, so that the chirality of the $Z^0  tu$  interactions is preferable left-handed. Unfortunately, in view of the smallness of ${\cal B}(t\to u Z^0)$, it seems impossible to test this prediction at the LHC.

We conclude the discussion of FCNC top-quark decays to $Z^0$ bosons by noting that in RS models with custodial symmetry, the observed correlations between the $t\to c (u) Z^0$, $Z^0 \to b\bar b$, and $b\to s Z^0$ transitions can be less pronounced. This is a consequence of the left-right exchange symmetry, which allows to suppress tree-level corrections to the left-handed $Z^0 b \bar b$ vertex, while at the same time leaving the $t\to c (u) Z^0$ and $b\to s Z^0$ transitions unprotected \cite{Agashe:2006at}. We thus expect that the RS predictions for ${\cal B} (t \to c (u) Z^0)$ are typically larger in these scenarios and captured by the plots on the left-hand side of Figures~\ref{fig:tcZplots} and \ref{fig:PLRplots}. As a result, the experimental prospects for observing the decays $t\to c (u) Z^0$ seem more favorable. Similar statements apply to the case of the $t\to c(u) h$ processes discussed in the next section.

\subsection{Flavor-Changing Higgs-Boson Couplings}
\label{sec:tch}

The general form of the interactions of fermions with the Higgs boson has been given in (\ref{eq:hff}). We will express the couplings $(g_h^q)_{mn}$ in this relation in units of quark masses divided by the Higgs vacuum expectation value. After adjusting the phase of the SM quark fields according to the standard CKM phase convention, we obtain for our default RS parameters
\beq\label{eq:ghuddefault}
\begin{split}
   (g_h^u)_{ij} 
   &= \frac{\sqrt{m_{u_i} m_{u_j}}}{v} \left(
    \begin{array}{ccc}
     1.000 & ~~ -0.00198\,e^{-i\,3.9^\circ} ~~
      & 0.0224\,e^{i\,0.7^\circ} \\
    -3.40\cdot 10^{-6}\,e^{i\,3.5^\circ} & 1.000
      & 0.00299\,e^{i\,31.5^\circ} \\
    -7.94\cdot 10^{-4}\,e^{-i\,61.7^\circ}
     & -0.00431\,e^{i\,85.2^\circ} & 0.923
    \end{array} \right)_{ij} , \\
   (g_h^d)_{ij} 
   &= \frac{\sqrt{m_{d_i} m_{d_j}}}{v} \left(
    \begin{array}{ccc}
     1.000 & -1.66\cdot 10^{-4}\,e^{i\,60.2^\circ}
      & 0.00476\,e^{i\,82.2^\circ} \\
     -9.98\cdot 10^{-6}\,e^{-i\,60.2^\circ} & 1.000
      & 0.00356\,e^{i\,43.0^\circ} \\
     4.27\cdot 10^{-6}\,e^{-i\,83.1^\circ}
      & 7.83\cdot 10^{-5}\,e^{-i\,44.2^\circ} & 0.998
    \end{array} \right)_{ij} , 
\end{split}
\eeq
with $m_{q_i}$ given in (\ref{eq:massesexact}). The elements $(g_h^{q})_{i3}$ deviate notably from the SM expressions $(g_h^q)_{33}=m_{q_3}/v$ and $(g_h^{q})_{13}=(g_h^{q})_{23}=0$. In particular, the $ht\bar t$ coupling can receive sizable corrections of order $-10\%$. A smaller $ht\bar t$ coupling would lead to a reduction of the rate for the gluon-fusion process $gg\to h$ via a top-quark loop, resulting in a suppression of the Higgs-boson production cross section. Using the di-photon or the $W^{\pm} W^{\mp}$, $Z^0 Z^0$ channels at the LHC, the theoretical estimate of the Higgs-boson production cross section \cite{Harlander:2002wh, Anastasiou:2002yz} may be confronted with experiment. As both experiment and theory are limited to an accuracy of about 10\%, it will be challenging to detect the predicted suppression effects of the $ht\bar t$ coupling. A detailed study of the Higgs-boson production cross section in warped 5D scenarios, as performed in \cite{Djouadi:2007fm, Falkowski:2007hz}, is beyond the scope of this article.

When kinematically accessible, the couplings of (\ref{eq:hff}) allow
for the flavor-changing decay $t\to c h$. Including terms up to first
order in the charm-quark mass, the corresponding branching ratio takes
the form
\begin{eqnarray}
   {\cal{B}}(t\to c h) 
   = \frac{2 \left( 1-r_h^2 \right)^2 r_W^2}%
          {\left( 1-r_W^2 \right)^2 \left( 1+2r_W^2 \right) g^2} 
    \left\{ \left| \left( g_h^u \right)_{23} \right|^2
    + \left| \left( g_h^u \right)_{32} \right|^2 
    + \frac{4r_c}{1-r_h^2}\,
    \mbox{Re}\big[ \left( g_h^u \right)_{23}
     \left( g_h^u \right)_{32} \big] \right\} , \hspace{8mm}
\end{eqnarray}
where as before $r_i\equiv m_i^{\rm pole}/m_t^{\rm pole}$, and $g$ is the $SU(2)_L$ gauge coupling. In our numerical analysis of the $t\to c h$ branching ratio we will use $r_h=0.87$, corresponding to a Higgs-boson mass $m_h=150$\,GeV. 

\begin{figure}[!t]
\begin{center} 
\hspace{-2mm}
\mbox{\includegraphics[height=2.85in]{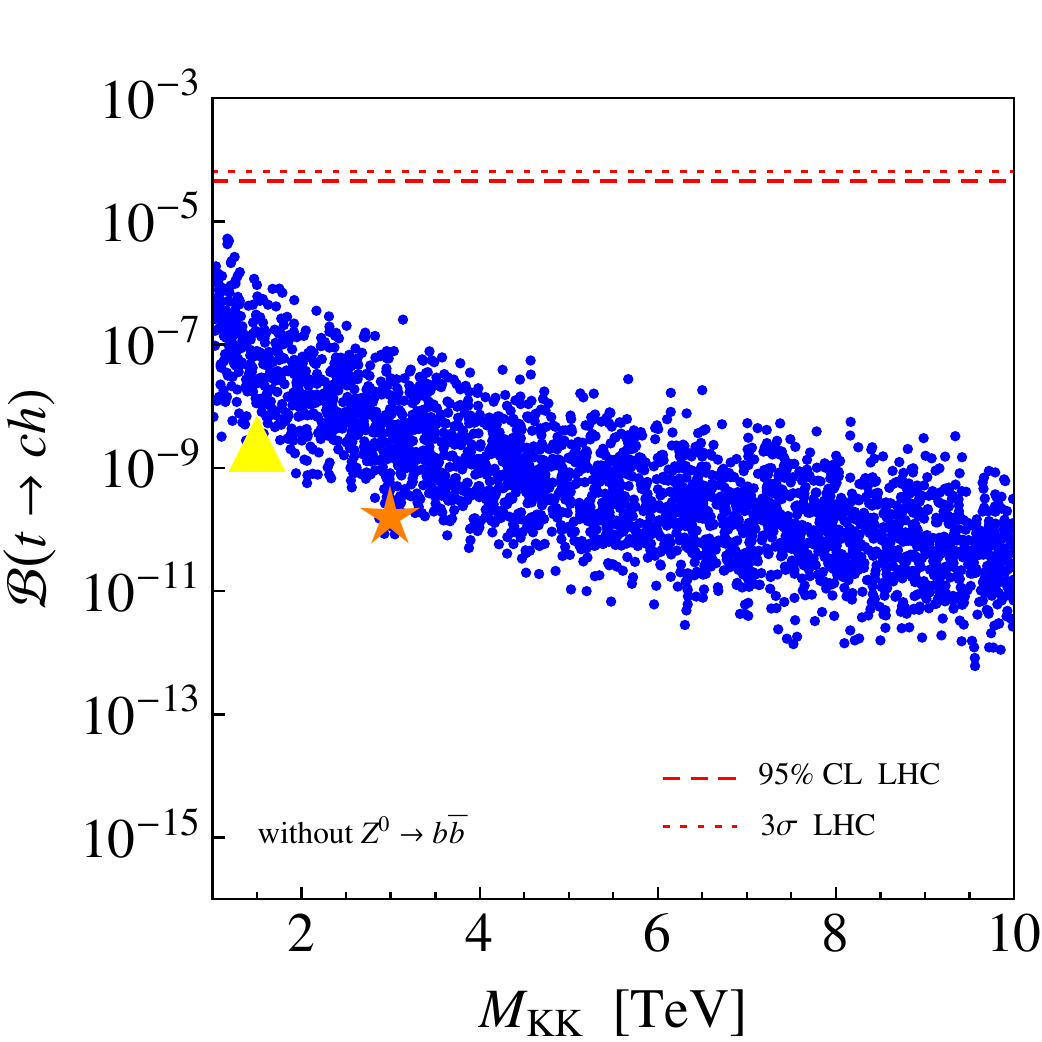}} 
\qquad 
\mbox{\includegraphics[height=2.85in]{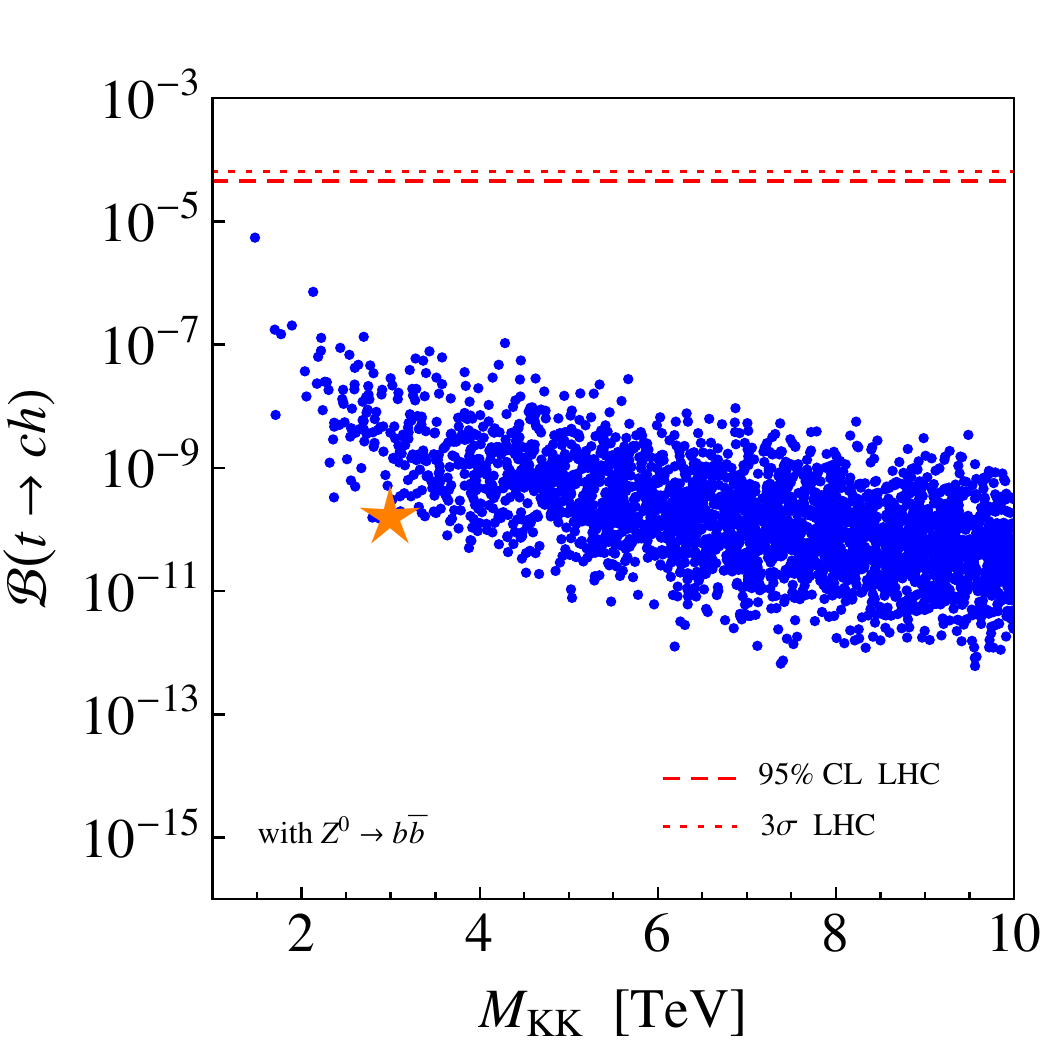}} 
\vspace{-2mm}
\parbox{15.5cm}{\caption{\label{fig:tchplots}
Branching ratio of $t\to c h$ as a function of $\Mkk$ in the RS model. The dashed and dotted red (dark gray) lines indicate the expected sensitivities of the LHC. In the left (right) plot points that violate the constraints from the $Z^0\to b\bar b$ ``pseudo observables'' are included (excluded). The yellow (light gray) triangle and the orange (medium gray) stars represent the results for our RS reference points. See text for details.}}
\end{center}
\end{figure}

The RS predictions for ${\cal B}(t\to c h)$ as a function of $\Mkk$ are shown in Figure~\ref{fig:tchplots} for 3000 randomly generated parameter points. The values of ${\cal B}(t\to c h)=2.4\cdot 10^{-9}$ and ${\cal B}(t\to c h)=1.7\cdot 10^{-10}$ corresponding to our default RS point for $\Mkk=1.5$\,TeV and $\Mkk=3$\,TeV are displayed by the triangle and stars. The LHC is expected to give $3\sigma$ evidence for ${\cal B}(t\to ch)$ larger than $6.5\cdot 10^{-5}$ or set a limit of $4.5\cdot 10^{-5}$ with 95\% CL if the decay is not observed \cite{AguilarSaavedra:2000aj}. These limits are indicated by the dashed and dotted lines in the panels. Since ${\cal B}(t\to c h)$ barely reaches values of $10^{-7}$ once the $Z^0\to b\bar b$ constraints are enforced, a detection of a RS signal for the $t\to c h$ decay will be taxing at the LHC. The branching ratio of $t\to u h$ is suppressed relative to the one of $t\to c h$ by more than an order of magnitude, so that a detection of $t \to uh$ at the LHC seems impossible in the minimal RS framework. Due to the weaker correlation between $t\to c(u) h$ and $Z^0\to b \bar b$ in RS models with custodial symmetry, the prospects for observing the decays $t\to c(u) h$ are likely to be better in warped scenarios with extended gauge symmetry.

\section{Conclusions and Outlook}
\label{sec:concl}

We have presented a detailed and comprehensive discussion of tree-level flavor-changing effects in the RS model with gauge and matter fields in the bulk and the Higgs sector localized on the IR brane. We have derived exact expressions for the masses of the gauge bosons and fermions and their KK excitations, as well as for the profiles of these fields along the extra dimension, by solving the bulk equations of motions with appropriate boundary conditions taking into account the couplings to the Higgs sector. In all cases analytical expressions are obtained for the bulk profiles of the fields in the effective 4D theory, including their normalization.

For the gauge fields we have discussed the KK decomposition in a covariant $R_\xi$ gauge. A single gauge-fixing term in the 5D action suffices to define the theory completely. The resulting Faddeev-Popov ghost Lagrangian contains a ghost field of the usual form for each KK mode. We have derived compact expressions accounting for sums over the KK towers of gauge bosons arising in tree-level diagrams at low energy. For fermions the KK decomposition is performed including the mixing between different generations sourced by the Yukawa couplings. Our exact results allow us to study flavor mixing not only among the lowest-lying (SM) fermions, but also among their KK excitations. We have also commented on dimensional-analysis constraints on the scale of the 5D Yukawa matrices, finding that these could in principle be larger than commonly assumed in the literature. Potential phenomenological implications of this observation will be discussed elsewhere \cite{mytalks,inprep}.

Our exact treatment of the field equations, which avoids a perturbative expansion in powers of $v^2/\Mkk^2$, extends previous analyses of flavor physics in the RS models in several ways. An important observation of our analysis is that the field equations in the fermion sector are incompatible with the naive orthonormality relations usually imposed on the bulk profiles. The properly generalized normalization conditions allow us to derive exact and compact expressions for flavor-changing effects arising from the mixing of $SU(2)_L$ doublet and singlet fermions via their KK excitations. These effects give rise to FCNC couplings of the $Z^0$ and Higgs bosons, which so far have not been studied systematically in the literature. 

The hierarchies observed in the fermion spectrum and the CKM matrix can be explained naturally in terms of anarchic 5D Yukawa matrices and wave-function overlap integrals. We have emphasized the invariance of the results for these masses and mixings under two types of reparametrization transformations of the 5D bulk mass parameters and Yukawa matrices. We have also pointed out that the KK excitations of the SM fermions have generically large mixings between different generations as well as between $SU(2)_L$ singlets and doublets. The reason is that without the Yukawa couplings the spectra of the KK towers of different fermion fields are nearly degenerate, so that even small Yukawa couplings can lead to large mixing effects, which could be an important source of flavor-changing loop effects. 

The heart of our analysis is a comprehensive, quantitative study of the structure of gauge and Higgs boson interactions with SM fermions and their KK excitations. In particular, we have investigated in detail the flavor-changing couplings of the $W^\pm$ and $Z^0$ bosons to SM fermions. They give rise to the leading flavor-changing effects in $\Delta F=1$ weak decay processes. The two main sources of non-standard flavor violations are the deviation of the $Z^0$-boson profiles along the extra dimension from a constant, and the $SU(2)_L$ singlet admixture in the wave function of the left-handed SM fermions. While the second effect is suppressed by two powers of light quark masses and has thus frequently been neglected in the literature, we have shown that it is parametrically and numerically as important as the first one. 

Performing a careful analysis of electroweak precision observables, including the $S$ and $T$ parameters and the $Z^0 b\bar b$ couplings, we have found that the simplest RS model containing only SM particles and their KK excitations is consistent with all experimental bounds for reasonably low KK scales if one allows for a heavy Higgs boson ($m_h\lesssim 1$\,TeV). In some regards, this model is even preferred over models with a custodial $SU(2)_R$ symmetry and an extended fermion spectrum. The reason is that the Higgs boson is naturally heavy in the RS framework, since radiative corrections to the Higgs-boson mass scale like the fourth power of the UV cutoff on the IR brane. This gives rise to a large negative contribution to the $T$ parameter, which can be compensated in the RS model without custodial symmetry by a large positive tree-level contribution.

We have concluded this work with a rather detailed study of tree-level 
flavor-changing effects, which includes analyses of the non-unitarity of the quark mixing matrix, anomalous right-handed couplings of the $W^\pm$ bosons, tree-level FCNC couplings of the $Z^0$ and Higgs bosons, the rare decays $t\to c(u) Z^0$ and $t\to c(u) h$, and the flavor mixing among KK fermions. The analytical and numerical results obtained in this paper form the basis for general calculations of flavor-changing processes in the RS model and its extensions, including loop-mediated decays. A comprehensive study of flavor effects in the $B$-, $D$-, and $K$-meson systems will be presented in a companion paper \cite{inprep}.

\subsubsection*{Acknowledgments}

We are grateful to Babis Anastasiou, Martin Bauer, Roberto Contino, Georgi Dvali, Christophe Grojean, Leonhard Gr\"under, Jos\'e Santiago, and Andreas Weiler for useful discussions, and to Donatello Dolce and Adam Falkowski for sharing unpublished notes and private communications. One of us (MN) likes to thank the Kavli Institute for Theoretical Physics at UC Santa Barbara for hospitality and support during two visits while this paper was being written.

\newpage
\begin{appendix}

\begin{landscape}

\section{Textures of Mixing Matrices}
\label{app:textures}

\renewcommand{\theequation}{A\arabic{equation}}
\setcounter{equation}{0}

It is useful to have explicit expressions for the various mixing matrices at hand, which include the relevant combinations of Yukawa couplings. We work in the ZMA and use the relations given in (\ref{ZMA1}), (\ref{ZMA2}), and (\ref{VRres}) along with the scaling relations derived in Section~\ref{sec:quark}. For simplicity, we evaluate polynomial factors involving the $c_i$ parameters by setting $c_i=-1/2$, which is a good approximation for all light fermions. It would be straightforward to reinstate the correct expressions if desired. 

To leading order in hierarchies, the matrices $\bm{V}_{L,R}$ parametrizing the couplings of the $W^\pm$ bosons to fermions in (\ref{VLVRdef}) are given by
\beq
\begin{split} 
   (V_L)_{ij}
   &\simeq \left( \begin{array}{ccc}
    1 & 
    \left[ \frac{(M_d)_{21}}{(M_d)_{11}} 
     - \frac{(M_u)_{21}}{(M_u)_{11}} \right] 
     \frac{F_{Q_1}}{F_{Q_2}} & 
    \left[ \frac{(M_u)_{31}}{(M_u)_{11}} 
     + \frac{(Y_d)_{13}}{(Y_d)_{33}} 
     - \frac{(M_u)_{21} (Y_d)_{23}}{(M_u)_{11} (Y_d)_{33}} \right] 
     \frac{F_{Q_1}}{F_{Q_3}} \\ 
    \left[ \frac{(M_u)_{21}^*}{(M_u)_{11}^*} 
     - \frac{(M_d)_{21}^*}{(M_d)_{11}^*} \right] 
     \frac{F_{Q_1}}{F_{Q_2}} & 
    1 &
    \left[ \frac{(Y_d)_{23}}{(Y_d)_{33}} 
     - \frac{(Y_u)_{23}}{(Y_u)_{33}} \right] 
     \frac{F_{Q_2}}{F_{Q_3}} \\ 
    \left[ \frac{(M_d)_{31}^*}{(M_d)_{11}^*} 
     + \frac{(Y_u)_{13}^*}{(Y_u)_{33}^*} 
     - \frac{(M_d)_{21}^* (Y_u)_{23}^*}{(M_d)_{11}^* (Y_u)_{33}^*}
     \right] \frac{F_{Q_1}}{F_{Q_3}} &
    \left[ \frac{(Y_u)_{23}^*}{(Y_u)_{33}^*} 
     - \frac{(Y_d)_{23}^*}{(Y_d)_{33}^*} \right] 
     \frac{F_{Q_2}}{F_{Q_3}} & 
    1 \\ 
   \end{array} \right)_{ij} \! , \\
   (V_R)_{ij}
   &\simeq \frac{m_{u_i} m_{d_j}}{2\Mkk^2}
    \left( \begin{array}{ccc}
    \frac{1}{F_{Q_1}^2} & 
     \frac{(M_d)_{21}}{(M_d)_{11}}\,\frac{1}{F_{Q_1} F_{Q_2}} & 
     \frac{(Y_d)_{13}}{(Y_d)_{33}}\,\frac{1}{F_{Q_1} F_{Q_3}} \\
    \frac{(M_u)_{21}^*}{(M_u)_{11}^*}\,\frac{1}{F_{Q_1} F_{Q_2}} & 
     \left[ 1 
      + \frac{(M_u)_{21}^* (M_d)_{21}^*}{(M_u)_{11}^* (M_d)_{11}^*} 
      \right] \frac{1}{F_{Q_2}^2} &
     \left[ \frac{(Y_d)_{23}}{(Y_d)_{33}} + 
      \frac{(Y_d)_{13} (M_u)_{21}^*}{(Y_d)_{33} (M_u)_{11}^*} \right]
      \frac{1}{F_{Q_2} F_{Q_3}} \\
   \frac{(Y_u)_{13}^*}{(Y_u)_{33}^*}\,\frac{1}{F_{Q_1} F_{Q_3}} & 
    \left[ \frac{(Y_u)_{23}^*}{(Y_u)_{33}^*} 
     + \frac{(Y_u)_{13}^* (M_d)_{21}}{(Y_u)_{33}^* (M_d)_{11}}
     \right] \frac{1}{F_{Q_2} F_{Q_3}} & 
    \left[ 1 + 
     \frac{(Y_u)_{13}^* (Y_d)_{13}}{(Y_u)_{33}^* (Y_d)_{33}}
     + \frac{(Y_u)_{23}^* (Y_d)_{23}}{(Y_u)_{33}^* (Y_d)_{33}}
     \right] \frac{1}{F_{Q_3}^2} - 1 + \frac{F_{Q_3}^2}{2} 
   \end{array} \right)_{ij} \! .
\end{split}
\eeq
The corresponding expressions for the matrices $\bm{\Delta}_A^{(\prime)}$ entering the $Z^0$-boson couplings to fermions in (\ref{gLR}) read
\beq  
\begin{split}
   (\Delta_Q^{(\prime)})_{ij}
   &\simeq \frac12 \left( \begin{array}{ccc}
    \left[ 1 + \frac{|(M_q)_{21}|^2}{|(M_q)_{11}|^2} 
     + \frac{|(M_q)_{31}|^2}{|(M_q)_{11}|^2} \right] F_{Q_1}^2  & 
    - \left[ \frac{(M_q)_{21}}{(M_q)_{11}} 
     + \frac{(Y_q)_{23}^* (M_q)_{31}}{(Y_q)_{33}^* (M_q)_{11}} \right]
     F_{Q_1} F_{Q_2} & 
    \frac{(M_q)_{31}}{(M_q)_{11}}\,F_{Q_1} F_{Q_3} \\ 
    - \left[ \frac{(M_q)_{21}^*}{(M_q)_{11}^*} 
     + \frac{(Y_q)_{23} (M_q)_{31}^*}{(Y_q)_{33} (M_q)_{11}^*} \right]
     F_{Q_1} F_{Q_2} & 
    \left[ 1 + \frac{|(Y_q)_{23}|^2}{|(Y_q)_{33}|^2} \right] 
     F_{Q_2}^2  & 
     - \frac{(Y_q)_{23}}{(Y_q)_{33}}\,F_{Q_2} F_{Q_3} \\ 
    \frac{(M_q)_{31}^*}{(M_q)_{11}^*}\,F_{Q_1} F_{Q_3} & 
    - \frac{(Y_q)_{23}^*}{(Y_q)_{33}^*}\,F_{Q_2} F_{Q_3} &
    F_{Q_3}^2 \\ 
   \end{array} \right)_{ij} \! , \\
   (\Delta_q^{(\prime)})_{ij}
   &\simeq \frac12\,e^{-i(\phi_i-\phi_j)}
    \left( \begin{array}{ccc}
    \left[ 1 + \frac{|(M_q)_{12}|^2}{|(M_q)_{11}|^2} 
     + \frac{|(M_q)_{13}|^2}{|(M_q)_{11}|^2} \right] F_{q_1}^2  & 
    - \left[ \frac{(M_q)_{12}^*}{(M_q)_{11}^*} 
     + \frac{(Y_q)_{32} (M_q)_{13}^*}{(Y_q)_{33} (M_q)_{11}^*}
     \right] F_{q_1} F_{q_2} & 
    \frac{(M_q)_{13}^*}{(M_q)_{11}^*}\,F_{q_1} F_{q_3} \\ 
    - \left[ \frac{(M_q)_{12}}{(M_q)_{11}} 
     + \frac{(Y_q)_{32}^* (M_q)_{13}}{(Y_q)_{33}^* (M_q)_{11}} \right]
     F_{q_1} F_{q_2} & 
    \left[ 1 + \frac{|(Y_q)_{32}|^2}{|(Y_q)_{33}|^2} \right] 
     F_{q_2}^2  & 
     - \frac{(Y_q)_{32}^*}{(Y_q)_{33}^*}\,F_{q_2} F_{q_3} \\ 
    \frac{(M_q)_{13}}{(M_q)_{11}}\,F_{q_1} F_{q_3} & 
    - \frac{(Y_q)_{32}}{(Y_q)_{33}}\,F_{q_2} F_{q_3} &
    F_{q_3}^2 \\ 
   \end{array} \right)_{ij} \! .
\end{split}
\eeq
Finally, the matrices $\bm{\delta}_A$ entering the same relations are given by
\beq
\begin{split}
   (\delta_Q)_{ij}
   &\simeq e^{-i(\phi_i-\phi_j)}\;  
    \frac{m_{q_i} m_{q_j}}{2\Mkk^2} \left( \begin{array}{ccc}
    \frac{1}{F_{q_1}^2}  & 
    \frac{(M_q)_{12}^*}{(M_q)_{11}^*}\,\frac{1}{F_{q_1} F_{q_2}} & 
    \frac{(Y_q)_{31}^*}{(Y_q)_{33}^*}\,\frac{1}{F_{q_1} F_{q_3}} \\ 
    \frac{(M_q)_{12}}{(M_q)_{11}}\,\frac{1}{F_{q_1} F_{q_2}} & 
    \left[ 1 + \frac{|(M_q)_{12}|^2}{|(M_q)_{11}|^2} \right] 
     \frac{1}{F_{q_2}^2}  & 
    \left[ \frac{(Y_q)_{32}^*}{(Y_q)_{33}^*}
     + \frac{(Y_q)_{31}^* (M_q)_{12}}{(Y_q)_{33}^* (M_q)_{11}}
     \right] \frac{1}{F_{q_2} F_{q_3}} \\ 
    \frac{(Y_q)_{31}}{(Y_q)_{33}}\,\frac{1}{F_{q_1} F_{q_3}} &
    \left[ \frac{(Y_q)_{32}}{(Y_q)_{33}}
     + \frac{(Y_q)_{31} (M_q)_{12}^*}{(Y_q)_{33} (M_q)_{11}^*}
     \right] \frac{1}{F_{q_2} F_{q_3}} & 
    \left[ 1 + \frac{|(Y_q)_{31}|^2}{|(Y_q)_{33}|^2}
     + \frac{|(Y_q)_{32}|^2}{|(Y_q)_{33}|^2} \right]
     \frac{1}{F_{q_3}^2} - 1 + \frac{F_{q_3}^2}{2} \\ 
   \end{array} \right)_{ij} \! , \\
   (\delta_q)_{ij}
   &\simeq \frac{m_{q_i} m_{q_j}}{2\Mkk^2} \left( \begin{array}{ccc}
    \frac{1}{F_{Q_1}^2}  & 
    \frac{(M_q)_{21}}{(M_q)_{11}}\,\frac{1}{F_{Q_1} F_{Q_2}} & 
    \frac{(Y_q)_{13}}{(Y_q)_{33}}\,\frac{1}{F_{Q_1} F_{Q_3}} \\ 
    \frac{(M_q)_{21}^*}{(M_q)_{11}^*}\,
     \frac{1}{F_{Q_1} F_{Q_2}} & 
    \left[ 1 + \frac{|(M_q)_{21}|^2}{|(M_q)_{11}|^2} \right] 
     \frac{1}{F_{Q_2}^2}  & 
    \left[ \frac{(Y_q)_{23}}{(Y_q)_{33}}
     + \frac{(Y_q)_{13} (M_q)_{21}^*}{(Y_q)_{33} (M_q)_{11}^*}
     \right] \frac{1}{F_{Q_2} F_{Q_3}} \\ 
    \frac{(Y_q)_{13}^*}{(Y_q)_{33}^*}\,
     \frac{1}{F_{Q_1} F_{Q_3}} &
    \left[ \frac{(Y_q)_{23}^*}{(Y_q)_{33}^*}
     + \frac{(Y_q)_{13}^* (M_q)_{21}}{(Y_q)_{33}^* (M_q)_{11}}
     \right] \frac{1}{F_{Q_2} F_{Q_3}} & 
    \left[ 1 + \frac{|(Y_q)_{13}|^2}{|(Y_q)_{33}|^2}
     + \frac{|(Y_q)_{23}|^2}{|(Y_q)_{33}|^2} \right]
     \frac{1}{F_{Q_3}^2} - 1 + \frac{F_{Q_3}^2}{2} \\ 
   \end{array} \right)_{ij} \! .
\end{split}
\eeq
In these formulae $F_{Q_i}\equiv F(c_{Q_i})$ and $F_{q_i}\equiv F(c_{q_i})$. The extra terms in the 33 entries of the matrices $\bm{V}_R$ and $\bm{\delta}_{Q,q}$ should be kept in cases where $F_{Q_3,q_3}=\ord(1)$. The phase factors $e^{i\phi_j}$ are defined in (\ref{eq:expphij}).

\section{Reference Values for SM Parameters}
\label{app:masses}

\renewcommand{\theequation}{B\arabic{equation}}
\setcounter{equation}{0}

The central values and errors of the quark masses used in our analysis are
\beq\label{eq:fitmasses}
\begin{aligned}
   m_u &= (1.5\pm 1.0)\,\mbox{MeV} \,, & \qquad 
   m_c &= (550\pm 40)\,\mbox{MeV} \,, & \qquad 
   m_t &= (140\pm 5)\,\mbox{GeV} \,, \\
   m_d &= (3.0\pm 2.0)\,\mbox{MeV} \,, & \qquad 
   m_s &= (50\pm 15)\,\mbox{MeV} \,, & \qquad 
   m_b &= (2.2\pm 0.1)\,\mbox{GeV} \,.
\end{aligned}
\eeq
They correspond to $\overline{\rm MS}$ masses evaluated at the scale
$\Mkk=1.5$\,TeV, obtained by using the low-energy values as compiled in \cite{Yao:2006px}. The central values and errors of the Wolfenstein parameters are taken from \cite{Charles:2004jd} and read
\beq\label{eq:fitwolf}
   \lambda = 0.2265\pm 0.0008 \,, \qquad 
   A = 0.807\pm 0.018 \,, \qquad
   \bar{\rho} = 0.141\,_{-0.017}^{+0.029} \,, \qquad 
   \bar{\eta} = 0.343\pm  0.016 \,.
\eeq
The central values and errors for the parameters entering our analysis of electroweak precision observables are \cite{LEPEWWG:2005ema, Group:2008nq}
\beq
\begin{aligned}
   \Delta\alpha^{(5)}_{\rm had}(m_Z) 
   &= 0.02758\pm 0.00035 \,, \qquad
   & m_Z &= (91.1875\pm 0.0021)\,\mbox{GeV} \,, \\
   \alpha_s(m_Z) &= 0.118\pm 0.003 \,, \qquad
   & m_t &= (172.6\pm 1.4)\,\mbox{GeV} \,.
\end{aligned}
\eeq
We refer to the central values for these quantities as SM reference values. Unless noted otherwise, the reference value for the Higgs-boson mass is $m_h=150$\,GeV.

\end{landscape}

\section{Mixing Matrices with Default Parameters}
\label{app:results}

\renewcommand{\theequation}{C\arabic{equation}}
\setcounter{equation}{0}

Here we collect results for some of the flavor-mixing matrices, which were not given in the main text. These results are obtained using the default parameters in (\ref{eq:cparameter}) and (\ref{eq:yukawas}), along with the KK scale $\Mkk=1.5$\,TeV. They refer to the standard CKM phase conventions.

The result for the right-handed analog of CKM matrix, which parametrizes the non-standard couplings of $W^\pm$ bosons to right-handed SM quark fields in (\ref{Wff}), is
\beq
   \bm{V}_R = \left(
    \begin{array}{ccc}
     1.79\cdot 10^{-9} & 8.40\cdot 10^{-10} & -3.46\cdot 10^{-8} \\
     -1.83\cdot 10^{-7}\,e^{-i\,38.6^\circ}
      & -5.08\cdot 10^{-7}\,e^{-i\,7.9^\circ} & 4.70\cdot 10^{-6} \\
     -5.33\cdot 10^{-6}\,e^{i\,12.7^\circ}
      & ~~ -1.17\cdot 10^{-5}\,e^{-i\,33.4^\circ} ~~ 
      & 1.20\cdot 10^{-3}
    \end{array}
   \right) .
\eeq
Potential effects of the largest coupling, $(V_R)_{33}$, have been discussed in Section~\ref{sec:631}.

The results for the matrices $\bm{\Delta}'$, which enter the $Z^0$-boson couplings to fermions in (\ref{Zff}), read
\beq
\begin{split}
   \bm{\Delta}_D' &= 10^{-3} \left(
    \begin{array}{ccc}
     0.298 & 0.233\,e^{-i\,5.6^\circ} & -0.754\,e^{i\,71.5^\circ} \\
     0.233\,e^{i\,5.6^\circ} & 7.854 & -4.175\,e^{-i\,9.4^\circ} \\
     -0.754\,e^{-i\,71.5^\circ} & ~~ -4.175\,e^{i\,9.4^\circ} ~~
      & 30.24
    \end{array}
    \right) , \\
   \bm{\Delta}_d' &= 10^{-4} \left(
    \begin{array}{ccc}
     0.00178 & ~~ 0.00182\,e^{-i\,31.6^\circ} ~~
      & 0.115\,e^{i\,72.5^\circ} \\
     0.00182\,e^{i\,31.6^\circ} & 0.0146 & 0.224\,e^{i\,70.7^\circ} \\
     0.115\,e^{-i\,72.5^\circ} & 0.224\,e^{-i\,70.7^\circ} & 12.12
    \end{array}
    \right) , \\
   \bm{\Delta}_U' &= 10^{-3} \left(
    \begin{array}{ccc}
     0.782 & 1.817\,e^{-i\,1.6^\circ} & -1.306\,e^{i\,24.2^\circ} \\
     1.817\,e^{i\,1.6^\circ} & 7.065 & -3.096\,e^{-i\,17.1^\circ} \\
     -1.306\,e^{-i\,24.2^\circ} & ~~ -3.096\,e^{i\,17.1^\circ} ~~
      & 29.02
    \end{array}
    \right) , \\
   \bm{\Delta}_u' &= 10^{-4} \left(
    \begin{array}{ccc}
     0.000435 & ~~ -0.0590\,e^{-i\,3.3^\circ} ~~
      & 0.208\,e^{-i\,67.9^\circ} \\
     -0.0590\,e^{i\,3.3^\circ} & 9.539
      & -31.41\,e^{-i\,82.5^\circ} \\
     0.208\,e^{i\,67.9^\circ} & -31.41\,e^{i\,82.5^\circ}
      & 3539.6
    \end{array}
    \right) .
\end{split}
\eeq
Based on the results (\ref{ZMA1}) obtained in the ZMA, these matrices are expected to be approximately equal to the corresponding matrices $\bm{\Delta}_A$. This expectation is supported by the comparison with the results given in Section~\ref{sec:632}.

For completeness, we also list the mixing matrices $\bm{\epsilon}_A^{(\prime)}$. They read
\beq
\begin{split}
   \bm{\epsilon}_D &= 10^{-4} \left(
    \begin{array}{ccc}
     0.117 & 0.223\,e^{i\,60.2^\circ} & -0.951\,e^{i\,82.2^\circ} \\
     0.223\,e^{-i\,60.2^\circ} & 2.042 & -2.889\,e^{i\,42.8^\circ} \\
     -0.951\,e^{-i\,82.2^\circ} & ~~ -2.889\,e^{-i\,42.8^\circ} ~~ 
      & 11.27    
    \end{array} \right) , \\
   \bm{\epsilon}_d &= 10^{-6} \left(
    \begin{array}{ccc}
     0.00202 & ~~ -0.00117\,e^{-i\,31.6^\circ} ~~
      & 0.0392\,e^{i\,80.4^\circ} \\
     -0.00117\,e^{i\,31.6^\circ} & 0.0200
      & 0.136\,e^{-i\,19.5^\circ} \\
     0.0392\,e^{-i\,80.4^\circ} & 0.136\,e^{i\,19.5^\circ} & 10.39    
    \end{array} \right) , \\
   \bm{\epsilon}_U &= 10^{-4} \left(
    \begin{array}{ccc}
     0.232 & 0.557\,e^{-i\,3.9^\circ} & -0.406\,e^{i\,0.7^\circ} \\
     0.557\,e^{i\,3.9^\circ} & 1.542 & -1.049\,e^{i\,29.5^\circ} \\
     -0.406\,e^{-i\,0.7^\circ} & ~~ -1.049\,e^{-i\,29.5^\circ} ~~
      & 9.641 \end{array} \right) , \\
    \bm{\epsilon}_u &= 10^{-6} \left(
     \begin{array}{ccc}
      0.000450 & ~~ -0.0446\,e^{-i\,4.7^\circ} ~~
       & 1.346\,e^{i\,61.8^\circ} \\
      -0.0446\,e^{i\,4.7^\circ} & 7.089 & 143.3\,e^{-i\,85.6^\circ} \\
      1.346\,e^{-i\,61.8^\circ} & 143.3\,e^{i\,85.6^\circ} & 38258.0    
     \end{array} \right) ,
\end{split}
\eeq
and
\beq
\begin{split}
   \bm{\epsilon}_D' &= 10^{-4}\left(
    \begin{array}{ccc}
     0.0853 & 0.162\,e^{i\,60.2^\circ} & -0.690\,e^{i\,82.2^\circ} \\
     0.162\,e^{-i\,60.2^\circ} & 1.490 & -2.103\,e^{i\,42.9^\circ} \\
     -0.690\,e^{-i\,82.2^\circ} & ~~ -2.103\,e^{-i\,42.9^\circ} ~~
      & 8.230
    \end{array} \right) , \\
   \bm{\epsilon}_d' &= 10^{-6} \left(
    \begin{array}{ccc}
     0.00150 & ~~ -0.000869\,e^{-i\,31.6^\circ} ~~
      & 0.0290\,e^{i\,80.4^\circ} \\
     -0.000869\,e^{i\,31.6^\circ} & 0.0149
      & 0.101\,e^{-i\,19.4^\circ} \\
     0.0290\,e^{-i\,80.4^\circ} & 0.101\,e^{i\,19.4^\circ} & 7.746 
    \end{array} \right) , \\
   \bm{\epsilon}_U' &= 10^{-4} \left(
    \begin{array}{ccc}
     0.168 & 0.403\,e^{-i\,3.9^\circ} & -0.293\,e^{i\,0.7^\circ} \\
     0.403\,e^{i\,3.9^\circ} & 1.119 & -0.762\,e^{i\,30.1^\circ} \\
     -0.293\,e^{-i\,0.7^\circ} & ~~ -0.762\,e^{-i\,30.1^\circ} ~~
      & 6.616
    \end{array} \right) , \\
   \bm{\epsilon}_u' &= 10^{-6} \left(
    \begin{array}{ccc}
     0.000333 & ~~ -0.0330\,e^{-i\,4.7^\circ} ~~
      & 0.997\,e^{i\,61.7^\circ} \\
     -0.0330\,e^{i\,4.7^\circ} & 5.263 & 106.2\,e^{-i\,85.4^\circ} \\
     0.997\,e^{-i\,61.7^\circ} & 106.2\,e^{i\,85.4^\circ} & 28513.5
    \end{array} \right) .
\end{split}
\eeq
The contributions of these matrices to the $Z^0$-boson couplings in (\ref{gLR}) are of $\ord(v^4/\Mkk^4)$ and thus should be dropped for consistency.

\end{appendix}

\end{document}